\documentclass[12pt,a4paper]{article}

\usepackage[OT2,T1]{fontenc}

\usepackage{amsmath,amsfonts,amssymb,amsthm,mathtools}
\usepackage{mathrsfs}
\usepackage{makecell}
\usepackage{diagbox}
\usepackage{pifont}

\usepackage{braket}
\usepackage{bm}
\usepackage{cancel}
\usepackage{verbatim}
\usepackage{graphicx}
\usepackage[rightcaption]{sidecap}
\usepackage{caption}
\usepackage{subcaption}
\usepackage{here}
\usepackage{authblk}

\usepackage[a4paper, bindingoffset=0cm, left=2cm, right=2cm, top=2.5cm, bottom=2.5cm, headsep=0.5cm, footskip=1cm]{geometry}

\usepackage{parskip}
\usepackage{tikz-cd}
\usepackage{cite}
\DeclareMathAlphabet{\mathscrbf}{OMS}{mdugm}{b}{n}
\usepackage{tikz}
\usetikzlibrary{decorations.markings}

\usepackage[most]{tcolorbox}
\newtcbox{\othermathbox}[1][]{nobeforeafter, math upper, tcbox raise base, enhanced, rounded corners, colback=black!5, colframe=black, left=0.7em, top=0.4em, right=0.7em, bottom=0.5em}

\usepackage{empheq}

\usepackage{enumerate}

\usepackage{titlesec}
\usepackage[titles]{tocloft}
\definecolor{MyYellow}{RGB}{248,199,82}

\numberwithin{equation}{section}

\let\OLDthebibliography\thebibliography
\renewcommand\thebibliography[1]{
  \OLDthebibliography{#1}
  \setlength{\parskip}{0pt}
  \setlength{\itemsep}{3pt plus 0.3ex}
}

\usepackage[colorlinks,allcolors=blue]{hyperref}

\definecolor{LightBrown}{RGB}{255,188,0}
\definecolor{MiddleBrown}{RGB}{199,146,0}
\definecolor{DarkBrown}{RGB}{143,104,0}
\definecolor{DarkerBrown}{RGB}{87,62,0}
\definecolor{Purple}{RGB}{255,0,188}

\DeclareSymbolFont{usualmathcal}{OMS}{cmsy}{m}{n}
\DeclareSymbolFontAlphabet{\mathcal}{usualmathcal}

\newcommand{\be}{\begin{equation}}
\newcommand{\ee}{\end{equation}}
\newcommand{\f}{\frac}

\newcommand{\bea}{\begin{eqnarray}}
\newcommand{\eea}{\end{eqnarray}}
\newcommand{\ba}{\begin{align}}
\newcommand{\ea}{\end{align}}

\newcommand{\beq}{\begin{equation}}
\newcommand{\eeq}{\end{equation}}

\newcommand{\AdS}{\text{AdS}}
\newcommand{\CFT}{\text{CFT}}

\DeclareMathOperator{\tr}{Tr}

\usepackage{pifont}

\begin{document}

\title{%
\vspace{-15mm}
Spatially Structured Entanglement from \\ Nonequilibrium
Thermal Pure States%
}%

\renewcommand\Authfont{\normalsize}
\author[1, 2]{%
Chen Bai\thanks{\,\texttt{baichen22@mails.ucas.ac.cn}}%
}%

\renewcommand\Affilfont{\footnotesize}
\affil[1]{%
Kavli Institute for Theoretical Sciences, University of Chinese Academy of Sciences, Beijing
100190, China
}%
\affil[2]{%
Department of Physics, Princeton University, Princeton, New Jersey, 08544, USA%
}%

\author[3]{%
Mao Tian Tan\thanks{\,\texttt{mao.tan@fmf.uni-lj.si}}%
}%

\author[2, 4]{%
Bastien Lapierre\thanks{\,\texttt{bastien.lapierre@phys.ens.fr}}%
}%

\renewcommand\Affilfont{\footnotesize}

\renewcommand\Affilfont{\footnotesize}
\affil[3]{%
Department of Physics, Faculty of Mathematics and Physics, University of Ljubljana, Jadranska 21, SI-1000 Ljubljana, Slovenia
}%

\author[2]{%
Shinsei Ryu\thanks{\,\texttt{shinseir@princeton.edu}}%
}%

\affil[4]{%
Philippe Meyer Institute, Physics Department, École Normale Supérieure (ENS), Université PSL, 24 rue Lhomond, F-75231 Paris, France
}%

\date{%
\vspace{-10pt}%
\normalsize%
\today%
\vspace{-10pt}%
}%

\maketitle

\begin{abstract}
We study quantum quench dynamics in (1+1)-dimensional critical systems, starting from thermal pure states called crosscap states, and evolving them under spatially inhomogeneous Hamiltonians. 
The spatial inhomogeneity is introduced through a deformation of the Hamiltonian,
expressed as linear combinations of the generators of the $\mathrm{SL}^{(q)}(2,\mathbb{R})$ subalgebra of the Virasoro algebra.
We analyze the free massless Dirac fermion theory and holographic conformal field theory as prototypical examples of integrable and non-integrable dynamics. 
Consistent with general expectations, ``M\"obius-type'' deformations lead to thermalization in the non-integrable case, and to periodic revivals in the integrable one.
In contrast, “sine-square–type” and “displacement-type” deformations prevent both thermalization and scrambling, instead producing late-time, graph-like entanglement patterns.
These patterns emerge from the interplay between the deformed Hamiltonian and the crosscap initial state and appear to be universal: they are determined solely by the deformation profile while remaining largely insensitive to microscopic details. 
Finally, we perform a holographic calculation in three-dimensional gravity using the $\AdS_3/\CFT_2$ correspondence, which reproduces the main features of our (1+1)-dimensional study.
\end{abstract}

\newpage
\setcounter{tocdepth}{2}
\tableofcontents

\newpage

\section{Introduction}
\label{sec:intro}

Quantum quenches provide a paradigmatic setting for studying non-equilibrium dynamics in quantum many-body systems. Various types of quenches have been investigated both theoretically and experimentally
\cite{Calabrese:2005in, Cheneau_2012,Langen_2015, Kaufman_2016, Brydges_2019}, particularly to reveal whether a quantum system thermalizes, fails to do so, or exhibits scrambling.
Motivated by recent advances in quantum simulators, which enable the engineering of spatially varying Hamiltonians~\cite{PhysRevLett.124.063601, Tajik_2023}, inhomogeneous quenches have recently attracted significant interest.
In particular, 
inhomogeneous quantum quenches generated by spatially deformed (1+1)d critical Hamiltonians 
have been recently investigated~\cite{Dubail_2017_inhCFT,Allegra_2016,Dubail_2017,Gaw_dzki_2018,Wen_2018_SSD,PhysRevLett.122.020201, Lapierre:2019rwj,cn3z-vfgr,Wen:2020wee,Moosavi_2021,Wen:2021mlv,Wen:2018agb,Fang:2025rie,Fan:2020orx,Liu:2023tiq,chen2025symmetryresolvedentanglemententropy,goto2023spatialdeformationmanybodyquantum} and were shown to feature a variety of phenomena, including entanglement localization~\cite{Fan:2019upv}, inhomogeneous scrambling~\cite{cn3z-vfgr}, 
and efficient ground state cooling~\cite{Wen:2022pyj}, to name a few.
A similar setting has also been investigated recently through the lens of holographic duality~\cite{Bai:2024azk,Miyata:2024gvr,Das:2024lra,deBoer:2023lrd,Nozaki:2023fkx,Goto:2021sqx,Goto:2023wai,Mao:2025hkp,Mao:2024cnm,Das:2023xaw,bernamonti2024boundaryinducedtransitionsmobiusquenches,erdmenger2025driveninhomogeneouscfttheory,de_Boer_2022}.
Crucially, this nonequilibrium process is analytically tractable in 
(1+1)d conformal field theory (CFT), as guaranteed by Virasoro symmetry~\cite{Blumenhagen:2009zz,DiFrancesco:1997nk}. Thus, this offers a platform for exploring quantum dynamics in the absence of translation invariance.

Several classes of initial states have been studied so far. In the case where the post-quench Hamiltonian is described by a conformal field theory (CFT), the prototypical example is the Calabrese–Cardy quench~\cite{Calabrese:2005in}. In this setting, the initial state is taken to be a regularized conformal boundary state, which effectively models a short-range entangled (or gapped) state.
Other examples include local quantum quenches,
such as the splitting-joining quench
\cite{Calabrese_2007}
and the local operator quench
\cite{PhysRevLett.112.111602},
where the initial conditions break translational invariance.
In this work, we instead investigate a distinct and comparatively less explored class of initial states—crosscap states—which introduce non-orientable topology into the system, enabling new dynamical behavior under time evolution.

Crosscap states can be viewed as close analogues of the conformal boundary states used in the Calabrese–Cardy quench.
Recent works have proposed that crosscap states represent thermal pure states in one-dimensional systems with periodic boundary conditions~\cite{Yoneta_2024,Chiba:2024dch}. Their lattice counterparts, 
antipodal pair (EAP) states
\cite{Chiba:2024dch,Yoneta_2024,mestyán2025crosscapstatestunableentanglement,Caetano:2021dbh,PhysRevLett.134.210403}, 
provide pairwise maximal antipodal entanglement and are in microscopic thermal equilibrium
\cite{Mori_2018} while remaining globally atypical due to their long-range EPR structure. 
Remarkably, these long-range entangled states naturally appear as eigenstates of various local many-body Hamiltonians~\cite{PhysRevLett.134.210403, mestyán2025crosscapstatestunableentanglement}. 
Both the crosscap and
EAP states are a natural choice of pure thermal state that does not require two copies of a quantum system.
They can be thought of as a single-copy counterpart of the 
thermofield double (TFD) state,
a canonical realization of a thermal pure state,
constructed from 
two copies of any quantum system~\cite{Takahashi:1996zn}.
In CFT, the TFD state can be thought of as a particular example of (smeared) conformal boundary states. 
In gravity, 
TFD corresponds to the two-sided BTZ black hole, while the crosscap corresponds to the single-copy counterpart 
(geon).

As a first step toward understanding the dynamics of thermal pure states, quantum quench protocols in (1+1)d systems starting from crosscap states were recently investigated~\cite{Chalas:2024yts, Wei:2024kkp}. In these works, the post-quench evolution was assumed to be translation invariant, and was applied in different settings, including random circuits, free fermions, and holography.
In the present work, 
we further explore the ``crosscap quench'' in (1+1)d CFT, 
by considering, in addition to 
the regular CFT Hamiltonian, 
inhomogeneous post-quench 
Hamiltonians consisting of 
the generators of the
$\mathrm{SL}(2, \mathbb{R})$
and $\mathrm{SL}^{(q)}(2,\mathbb{R})$
subalgebra of the Virasoro algebra.
One of our main findings is that the interplay between the crosscap state and the inhomogeneous time evolution leads, at late times, to a graph-like, long-ranged, spreading of entanglement in the system. In fact, 
with inhomogeneous evolution, the velocity of the EPR entangled pairs becomes position-dependent~\cite{Goto:2021sqx,Nozaki:2023fkx}.
Consequently, spatial locations where the local energy density vanishes act as sinks for EPR pairs.
When starting from the crosscap state, such entanglement hotspots give rise to a long-ranged and localized entanglement structure that is largely insensitive to microscopic details.

Besides generalizing nonequilibrium thermal pure state dynamics to inhomogeneous time evolution, our aim is to contrast integrable (rational) and holographic (``chaotic'') critical systems. 
As a representative case of the former, we consider the (1+1)d free massless Dirac fermion theory.
The entanglement evolution of this integrable model is expected to be well-described by the quasiparticle picture. We demonstrate that both the free Dirac fermion theory and the quasiparticle picture predict the same graph-like entanglement structure, in perfect agreement with each other.
On the other hand,
as a non-integrable archetype, we also consider a (1+1)d holographic CFT~\cite{Brown:1986nw,Hartman_2014}: a strongly interacting, large central charge theory with a sparse light spectrum and a $(2+1)$d gravity dual, known for rapid entanglement growth and efficient mixing~\cite{Hartman:2013qma,PhysRevLett.112.011601}. While this nonintegrable setting is known not to be described by the quasiparticle picture but by a membrane tension picture~\cite{PhysRevX.7.031016, jonay2018coarsegraineddynamicsoperatorstate, Mezei_2018, Zhou_2019, 
Kudler_Flam_2020}, we find that, in some cases, the emergent entanglement structure nevertheless retains a graph-like organization predicted by the quasiparticle picture, suggesting a robustness beyond the integrable scenarios.

The remainder of the paper is organized as follows. In Sec.~\ref{sec:EE-QP}, we introduce the EAP and crosscap states, define the class of spatially deformed $\text{SL}^{(q)}(2,\mathbb{R})$ Hamiltonians under consideration, and summarize the analytical methods employed in this work. In Sec.~\ref{sec:uniformcrosscap}, we analyze the uniform time evolution of the crosscap state as a preliminary step toward understanding inhomogeneous crosscap quenches. In Sec.~\ref{sec:Inh-Quench} we analyze the dynamics of entanglement entropy and mutual information for the different classes of $\text{SL}^{(q)}(2,\mathbb{R})$ quench Hamiltonians, and contrast their late-time behaviors. 
Furthermore, we discuss how graph-like entanglement patterns emerge.
In Sec.~\ref{sec:QP-EP}, we recover these entanglement patterns through the lens of the quasiparticle picture, and we show their relation to circulant graphs.
In Sec.~\ref{sec:Holo-Dual}, we carry out gravitational computations of entanglement entropy and mutual information in the $\AdS_3$ geon spacetime. 
Finally, in Sec.~\ref{sec:Conclusion} we summarize our results and outline directions for future work.

\section{Setting and Methods\label{sec:EE-QP}}
\begin{figure}[htbp]
  \centering
  \begin{subfigure}[b]{0.35\textwidth}
    \includegraphics[width=\textwidth]{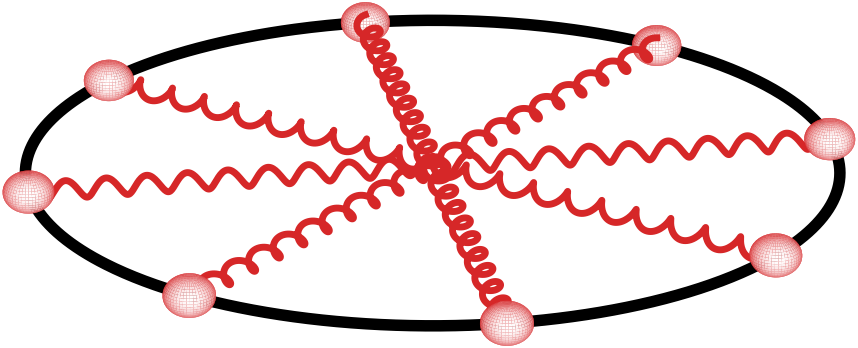}
    \caption{EAP state with $8$ sites}
    \label{fig:crosscap_lattice_3d}
  \end{subfigure}
    \hfill
  \begin{subfigure}[b]{0.6\textwidth}
    \includegraphics[width=\textwidth]{Figures_Draft/Schematic_Figures/cylinder_crosscap.pdf}
    \caption{Klein bottle}
    \label{fig:cylinder_crosscap}
  \end{subfigure}
  
  \caption{Schematics of EAP state (left panel) and path-integral construction of the Klein bottle (right panel). Left panel: red dots and lines denote the qubits and their EPR links, respectively; right panel: crosses represent the crosscap boundary conditions. 
  }
  \label{fig:Klein Bottle}
\end{figure}

We consider a one-dimensional critical system of size $L$ with periodic boundary conditions (PBC), which is described by $(1+1)$-dimensional conformal field theory ((1+1)d CFT) in the low-energy continuum regime. 
Besides the familiar Cardy boundary states that preserve half of the Virasoro symmetry~\cite{Cardy:1984bb,Ishibashi:1988kg}, (1+1)d CFTs also admit crosscap states $\{\ket{C}\}$ associated with orientation–reversing identifications, which also preserve half of the Virasoro symmetry via the condition $\left(L_q-(-1)^q\bar{L}_{-q}\right)\ket{C}=0$~\cite{Ishibashi:1988kg}. In the discretized lattice description, the crosscap states are supposed to be directly related to the entangled antipodal pair (EAP) state. For example, in a spin-$\frac{1}{2}$ lattice model, the EAP state is~\cite{Yoneta_2024,Wei:2024kkp,Chiba:2024dch} 
\be
    \ket{\text{EAP}}=\bigotimes_{i=1}^{\frac{L}{2}}\frac{1}{\sqrt{2}}\left(\ket{1}_i\ket{1}_{i+\frac{L}{2}}+\ket{0}_i\ket{0}_{i+\frac{L}{2}}\right),
\ee
where $\ket{0}_i$ and $\ket{1}_i$ denote the spin-down and spin-up states at the $i$-th site; see Fig.~\ref{fig:crosscap_lattice_3d} for a schematic. In the transverse-field Ising model (TFIM), the EAP states have been shown to flow, in the low-energy continuum (scaling) limit, to the crosscap states within the Neveu-Schwarz sector of the Ising CFT~\cite{Zhang:2024rnh}. EAP and crosscap states are highly entangled yet athermal (out of equilibrium), and therefore can further thermalize. They thus serve as an example of pure states with volume law entanglement, whose thermalization properties after a quantum quench can be nontrivial.

In this work, we mainly focus on (1+1)d CFT and crosscap state $\ket{C}$, which are analytically tractable. We consider the regularized (by an Euclidean time evolution) crosscap state as our initial state~\cite{Wei:2024kkp}:
\be
\begin{split}
    &\ket{\Psi(t=0)}=\mathcal{N}_Ce^{-\frac{\beta}{4}H_0}\ket{C},
    \quad \mathcal{N}_C^{-2}=\langle C|e^{-\frac{\beta}{4}H_0}|C\rangle,
\end{split}
\ee
where $H_0$ denotes the uniform CFT Hamiltonian, factor $\beta/4$ plays the role of extrapolation (Euclidean smearing) length~\cite{Calabrese:2005in,Calabrese:2016xau,Calabrese:2009qy,Asplund:2015eha}. At high temperature, i.e., when $\beta$ is much smaller than any subsystem size, to leading order, entanglement entropies are crosscap-state independent: different crosscap states contribute only an $O(1)$ ``crosscap entropy'' (the analogue of boundary entropy)~\cite{Caetano:2021dbh,Wei:2024zez,Calabrese:2005in,Calabrese:2016xau,Calabrese:2009qy, Asplund:2015eha}. Therefore, we work in the high-temperature regime throughout. The mutual information is likewise crosscap-state insensitive, but its time-dependence is non-universal, set by the evolution and operator content, with revivals in integrable/free CFTs and, characteristically, a vanishing mutual information at late times in holographic large-$c$ theories. In this work, we identify the vanishing mutual information as a hallmark of thermalization and scrambling. 

The regularized crosscap state $\ket{\Psi(0)}$ encodes long-range antipodal correlations, which exhibit a characteristic decay: fast in chaotic regimes, but with recurrent revivals in integrable ones~\cite{Wei:2024kkp,Chalas:2024yts}. To isolate features intrinsic to the crosscap antipodal pairing from artifacts of homogeneous time evolution, we evolve the system with a spatially inhomogeneous CFT Hamiltonian—modulating local light-cone velocities to test whether thermalization/scrambling indicators (entanglement growth and late-time vanishing mutual-information) persist once translation-invariance is broken. 
In this work, we consider the following inhomogeneous Hamiltonian 
\be
    H_{1}=\int_0^Ldx~f(x)T_{00}(x)=\int_0^L\frac{dx}{2\pi}~f(x)\left[T(x)+\bar{T}(x)\right],
\ee
where $f(x)$ is a smooth envelope function modulating the energy density $T_{00}(x)$ (equivalently, the chiral/non-chiral components $T,\bar{T}$). When $f(x)=1$, $H_1$ reduces to the uniform Hamiltonian $H_0$ corresponding to the uniform crosscap quench, see~\cite{Wei:2024kkp} and Sec.~\ref{sec:uniformcrosscap}. The quenched state we study throughout is
\be
    \ket{\Psi(t)}=\mathcal{N}_C\, e^{-it H_1}\, e^{-\frac{\beta}{4}H_0}\ket{C},
\ee
where we assume $\beta<\frac{L}{2}$ throughout.
Here, we are interested in the $\text{SL}^{(q)}(2,\mathbb{R})$ deformation~\cite{Han:2020kwp}, i.e.,
\be
    f(x)=\sigma^0+\sigma^+\cos\left(\frac{2q\pi x}{L}\right)+\sigma^-\sin\left(\frac{2q\pi x}{L}\right),
    \quad q\in\mathbb{Z},
    \quad \sigma^0,\sigma^{\pm}\in\mathbb{R},
\ee
such that $H_1$ and $H_0$ can be written as
\be\label{eq:SL2-Hamiltonian}
\begin{split}
    &H_0=\frac{2\pi}{L}\left(L_0+\bar{L}_0-\frac{c}{12}\right),\quad L_{q,+}=\frac{L_q+L_{-q}}{2},
    \quad L_{q,-}=\frac{L_q-L_{-q}}{2i},\\
    &H_1=\frac{2\pi}{L}\left[\sigma^0(L_0+\bar{L}_0)+\sigma^+(L_{q,+}+\bar{L}_{q,+})+\sigma^-(L_{q,-}+\bar{L}_{q,-})\right]-\frac{c\pi}{6L},
\end{split}
\ee
where $L_q = \frac{c}{24}\delta_{q,0} + \frac{L}{2\pi}\int_0^L \frac{dx}{2\pi}, e^{i\frac{2\pi q x}{L}} T(x)$ are the Virasoro generators, and $c$ is the central charge of the (1+1)d CFT. Eq.~\eqref{eq:SL2-Hamiltonian} generates M\"obius transformations actings on the complex coordinates as $z^q\to\frac{A(t)z^q+B(t)}{C(t)z^q+D(t)}$ and $\bar{z}^q\to\frac{\bar{A}(t)\bar{z}^q+\bar{B}(t)}{\bar{C}(t)\bar{z}^q+\bar{D}(t)}$, where $\left(z=e^{\f{2\pi w}{L}},\bar{z}=e^{\f{2\pi\bar{w}}{L}}\right)$~\cite{Fan:2020orx}. Different choices of $\{\sigma^0,\sigma^{\pm}\}$ can be classified into three distinct types of Hamiltonian based on the quadratic Casimir invariant~\cite{Ishibashi:2015jba,Ishibashi:2016bey}:
\be
\begin{split}
    &\Delta^{(2)}:=-(\sigma^0)^2+(\sigma^+)^2+(\sigma^-)^2~\begin{cases}
        <0~&\text{Non-heating Phase (Elliptic)}\\
        =0~&\text{Critical Point (Parabolic)}\\
        >0~&\text{Heating Phase (Hyperbolic)}\\
    \end{cases},
\end{split}
\ee
where hyperbolic, parabolic, and elliptic refer to the classification of the corresponding M\"obius transformations acting on $(z^q,\bar{z}^q)$. The associated dynamical phases—heating, critical, and non-heating—are diagnosed by, respectively, linear-in-time growth, logarithmic-in-time growth, and temporal oscillations of the energy~\cite{Wen:2018agb}. We note that this classification applies not only to static Hamiltonians, but also to time-periodic Hamiltonians. In fact, by designing a time-dependent Hamiltonian with the same underlying algebra, one can realize the different dynamical phases mentioned above, which are then classified by the (static) Floquet Hamiltonian~\cite{Fan:2019upv,Lapierre:2019rwj,Wen:2020wee,deBoer:2023lrd,Lapierre:2020ftq,Han:2020kwp,cn3z-vfgr}.  
In the critical and heating phase cases, $f(x)$ has roots, which correspond to fixed points of the evolution operator, and which act as sources or sinks for quasiparticles depending on the stability of the fixed point. 
More generally, inhomogeneous CFT evolution can be interpreted as evolution in a curved spacetime, which has been used to analyze transport and entanglement spreading with spatially varying velocity~\cite{Wen:2018agb, Moosavi_2021, Lapierre:2019rwj, Nozaki:2023fkx,Goto:2021gve,Goto:2021sqx,Goto:2023wai,Bai:2024azk,Jiang:2024hgt, erdmenger2025driveninhomogeneouscfttheory}. 
Concretely, throughout this work, we consider three representative examples for the three classes, namely
\be\label{eq:Envelope-Fn}
\begin{split}
    &f(x)=\begin{cases}
        1-\tanh(2\theta)\cos\left(\frac{2q\pi x}{L}\right)~&q\text{-M\"obius Hamiltonian in elliptic class},\\
        2\sin^2\left(\frac{q\pi x}{L}\right)~&q\text{-SSD Hamiltonian in parabolic class},\\
        \sin\left(\frac{2q\pi x}{L}\right)~&q\text{-Displacement Hamiltonian in hyperbolic class}.\\
    \end{cases}
\end{split}
\ee

Our goal is to study the non-equilibrium dynamics after an inhomogeneous quantum quench starting from a crosscap state, for both holographic CFTs (chaotic) with large central charge $c\gg 1$ and free Dirac fermion CFT (integrable) with $c=1$. We will analytically derive the time evolution of both entanglement entropy and mutual information through different angles, using the twist-field formalism~\cite{Calabrese:2004eu,Cardy:2007mb,Calabrese:2009qy}, the quasiparticle picture~\cite{Calabrese:2009qy,Calabrese:2005in,Calabrese:2016xau} and the Ryu-Takayanagi/Hubeny-Rangamani-Takayanagi (RT/HRT) formula~\cite{Ryu:2006bv,Hubeny:2007xt}.

\subsection{Twist-field Calculations of Entanglement Entropy and Mutual Information}
From the Euclidean path integral perspective, the system is defined on a cylinder of circumference $L$ (labeled by the spatial coordinate $x\sim x+L$) and length $\frac{\beta}{2}$ (labeled by the Euclidean time $t_E\in[0,\beta/2]$), and two crosscaps are inserted at $t_E=0, \frac{\beta}{2}$ as shown in Fig.~\ref{fig:cylinder_crosscap}. Such a manifold is thus non-orientable: it is the Klein bottle $\mathbb{K}^2$ with a moduli parameter $\tau_{\text{mod.}}=i\frac{\beta}{L}$~\cite{Wei:2024zez,Blumenhagen:2009zz,Wei:2024kkp}. Thereafter, we introduce the cylinder coordinates as $w=t_E+ix,\bar{w}=t_E-ix$. Note that $t_E\neq\tau= it$ in general, as $t_E$ is generated by $H_0$ while $\tau= it$ is generated by $H_1$.

For primary operators $\mathcal{O}_i$ with conformal dimension $\left(h_{\mathcal{O}_i},\bar{h}_{\mathcal{O}_i}\right)$, one finds
\be\label{eq:Heisenberg-Evolution-K2}
    \bra{\Psi(t)}\prod_{i}\mathcal{O}_i(w_i,\bar{w}_i)\ket{\Psi(t)}=\prod_{i}\left(\frac{dw_i^{\text{new}}}{dw_i}\right)^{h_{\mathcal{O}_i}}\left(\frac{d\bar{w}_i^{\text{new}}}{d\bar{w}_i}\right)^{\bar{h}_{\mathcal{O}_i}}\left\langle\prod_i\mathcal{O}_i\left(w_i^{\text{new}}+\frac{\beta}{4},\bar{w}_i^{\text{new}}+\frac{\beta}{4}\right)\right\rangle_{\mathbb{K}^2},
\ee
where $\langle\cdots\rangle_{\mathbb{K}^2}$ denotes correlators on 
the Klein bottle $\mathbb{K}^2$, 
and the post-quench coordinates $(w^{\text{new}},\bar{w}^{\text{new}})$ are defined by
\be\label{eq:Post-Quench-Coordinate}
    e^{itH_1}\mathcal{O}(w,\bar{w})e^{-itH_1}=\left(\frac{dw^{\text{new}}}{dw}\right)^{h_{\mathcal{O}}}\left(\frac{d\bar{w}^{\text{new}}}{d\bar{w}}\right)^{\bar{h}_{\mathcal{O}}}\mathcal{O}(w^{\text{new}},\bar{w}^{\text{new}}),
\ee 
and we further define $w^{\text{new}}_x=ix,\bar{w}^{\text{new}}_x=-ix$. Using the replica trick and the twist-field formalism~\cite{Calabrese:2004eu,Cardy:2007mb,Calabrese:2009qy}, the $n$th R\'enyi entanglement entropy of subsystem $A=[X_2,X_1] \subset[0,L]$ is given by
\be\label{eq:Single-Renyi-EE}
\begin{split}
    S_{A}^{(n)}(t)&=\frac{1}{1-n}\log\left\langle\Psi(t)\left|\,
    \sigma_n(t=0,X_2)\,
    \bar{\sigma}_n(t=0,X_1)\, \right|\Psi(t)\right\rangle.\\
\end{split}
\ee
The von Neumann entanglement entropy for the subsystem $A$ is then obtained by taking the replica limit $S_{A}(t)=\lim_{n\to 1}S_{A}^{(n)}(t)$. Here, $\sigma_n,\bar{\sigma}_n$ are non-chiral primary twist-field operators with conformal dimension $\left(h_n=\frac{c(n^2-1)}{24n},\bar{h}_n=h_n\right)$. 
On the other hand, for subsystems of the form $A\cup B$ with $A=[X_2,X_1],B=[Y_2,Y_1]$ and $0<X_2<X_1<Y_2<Y_1<L$, the entanglement entropy reads
\be\label{eq:Double-Interval-EE}
\begin{split}
    &S_{A\cup B}^{(n)}(t)=\frac{1}{1-n}\log\left\langle\Psi(t)\left|\sigma_n(0,X_2)\bar{\sigma}_n(0,X_1)\sigma_n(0,Y_2)\bar{\sigma}_n(0,Y_1)\right|\Psi(t)\right\rangle,\\
    &S_{A\cup B}(t)=\lim_{n\to 1}S_{A\cup B}^{(n)}(t).
\end{split}
\ee
We define the mutual information and its $n$th R\'enyi generalization between $A$ and $B$ as follows:
\be\label{eq:Mutual-Information}
\begin{split}
    I_{A,B}^{(n)}(t)=S_{A}^{(n)}(t)+S_{B}^{(n)}(t) - S_{A\cup B}^{(n)}(t),
    \quad I_{A,B}(t)=S_{A}(t)+S_{B}(t) - S_{A\cup B}(t).
\end{split}
\ee
For our purposes, the antipodally symmetric subsystem is of particular interest, i.e., $B=\overline{A}$, where $\overline{A}\equiv[X_4,X_3]$ with $X_3=X_1+L/2,~X_4=X_2+L/2$ is the antipodal counterpart of $A$.

It is worth mentioning that whenever $H_1$ is of the form~\eqref{eq:SL2-Hamiltonian}, the post-quench complex coordinate $(w^{\text{new}},\bar{w}^{\text{new}})$ simply reads~\cite{Han:2020kwp,Wen:2020wee,Fan:2020orx}
\be\label{eq:Post-Quench-Coordinate-Form}
\begin{split}
    &w^{\text{new}}=\frac{L}{2q\pi}\log\left(\frac{A(t)z^q+B(t)}{C(t)z^q+D(t)}\right),
    \quad \bar{w}^{\text{new}}=\frac{L}{2q\pi}\log\left(\frac{\bar{A}(t)\bar{z}^q+\bar{B}(t)}{\bar{C}(t)\bar{z}^q+\bar{D}(t)}\right),\\
\end{split}
\ee
where $\left(z=e^{\frac{2\pi w}{L}},\bar{z}=e^{\frac{2\pi \bar{w}}{L}}\right)$. Functions $A(t),B(t),C(t),D(t)$ and their anti-holomorphic partners encode all dynamical information.

\subsection{Quasiparticle Picture\label{sec:QP}}

The entanglement dynamics of integrable systems can be quantitatively captured by the quasiparticle picture~\cite{Calabrese:2009qy,Calabrese:2005in,Calabrese:2016xau}. Recently, this method has been generalized to the inhomogeneous evolution~\cite{Goto:2021sqx,Goto:2023wai,Nozaki:2023fkx,Goto:2021gve} and to the uniform crosscap quench~\cite{Chalas:2024yts}. In the following, we apply inhomogeneous evolution to the crosscap quench and thereby extend the quasiparticle picture to inhomogeneous crosscap quenches in (1+1)d integrable systems. 

\begin{figure}[htbp]
  \centering
  \begin{subfigure}[b]{0.4\textwidth}
\centering  

\includegraphics[width=0.8\textwidth]{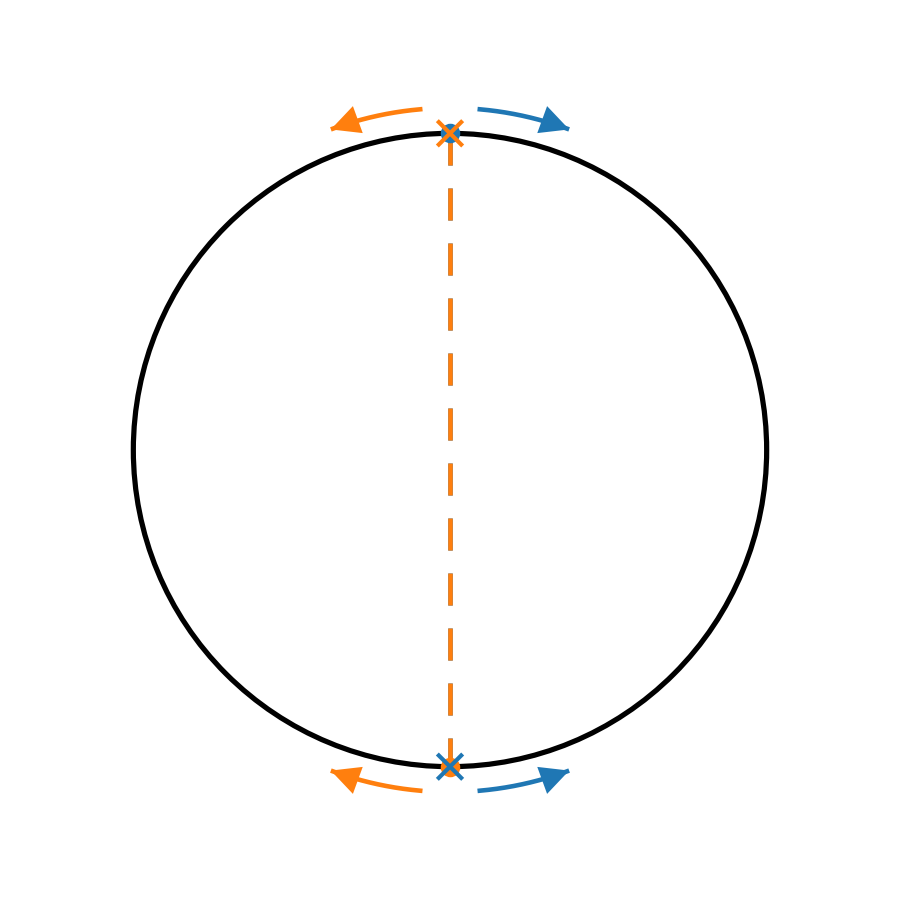}
    \caption{Antipodal entangled pair at $t=0$}
    \label{fig:Schematic QP1}
  \end{subfigure}
  \begin{subfigure}[b]{0.4\textwidth}
  \centering
    \includegraphics[width=0.8\textwidth]{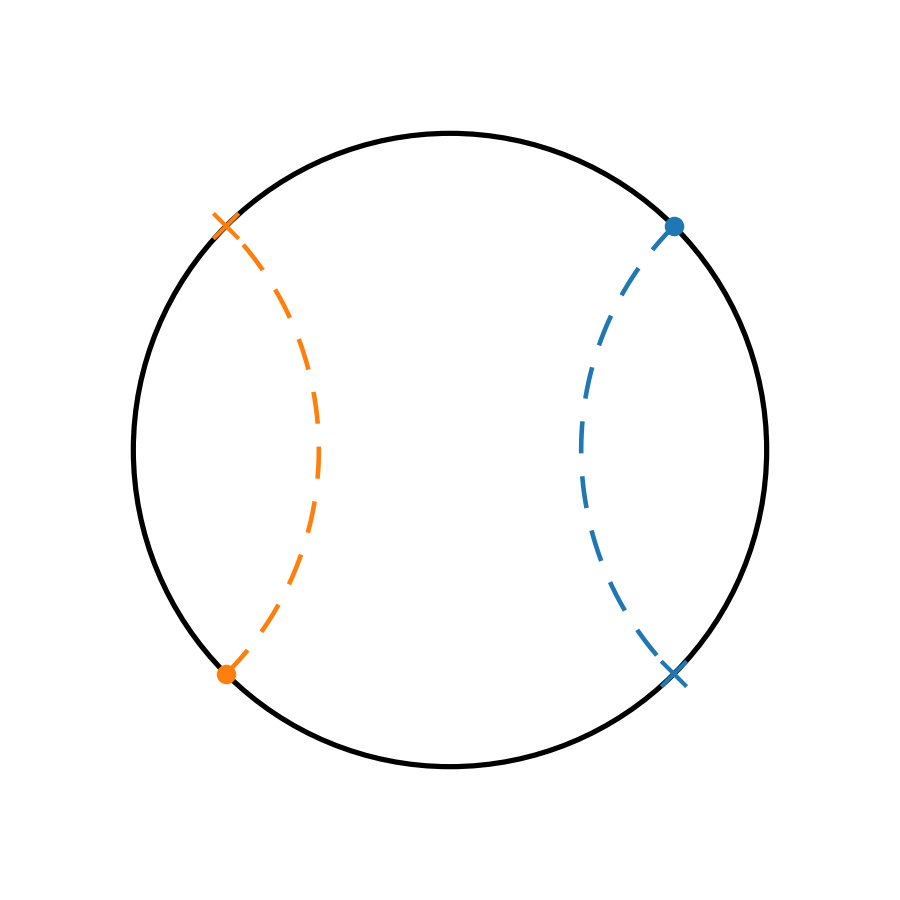}
    \caption{Positions of quasiparticles at $t>0$}
    \label{fig:Schematic QP2}
  \end{subfigure}
  
  \caption{Schematic of the quasiparticle motion. Throughout this work, the crosses and dots represent the right- and left-movers, respectively. The dashed lines indicate the EPR links between quasiparticle pairs. The coloring does not carry any physical meaning; it merely labels distinct quasiparticle pairs.}
  \label{fig:Schematic QP1-2}
\end{figure}

The quasiparticle picture for crosscap quenches can be understood as follows (see schematic Fig.~\ref{fig:Schematic QP1-2}): the initial state $\ket{\Psi(0)}$ in general is a state with finite energy density 
and thus acts as a source of quasiparticle pair creation. At $t=0$, the quantum quench uniformly generates entangled pairs of quasiparticle across the entire system.
Two EPR pairs emitted from the antipodal points $x$ and $x+L/2$ each consist of a left- and a right-mover, whereas excitations from non-antipodal locations are uncorrelated~\cite{Chalas:2024yts}. At any position $x$, one pair contributes a left-mover and the other contributes a right-mover. Once created, the number of quasiparticles remains fixed, and the pairs move ballistically. In contrast to the uniform quench, where quasiparticles propagate at a constant light speed $v=1$, inhomogeneous dynamics leads to a position-dependent group velocity, such that $v(x)=\pm f(x)$~\cite{Goto:2021sqx,Goto:2023wai,Nozaki:2023fkx,Goto:2021gve}.

The quasiparticle picture can thus be employed to predict the entanglement time evolution. For a generic subsystem $\mathcal{V}$ (which can be, e.g., $A$ or $A\cup B$), the entanglement entropy of $\mathcal{V}$, denoted as $S_{\mathcal{V}}(t)$, is given by the number of entangled quasiparticle pairs shared by $\mathcal{V}$ and its complement $\mathcal{V}^c$. Suppose that $x_{0;i=L,R}(t,x)$ is the initial location of a quasiparticle (moving from right, left) at position $(x,t)$. Moreover, let $x_{0;i=R,L}(t,\mathcal{V}),x_{0;i=R,L}(t,\mathcal{V}^c)$ be the images of the functions $x_{0;i=R,L}$ supported on the subsystem $\mathcal{V}$ and on its complement $\mathcal{V}^c$, respectively. Then, the right(left)-movers inside $\mathcal{V}$ 
at a given time $t$ are supported on 
$\overline{x_{0;R(L)}(t,\mathcal{V})}$ and on $x_{0;L(R)}(t,\mathcal{V}^c)$, where the overline indicates the antipodal counterpart.
As a result, the $n$th R\'enyi entanglement entropy is approximately given by 
\begin{equation}\label{eq:QP-EE}
    S_{\mathcal{V}}^{(n)}(t) = \rho_0^{(n)}\left\{l\left[x_{0;L}(t,\mathcal{V}^c)\cap \overline{x_{0;R}(t,\mathcal{V})}\right]+l\left[x_{0;R}(t,\mathcal{V}^c)\cap \overline{x_{0;L}(t,\mathcal{V})}\right]\right\},
\end{equation}
where $\rho_0^{(n)}=\rho_{0,L}^{(n)}=\rho_{0,R}^{(n)}$ is the initial density of quasiparticles, $\rho_{0;i=L,R}^{(n)}$ are densities for left and right movers and $l[\cdots]$ represents the size of subsystem ``$\cdots$''. If the local Hilbert space dimension is finite, 
e.g., two-dimensional,
each quasiparticle pair contributes to a factor $\log 2$ to the entanglement entropy. Since our framework is that of a continuum field theory whose local Hilbert space is infinite-dimensional, we do not count quasiparticles exactly. Instead, we define the initial quasiparticle density $\rho_0^{(n)}$ at $t=0$ through the thermodynamic entropy of $\mathcal{V}$,
\be\label{eq:QP-density}
\begin{split}
    &S_{\mathcal{V};\text{thermal}}^{(n)} = \frac{c}{6}\frac{n+1}{n}\log\left( \frac{\beta}{\pi}\sinh\frac{\pi\cdot l_{\text{min}}[\mathcal{V}]}{\beta}\right) \approx \frac{c}{6} \frac{n+1}{n}\frac{\pi l_{\text{min}}[{\mathcal{V}}]}{\beta},\\
    &\rho_0^{(n)}=\rho_{0;L}^{(n)}=\rho_{0;R}^{(n)}=\frac{S_{\mathcal{V};\text{thermal}}^{(n)}}{2l[{\mathcal{V}}]}=\frac{c}{12} \frac{n+1}{n}\frac{\pi }{\beta},
\end{split}
\ee
where $l_{\text{min}}[\mathcal{V}]=\text{Min}\left\{ l[{\mathcal{V}}],L-l[{\mathcal{V}}]\right\}$.
One can also study the mutual information between two subsystems $A$ and $B$ by counting the number of EPR pairs shared by $A$ and $B$. This leads to
\begin{equation}\label{eq:QP-MI}
    I_{A,B}^{(n)}(t) =  2\rho_0^{(n)}\left\{l\left[x_{0;L}(t,B)\cap \overline{x_{0;R}(t,A)}\right]+l\left[x_{0;R}(t,B)\cap \overline{x_{0;L}(t,A)}\right]\right\}.
\end{equation}
We stress that the quasiparticle approximation yields entanglement entropies and mutual information of order $\beta^{-1}$. Consequently, it does not capture behavior beyond or below the $\beta^{-1}$-scale.
\section{Warm-Up: Uniform Crosscap Quench\label{sec:uniformcrosscap}}
As a first step, we investigate the uniform crosscap quench with $f(x)=1$ and $H_1=H_0$~\cite{Wei:2024kkp}.
In this case, the uniform time evolution maps $(w,\bar{w})$ to $\left(w^{\text{new}}=w+it,~\bar{w}^{\text{new}}=\bar{w}+it\right)$, such that $\frac{dw^{\text{new}}}{dw}=\frac{d\bar{w}^{\text{new}}}{d\bar{w}}=1$. Therefore, by~\eqref{eq:Heisenberg-Evolution-K2},~\eqref{eq:Single-Renyi-EE} and~\eqref{eq:Double-Interval-EE}, the entanglement entropies are given by
{\small
\be\label{eq:Uniform-EE-MI}
\begin{split}
    S_A(t)&=\lim_{n\to 1}\frac{1}{1-n}\log\left\langle\sigma_n\left(it+\frac{\beta}{4}+iX_1,it+\frac{\beta}{4}-iX_1\right)\bar{\sigma}_n\left(it+\frac{\beta}{4}+iX_2,it+\frac{\beta}{4}-iX_2\right)\right\rangle_{\mathbb{K}^2},\\
    S_{A\cup B}(t)&=\lim_{n\to 1}\frac{1}{1-n}\log\left\langle\sigma_n\left(it+\frac{\beta}{4}+iX_1,it+\frac{\beta}{4}-iX_1\right)\bar{\sigma}_n\left(it+\frac{\beta}{4}+iX_2,it+\frac{\beta}{4}-iX_2\right)\right.\\
    &~~~~~~~~~~~~~~~~~~~~~~~\times\left.\sigma_n\left(it+\frac{\beta}{4}+iY_1,it+\frac{\beta}{4}-iY_1\right)\bar{\sigma}_n\left(it+\frac{\beta}{4}+iY_2,it+\frac{\beta}{4}-iY_2\right)\right\rangle_{\mathbb{K}^2},\\
\end{split}
\ee}and the mutual information between $A$ and $B$ is given by~\eqref{eq:Mutual-Information}.

\begin{figure}[htbp]
  \centering

  \begin{subfigure}[b]{0.97\textwidth}
    \centering  
    \includegraphics[width=\textwidth]{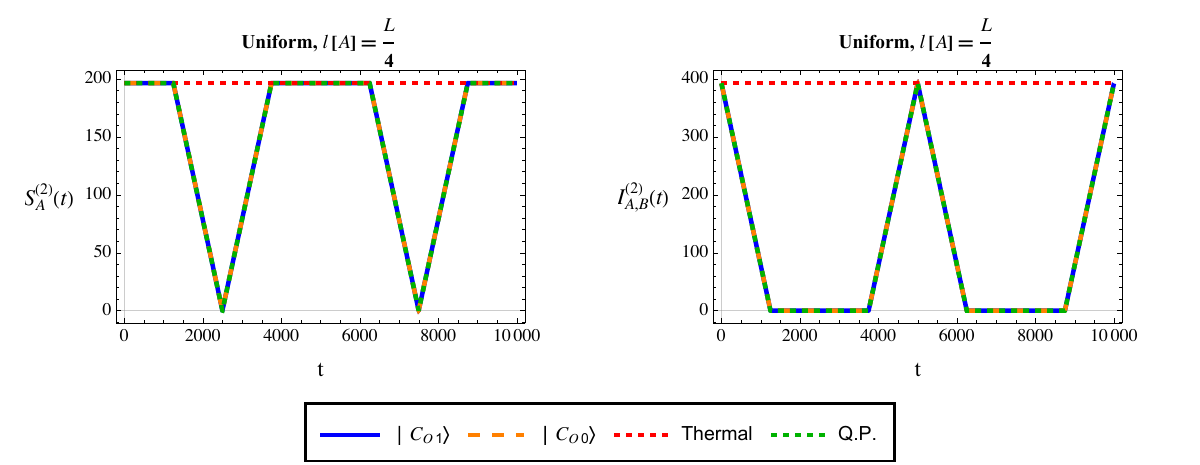}
    \caption{Free Dirac Fermion CFT.}
    \label{fig:FFCFT-Uniform-EE}
  \end{subfigure}

  \vspace{0.5cm}

     \begin{subfigure}[b]{0.97\textwidth}
    \centering  
    \includegraphics[width=\textwidth]{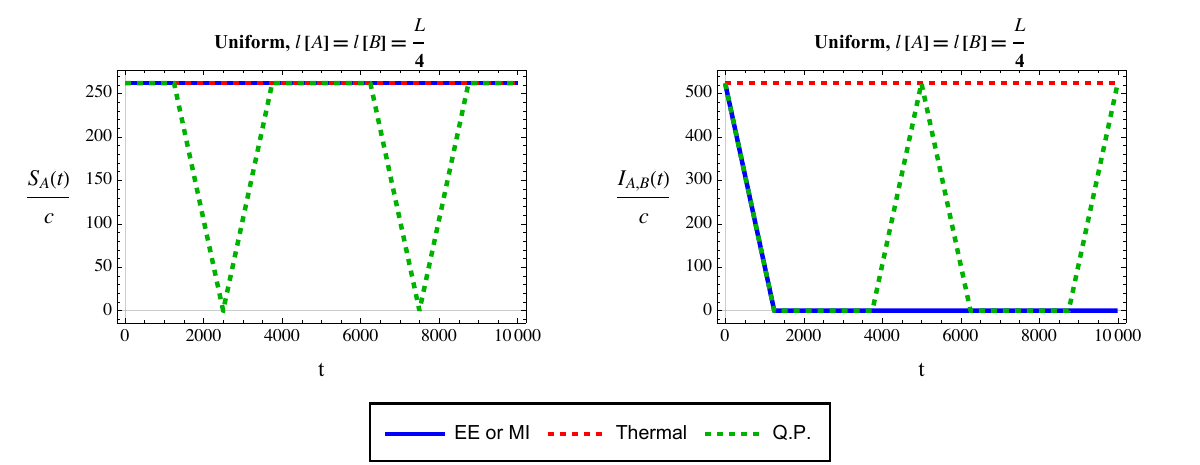}
    \caption{Holographic CFT.}
    \label{fig:HCFT-Uniform-EE}
  \end{subfigure}
  
  \caption{   
  Entanglement entropy (EE) and mutual information (MI) following a uniform quench in free fermionic and holographic theories.
  Upper panels: EE and MI for the free Dirac fermion theory. Hereafter, we set $L=10000$ for all plots. The solid blue and dashed orange curves show the EE/MI corresponding to the two crosscap states $\ket{C_{O1}}$ and $\ket{C_{O0}}$, respectively, which are defined in~\eqref{eq:Compact-Boson-Crosscap-States}. The entanglement evolution is explicitly compared with the quasiparticle prediction (dashed green), showing excellent agreement. Lower panels: EE and MI in holographic CFT. Both quantities clearly deviate from quasiparticle picture predictions; the EE saturates at the  subsystem thermal entropy (dashed red) given by~\eqref{eq:QP-density}, and the MI decays to zero.}
  \label{fig:Uniform-Crosscap-Quench}
\end{figure}

If we consider the case of holographic CFTs, i.e., the large central charge CFT $c\gg 1$, the one-point function in $\mathbb{K}^2$ is equivalent to the square-root of a two-point function defined on the torus,
\be\label{eq:Doubling-Trick}
    \langle\mathcal{O}(w,\bar{w})\rangle_{\mathbb{K}^2}=\left[\langle\mathcal{O}(w,\bar{w})\mathcal{O}^I(\beta-\bar{w}-iL/2,\beta-w+iL/2)\rangle_{\mathbb{T}^2}\right]^{\frac{1}{2}},
\ee
where the torus expectation value is the thermal expectation $\langle\cdots\rangle_{\mathbb{T}^2}=\frac{\tr(\cdots e^{-\beta H_0})}{\tr(e^{-\beta H_0})}$, and $\mathcal{O}^I$ denotes the imaged operator with the same conformal dimension as $\mathcal{O}$.\footnote{This is true for non-chiral $\mathcal{O}$, but it fails for chiral cases.} This follows from the doubling trick for the Klein bottle~\cite{Wei:2024kkp,Tsiares:2020ewp}: for a $n$-point function defined on the Klein bottle, one first computes the $2n$-point correlation function on its double cover—the torus—and then takes the square root. This is in direct analogy with the method-of-images calculation for Cardy boundary states. The entanglement entropy of $A$ and the mutual information between $A$ and $B=\overline{A}$ are then given by~\cite{Wei:2024kkp}
\be
\begin{split}
    &S_A(t)=\begin{cases}
        \frac{c}{3}\log\left(\frac{\beta}{\pi}\sinh\left[\frac{\pi\cdot l[A]}{\beta}\right]\right)~&\text{for}~l[A]<\frac{L}{2}\\
        \frac{c}{3}\log\left(\frac{\beta}{\pi}\sinh\left[\frac{\pi}{\beta}(L-l[A])\right]\right)~&\text{for}~L>l[A]>\frac{L}{2}\\
    \end{cases},\\
    &I_{A,B}(t)=\begin{cases}
        2S_A(t)-\frac{2c}{3}\log\left(\frac{\beta}{\pi}\cosh\left[\frac{2\pi t}{\beta}\right]\right)~&\text{for}~0<t\lesssim\frac{1}{2}\text{Min}\left\{l[A],L-l[A]\right\}\\
        0~&\text{for}~t\gtrsim \frac{1}{2}\text{Min}\left\{l[A],L-l[A]\right\}\\
    \end{cases},
\end{split}
\ee
where $l[A]=X_1-X_2$. The single interval entanglement entropy $S_A(t)$ is time-independent and fixed at the thermal value, while $I_{A,B}(t)$ falls monotonically from its initial maximum to zero, highlighting scrambling, which is the hallmark of holographic CFTs~\cite{Asplund:2015eha}.

In the case of the free Dirac fermion, we do not use the doubling trick but instead calculate the correlation functions of vertex operators $\left\{V_{k^{(j)}_L,k^{(j)}_R}\right\}$ on $\mathbb{K}^2$, where $k_{i=L,R}^{(j=1,2,3,4)}=\pm k$ label the replica modes, and the plus (minus) sign depends on the choice of crosscap state $\ket{C}$. The computation of these correlators is lengthy, and the details are therefore presented in Appendix~\ref{app:EEFDF}. After some algebra, one finds that the single and double interval entanglement entropies read
{ \be\label{eq:Uniform-Crosscap-Quench-Fermion-EE}
\begin{split}
    &S_{A}(t)=\frac{1}{6}\log\left[\left(\frac{L}{2\pi}\right)^2\left|\frac{\theta_1\left(i\frac{w_2-w_1}{L}|i\frac{\beta}{L}\right)\theta_1\left(i\frac{\bar{w}_2-\bar{w}_1}{L}|i\frac{\beta}{L}\right)\prod_{j=1}^2\theta_2\left(i\frac{w_j+\bar{w}_j}{L}+\frac{i\beta}{2L}|i\frac{\beta}{L}\right)}{\eta^6\left(i\frac{\beta}{L}\right)\theta_2\left(i\frac{w_1+\bar{w}_2}{L}+\frac{i\beta}{2L}|i\frac{\beta}{L}\right)\theta_2\left(i\frac{\bar{w}_1+w_2}{L}+\frac{i\beta}{2L}|i\frac{\beta}{L}\right)}\right|\right],\\
    &S_{A\cup B}(t)=\\
    &\qquad\frac{1}{6}\log\left[\left(\frac{L}{2\pi}\right)^2\left| \frac{\left[\prod_{j=1}^4\theta_2\left(i\frac{w_j+\bar{w}_j}{L}+\frac{i\beta}{2L}|i\frac{\beta}{L}\right)\right]\left[\prod_{j=1}^3\theta_1\left(i\frac{w_{j+1}-w_j}{L}|i\frac{\beta}{L}\right) \theta_1\left(i\frac{\bar{w}_{j+1}-\bar{w}_j}{L}|i\frac{\beta}{L}\right)\right]}{\eta^{12}\left(i\frac{\beta}{L}\right)\left[\prod_{j=1}^3\theta_2\left(i\frac{\bar{w}_{j+1}+w_j}{L}+\frac{i\beta}{2L}|i\frac{\beta}{L}\right) \theta_2\left(i\frac{w_{j+1}+\bar{w}_j}{L}+\frac{i\beta}{2L}|i\frac{\beta}{L}\right)\right]}\right.\right.\times\\
    &\qquad\left.\left.\frac{\left[\prod_{j=1}^2\theta_2\left(i\frac{\bar{w}_{j+2}+w_j}{L}+\frac{i\beta}{2L}|i\frac{\beta}{L}\right) \theta_2\left(i\frac{w_{j+2}+\bar{w}_j}{L}+\frac{i\beta}{2L}|i\frac{\beta}{L}\right)\right]\left[\theta_1\left(i\frac{w_4-w_1}{L}|i\frac{\beta}{L}\right) \theta_1\left(i\frac{\bar{w}_4-\bar{w}_1}{L}|i\frac{\beta}{L}\right)\right]}{\left[\prod_{j=1}^2\theta_1\left(i\frac{w_{j+2}-w_j}{L}|i\frac{\beta}{L}\right) \theta_1\left(i\frac{\bar{w}_{j+2}-\bar{w}_j}{L}|i\frac{\beta}{L}\right)\right]\left[\theta_2\left(i\frac{\bar{w}_4+w_1}{L}+\frac{i\beta}{2L}|i\frac{\beta}{L}\right) \theta_2\left(i\frac{w_4+\bar{w}_1}{L}+\frac{i\beta}{2L}|i\frac{\beta}{L}\right)\right]}\right|\right],
\end{split}
\ee}which involve the Dedekind eta function $\eta(\tau)$ and Jacobi theta functions $\theta_{1,2}(z|\tau)$~\cite{DiFrancesco:1997nk}. Substituting $S_A(t)$ and $S_{A\cup B}(t)$ into~\eqref{eq:Mutual-Information}, we obtain the mutual information between $A$ and $B$. Since $|\theta_{1,2}(z+1|\tau)|=|\theta_{1,2}(z|\tau)|$, entanglement entropy and mutual information are $\frac{L}{2}$ periodic in time for the uniform crosscap quench in a free Dirac CFT. Additionally, as the free Dirac fermion theory is integrable, its entanglement dynamics is well-captured by quasiparticle picture, as shown in Fig.~\ref{fig:Uniform-Crosscap-Quench}.

We close this warm-up section with a comparison between the uniform crosscap quench and the global (boundary state) quench\footnote{The global quench corresponds to studying the time-evolved state $\ket{\Psi_B(t)}=\mathcal{N}_Be^{-it H_0}e^{-\frac{\beta}{4}H_0}\ket{B}$, where $\mathcal{N}_B=\bra{B}e^{-\frac{\beta}{2}H_0}\ket{B}^{-\frac{1}{2}}$.}~\cite{Calabrese:2005in,Calabrese:2016xau,Calabrese:2009qy,Hartman:2013qma}. In holographic CFTs, a global quench leads to linear entanglement growth, followed by saturation, for both $S_A(t)$ and $S_{A\cup B}(t)$. By contrast, after a uniform crosscap quench, only $S_{A\cup B}(t)$ grows linearly and then saturates, while $S_A(t)=S_{A;\text{thermal}}$ remains fixed at its thermal value. 
In the free Dirac fermion theory, in order to obtain the exact global quench expressions for $S_{A}(t)$ and $S_{A\cup B}(t)$, we simply 
replace
all $\theta_2$ with $\theta_1$ in~\eqref{eq:Uniform-Crosscap-Quench-Fermion-EE}. Therefore, in both cases, the entanglement entropy and mutual information are periodic in time, and thermalization is absent. Because the state $e^{-\frac{\beta}{4}H_0}\ket{B}$ is short-range entangled and has vanishing real-space entanglement (up to a divergent term)~\cite{Miyaji:2014mca}, $S_A(t)$ initially grows linearly under a global quench. In contrast, for the crosscap state, entanglement entropy starts from its maximal value and subsequently decreases, before finite-size revivals.

\section{Inhomogeneous Crosscap Quenches\label{sec:Inh-Quench}}
In this section, we investigate the time dependence of the entanglement entropy $S_{A}(t)$ and $S_{A}^{(2)}(t)$, as well as mutual information $I_{A,B}(t)$ and $I_{A,B}^{(2)}(t)$ with a spatially inhomogeneous quench. We examine the time evolution individually for the three cases discussed in Sec.~\ref{sec:EE-QP}: the non-heating $q$-M\"obius quench, the critical $q$-SSD quench, and the heating $q$-Displacement quench. Since the Jacobian factors $\frac{dw^{\text{new}}}{dw},\frac{d\bar{w}^{\text{new}}}{d\bar{w}}$ depend explicitly on time $t$, the $n$th R\'enyi entanglement entropies are given by
{\small \be\label{eq:REE-Inhomo-Expression}
\begin{split}
    S_{A}^{(n)}(t)&=\frac{h_n}{1-n}\log\left(\prod_{j=1}^2\left[\frac{dw^{\text{new}}_{j}}{dw_{X_j}}\frac{d\bar{w}^{\text{new}}_{j}}{d\bar{w}_{X_j}}\right]\right)+\frac{1}{1-n}\log\left\langle \sigma_n\left(w^{\text{new}}_{2;\frac{\beta}{4}},\bar{w}^{\text{new}}_{2;\frac{\beta}{4}}\right)\bar{\sigma}_n\left(w^{\text{new}}_{1;\frac{\beta}{4}},\bar{w}^{\text{new}}_{1;\frac{\beta}{4}}\right)\right\rangle_{\mathbb{K}^2},\\
    S_{A\cup B}^{(n)}(t)&=\frac{h_n}{1-n}\log\left(\prod_{j=1}^4\left[\frac{dw^{\text{new}}_{j}}{dw_{X_j}}\frac{d\bar{w}^{\text{new}}_{j}}{d\bar{w}_{X_j}}\right]\right)\\
    &+\frac{1}{1-n}\log\left\langle \sigma_n\left(w^{\text{new}}_{2;\frac{\beta}{4}},\bar{w}^{\text{new}}_{2;\frac{\beta}{4}}\right)\bar{\sigma}_n\left(w^{\text{new}}_{1;\frac{\beta}{4}},\bar{w}^{\text{new}}_{1;\frac{\beta}{4}}\right) \sigma_n\left(w^{\text{new}}_{4;\frac{\beta}{4}},\bar{w}^{\text{new}}_{4;\frac{\beta}{4}}\right)\bar{\sigma}_n\left(w^{\text{new}}_{3;\frac{\beta}{4}},\bar{w}^{\text{new}}_{3;\frac{\beta}{4}}\right)\right\rangle_{\mathbb{K}^2},\\
\end{split}
\ee}where $w^{\text{new}}_{j;\frac{\beta}{4}}=w^{\text{new}}_{j}+\frac{\beta}{4}$ and coordinates $(w^{\text{new}}_{j},\bar{w}^{\text{new}}_{j})$ are obtained from $(w_{X_j},\bar{w}_{X_j})$ via~\eqref{eq:Post-Quench-Coordinate}; moreover, $X_3=Y_1=X_1+L/2,~X_4=Y_2=X_2+L/2$. Inhomogeneous time evolution leaves twist-field–independent terms nonzero and time-dependent, unlike the uniform case. In the holographic theory, the twist-field correlation functions in~\eqref{eq:REE-Inhomo-Expression} are once again evaluated using the Klein-bottle doubling trick (i.e., by computing them on the torus double cover and taking the square root). By contrast, for the free Dirac fermion, they are obtained from vertex-operator correlation functions on the Klein bottle, in a similar spirit as Sec.~\ref{sec:uniformcrosscap}. Once again, technical details are deferred to Appendix~\ref{app:EEFDF}. Specifically, we analyze the von Neumann entanglement entropy $S_A(t)$ and the mutual information $I_{A,B}(t)$ in holographic CFTs, as well as the second R\'enyi entanglement entropy $S_A^{(2)}(t)$ and its associated mutual information $I_{A, B}^{(2)}(t)$ in the free Dirac fermion theory. Furthermore, we use the quasiparticle picture to benchmark our analytical results. Although the quasiparticle picture is generally valid for free Dirac fermions and fails for holographic CFTs, we find that inhomogeneous time evolution significantly extends its applicability in the holographic setting (see Table~\ref{tab:QP-HOLO}).

\begin{table}[h!]
\centering
\label{tab:QP_description}
\begin{tabular}{|c|c|c|}
\hline
Hamiltonian & Not described by QP & Described by QP \\
\hline
Non-heating        & Always & Never \\
\hline
Critical       & $q\leq2$, or $A\cup B$ includes only one or all fixed points & All other cases \\
\hline
Heating  & $q=1$, or $A\cup B$ includes all fixed points & All other cases \\
\hline
\end{tabular}
\caption{Summary of the applicability of the quasiparticle picture (QP) to describe the mutual information evolution for crosscap quenches in holographic CFTs, for the three classes of inhomogeneous Hamiltonians studied in this section.}
\label{tab:QP-HOLO}
\end{table}

\subsection{$q$-M\"obius (Non-Heating) Crosscap Quench}
We start with the $q$-M\"obius Hamiltonian, for which the deformation $f(x)$ is given by~\eqref{eq:Envelope-Fn}, which corresponds to setting $\sigma^0=1,\sigma^+=-\tanh(2\theta),\sigma^-=0$ in~\eqref{eq:SL2-Hamiltonian}. Accordingly, the post-quench coordinates~\eqref{eq:Post-Quench-Coordinate-Form} are determined by
\be\label{eq:Post-Quench-Coordinate-Coe-Mobius}
\begin{split}
    &A(t)=\bar{A}(t)=[1-\lambda(t)]\cosh(2\theta)-[1+\lambda(t)],
    \quad B(t)=\bar{B}(t)=-\left[1-\lambda(t)\right]\sinh(2\theta),\\
    &C(t)=\bar{C}(t)=[1-\lambda(t)]\sinh(2\theta),
    \quad D(t)=\bar{D}(t)=-\left([1-\lambda(t)]\cosh(2\theta)+[1+\lambda(t)]\right),
\end{split}
\ee
where $\lambda(t)=\exp\left(\frac{2q\pi it}{L_{\text{eff}}}\right),~L_{\text{eff}}=L\cosh(2\theta)$. Obviously, $A(t),B(t),C(t),D(t)$ are periodic functions with period $\frac{L_{\text{eff}}}{q}$. Hence, we expect entanglement entropy and mutual information to become time-periodic\footnote{Note that the period of $S_A(t)$ and $I_{A,B}(t)$ is not $L_{\text{eff}}/q$, since Eq.~\eqref{eq:Post-Quench-Coordinate-Form} is not single-valued in time. To resolve this multi-valuedness, we introduce an angular coordinate $\left(w^{\text{new}}=\frac{iL\varphi}{2q\pi},\bar{w}^{\text{new}}=-\frac{iL\bar{\varphi}}{2q\pi}\right)$ with an appropriate branch cut. Further details are provided in Appendix~\ref{app:EEHCFT}.} in both the holographic theory and the free Dirac fermion.

\begin{figure}[htbp]
  \centering
  \begin{subfigure}[b]{0.97\textwidth}
    \centering  
    \includegraphics[width=\textwidth]{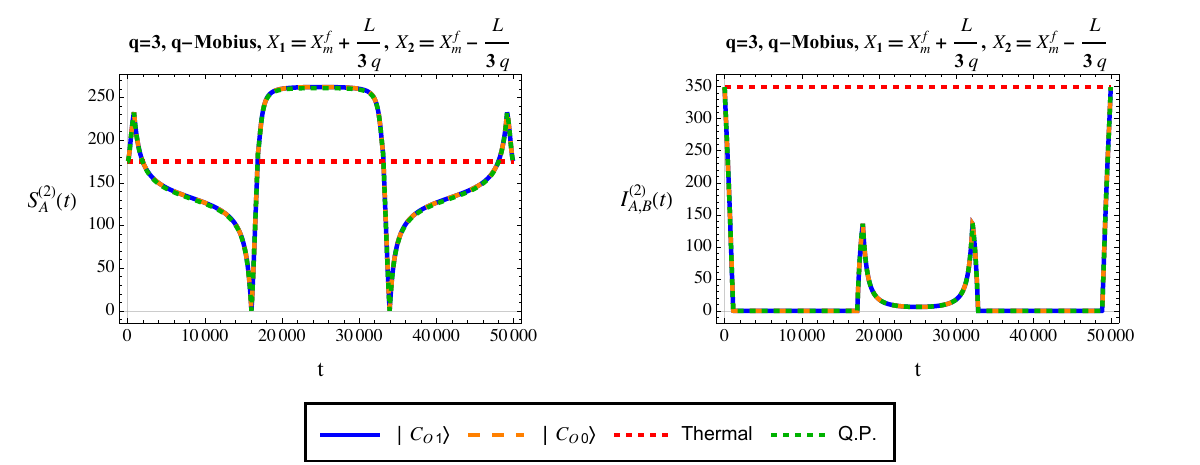}
    \caption{Free Dirac Fermion CFT.}
    \label{fig:FFCFT-qMobius-EE}
  \end{subfigure}

  \vspace{0.5cm}

  \begin{subfigure}[b]{0.97\textwidth}
    \centering  
    \includegraphics[width=\textwidth]{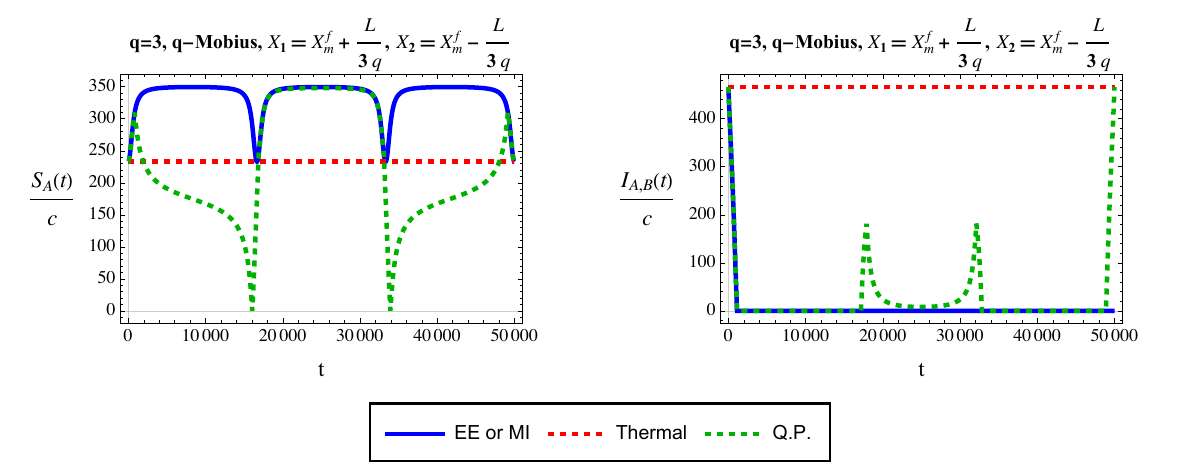}
    \caption{Holographic CFT.}
    \label{fig:HCFT-qMobius-EE}
  \end{subfigure}  
  
  \caption{
  EE and MI time evolution for $q$-M\"obius quench. Upper panels: Evolution of EE and MI in free Dirac fermion CFT. In this setup, both the EE and the MI are well described by the quasiparticle picture and exhibit periodic revivals. In contrast to the uniform quench in Fig.~\ref{fig:Uniform-Crosscap-Quench}, the EE plateau–decay behavior and the MI revival around $t\approx\frac{L}{2}$ are significantly altered by the inhomogeneous quench, where $L=10000$; lower panels: EE and MI in holographic CFT. Although the EE shows residual oscillations, the system nonetheless thermalizes and scrambles: both the EE and the MI no longer follow the quasiparticle predictions, and the MI decays monotonically to zero.}
  \label{fig:qMobius-Crosscap-Quench}
\end{figure}

For any (1+1)d holographic CFT, $S_A(t)$ exhibits a periodic decay–plateau–growth (or growth-plateau-decay) structure, as plotted in Fig.~\ref{fig:HCFT-qMobius-EE}. However, this does not imply the absence of thermalization, for two reasons. First, the uniform crosscap quench, $\theta=0$, is a special case of the $q$-M\"obius Hamiltonian which has apparent thermalizing features. Second, for finite $\theta$, we observe that $I_{A,B}(t)$ decreases monotonically to zero and does not exhibit any revival, as shown in Fig.~\ref{fig:HCFT-qMobius-EE}.
Information about the initial state is thus fully scrambled; consequently, under time evolution generated by a non-heating Hamiltonian, the (regularized) crosscap state thermalizes at late times.

In the case of (1+1)d free Dirac fermion, $S_A^{(2)}(t)$ and $I_{A,B}^{(2)}(t)$ not only exhibit periodic evolution, but also can be well-approximated using the quasiparticle picture, as shown in Fig.~\ref{fig:FFCFT-qMobius-EE}. Thus, by tracking quasiparticle trajectories, one can account for the system’s entanglement dynamics. The quasiparticles travel with the local velocity
\begin{equation}\label{eq:QuasiparticleVelocity}
    \frac{dx}{dt}=v(x) = \sigma f(x),
\end{equation}
where $\sigma=\pm1$ applies to right/left-movers. By integrating this equation for a quasiparticle with initial location $(t_0,x_0)$, we find that the trajectory under the $q$-M\"obius evolution is given by
\begin{equation}\label{MobiusTrajectory}
    \sigma(t-t_0) = \frac{L\cosh{2\theta}}{\pi q}\left(\tan^{-1}\left(e^{2\theta}\tan\frac{\pi qx}{L}\right)-\tan^{-1}\left(e^{2\theta}\tan\frac{\pi qx_0}{L}\right)\right).
\end{equation}
Next, we define
\begin{equation}\label{Quotient}
    k_\sigma = \left\lfloor\frac{1}{\pi}\cdot\left(\frac{\pi q (t-t_0)}{L\cosh{2\theta}}-\sigma\tan^{-1}\left(e^{2\theta}\tan\frac{\pi q x}{L}\right)+\frac{\pi}{2}\right)\right\rfloor,
\end{equation}
where $\lfloor\cdots\rfloor$ denotes the floor function.
The initial position for a quasiparticle at $x$ and time $t$ then reads
\begin{equation}\label{MobiusTrajectory03}
\begin{split}
    x_{0;\sigma}(x,t) &= \frac{L}{\pi q}\left(\tan^{-1}\left\{e^{-2\theta}\tan\left[\tan^{-1}\left(e^{2\theta}\tan\frac{\pi qx}{L}\right)-\frac{\pi q\sigma(t-t_0)}{L\cosh{2\theta}}\right]\right\}-\sigma k_\sigma\pi\right)\\
    &+\frac{L}{q}\left\lfloor \frac{qx}{L}+\frac{1}{2}\right\rfloor,
\end{split}
\end{equation}
where $x_0(x,t=L_{\text{eff}})=x$. In particular, the quasiparticle returns to its original position after time $L_{\text{eff}}$.
Substituting $x_{0;\sigma}(x,t)$ back into~\eqref{eq:QP-EE} and~\eqref{eq:QP-MI} with $t_0=0$, we obtain the green dashed curves shown in Fig.~\ref{fig:qMobius-Crosscap-Quench}. Consequently, under non-heating evolution, quasiparticles move periodically with period $L_{\text{eff}}$, thus the regularized crosscap state does not thermalize, and in the (1+1)d free Dirac theories, both $S_{A}^{(2)}(t),~ I_{A,B}^{(2)}(t)$ exhibit periodic revivals to their initial values.

\subsection{$q$-SSD (Critical) Crosscap Quench}
We now consider the $q$-SSD evolution, generated by the deformation $f(x)=2\sin^2\left(\frac{q\pi x}{L}\right)$. This Hamiltonian is the so-called SSD limit ($\theta\to\infty$) of the $q$-M\"obius Hamiltonian~\cite{Okunishi:2016zat}, and corresponds to parameters $\sigma^0=1,\sigma^+=-1,\sigma^-=0$ in~\eqref{eq:SL2-Hamiltonian}. For $q$-SSD time evolution,~\eqref{eq:Post-Quench-Coordinate-Form} reads
\be
\begin{split}
    &A(t)=\bar{A}(t)=L+iq\pi t,\quad B(t)=\bar{B}(t)=-iq\pi t,
    \nonumber \\
    &
    C(t)=\bar{C}(t)=iq\pi t,
    \quad D(t)=\bar{D}(t)=L-iq\pi t.
\end{split}
\ee
It immediately follows that the coordinates $(w^{\text{new}},\bar{w}^{\text{new}})$ are monotonic functions of time. Therefore, it is enough to focus on the late-time ($t\gg L$) evolution of entanglement entropy and mutual information as diagnostics of thermalization. Crucially, in the $q$-SSD case, there exist special fixed points at which the energy density vanishes, i.e., $f(x=X_m^f)=0$ with $X_m^f=\frac{mL}{q},m=0,1,\cdots,q-1$. 

\begin{figure}[htbp]
  \centering
  \begin{subfigure}[b]{0.97\textwidth}
    \centering  
    \includegraphics[width=\textwidth]{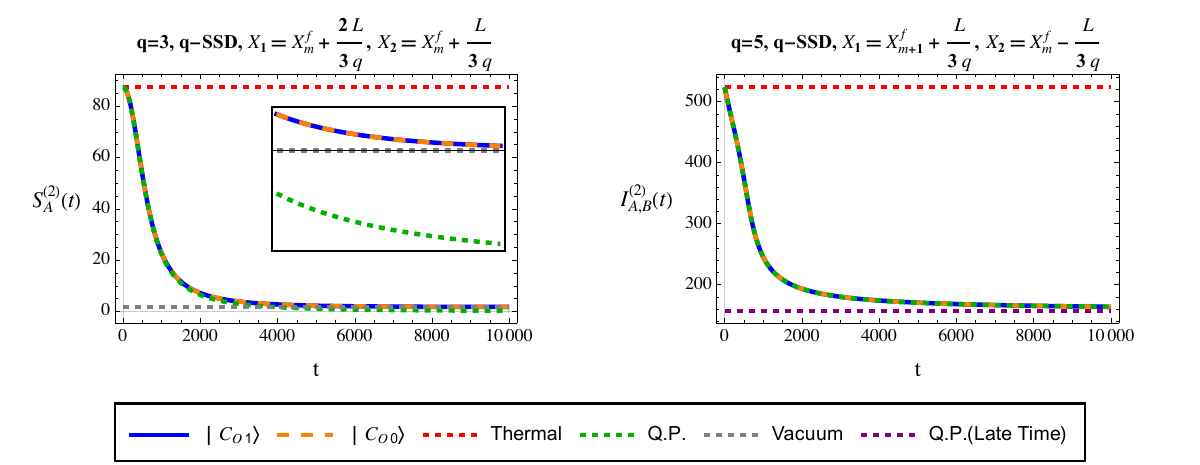}
    \caption{Free Dirac Fermion CFT.}
    \label{fig:FFCFT-qSSD-EE}
  \end{subfigure}

  \vspace{0.5cm}

  \begin{subfigure}[b]{0.97\textwidth}
    \centering  
    \includegraphics[width=\textwidth]{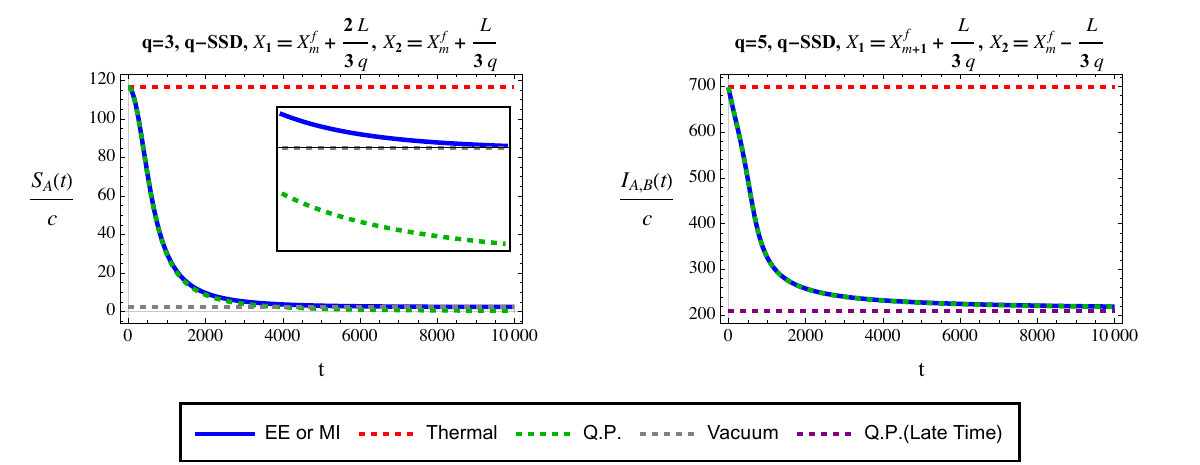}
    \caption{Holographic CFT.}
    \label{fig:HCFT-qSSD-EE}
  \end{subfigure}    
  
  \caption{
  EE and MI time evolution for $q$-SSD quench. Upper panels: EE and MI in free Dirac fermion CFT; lower panels: EE and MI in holographic CFT; left panels: conformal cooling drives $S_{A}^{(2)}(t)$ to the vacuum entanglement value (gray dashed line) given in~\eqref{eq:Conformal-Cooling-Vacuum} when $A$ contains no fixed points, in both free-fermion and holographic CFTs. The insets detail how the entanglement entropy (EE) approaches the vacuum value; the quasiparticle picture fails to capture this behavior, as its validity is restricted to order $\beta^{-1}$. Right panels: in both free-fermion and holographic CFTs, the late-time mutual information matches the graph-like pattern prediction (purple dashed line),~\eqref{eq:qSSD-Asym-MI}.}
  \label{fig:qSSD-Crosscap-Quench}
\end{figure}

Under the $q$-SSD time evolution with $q>4$, the holographic and free Dirac fermion theories share similar entanglement dynamics. If $A$ contains no fixed point, $S_A(t)$ in the holographic CFT and $S_A^{(2)}(t)$ in free Dirac theory both decrease and, at late times, asymptotically approach the $q$-SSD vacuum value (equivalently that of a CFT of size $L/q$)
\be\label{eq:Conformal-Cooling-Vacuum}
\begin{split}
    &S_{A}(t)\overset{t\gg L}{\approx} \frac{c}{3}\log\left(\frac{L}{q\pi}\sin\left[\frac{q\pi(X_1-X_2)}{L}\right]\right),\\
    &S_{A}^{(2)}(t)\overset{t\gg L}{\approx}  \frac{1}{4}\log\left(\frac{L}{q\pi}\sin\left[\frac{q\pi(X_1-X_2)}{L}\right]\right),\\
\end{split}
\ee
where $S_A(t)$ is computed analytically in Appendix~\ref{app:EEHCFT}, and $S_{A}^{(2)}(t)$ is verified numerically; see the sub-panels in Fig.~\ref{fig:qSSD-Crosscap-Quench}.
This phenomenon corresponds to the conformal cooling effect induced by the critical and heating-phase Hamiltonians~\cite{Wen:2022pyj,Goto:2021sqx,Jiang:2024hgt,Nozaki:2023fkx, cn3z-vfgr}. This cooling is a robust feature in both holographic CFTs~\cite{Goto:2021sqx,Jiang:2024hgt,Nozaki:2023fkx} and free-fermion theories~\cite{Wen:2022pyj,Nozaki:2023fkx}. In particular, it is not captured by the quasiparticle picture, since in the high-temperature regime the vacuum entanglement is parametrically below $O(\beta^{-1})$; see sub-panels in Fig.~\ref{fig:qSSD-Crosscap-Quench}. On the other hand, when $A$ includes $p$ fixed points with $1\leq p< q$, the late time expressions of $S_A(t)$ and $S_A^{(2)}(t)$ are proportional to $\log(t)$. We stress that entanglement growth here is theory-independent, and is \textit{not} a sign of thermalization. 
In fact, the $\log(t)$ terms originate from the Jacobian factor $\prod_{j=1}^2\left[\frac{dw^{\text{new}}_{j}}{dw_{X_j}}\frac{d\bar{w}^{\text{new}}_{j}}{d\bar{w}_{X_j}}\right]$ in~\eqref{eq:REE-Inhomo-Expression}, whereas thermalization and scrambling are instead encoded in the correlators themselves~\cite{Asplund:2015eha}.
We depict the time evolutions $S_A(t)$ and $S_A^{(2)}(t)$ in Fig.~\ref{fig:qSSD-Crosscap-Quench}. 
As we will now show, for $q$-SSD dynamics, both $I_{A,B}(t)$ and $I_{A,B}^{(2)}(t)$ generally approach finite late-time values, which are captured by the quasiparticle picture.

If $A$ and its antipodal partner $B=\overline{A}$ contain less than three fixed points, the mutual information either monotonically decreases and remains close to zero at late times, or grows logarithmically at late times. Alternatively, if $A\cup B$ includes no less than three fixed points, $I_{A,B}(t)$ and $I_{A,B}^{(2)}(t)$ in the large time limit are approximately given by
\be\label{eq:qSSD-Asym-MI}
\begin{split}
    &I_{A,B}(t)\overset{t\gg L}{\approx}\frac{c\pi}{3\beta}\cdot\frac{L}{q}\cdot\begin{cases}
        (2p-1)~&\text{for odd }q\\
        (2p-2)~&\text{for even }q\\
    \end{cases},\\
    &I_{A,B}^{(2)}(t)\overset{t\gg L}{\approx}\frac{\pi}{4\beta}\cdot\frac{L}{q}\cdot\begin{cases}
        (2p-1)~&\text{for odd }q\\
        (2p-2)~&\text{for even }q\\
    \end{cases},\\
    &p=\text{Min}\left\{\text{number of fixed points in $A$}, \text{number of fixed points in $B$}\right\}.
\end{split}
\ee
The asymptotic value of $I_{A,B}(t)$ can be analytically derived (see Appendix~\ref{app:EEHCFT}) and is correctly predicted by the quasiparticle picture; see Fig.~\ref{fig:HCFT-qSSD-EE}. By contrast, although the quasiparticle picture gives an intuitive prediction for $I_{A,B}^{(2)}(t)$, its late-time value is difficult to obtain directly from~\eqref{eq:REE-Inhomo-Expression-FFCFT}. The agreement between the analytical result and the quasiparticle prediction is manifest in Fig.~\ref{fig:FFCFT-qSSD-EE}. 

The trajectory and initial location of a quasiparticle at position $(x,t)$ are both obtained by taking the SSD limit in~\eqref{MobiusTrajectory} and~\eqref{MobiusTrajectory03}, respectively, i.e.,\footnote{We take the modulus with respect to $\pi$ to make the $\cot^{-1}$ function continuous over the range $(0,\pi)$, and the quotient term in~\eqref{SSDTrajectory} was added so that $m\frac{L}{q}<x<(m+1)\frac{L}{q}$ implies that $m\frac{L}{q}<x_0<(m+1)\frac{L}{q}$, which has to be the case since quasiparticles can never cross the fixed point. We use modulus for $\cot^{-1}$ but not for $\tan^{-1}$ because the former is discontinuous but the latter is not.}
\be\label{SSDTrajectory}
\begin{split}
    &\cot \frac{\pi q x_0}{L} = \cot \frac{\pi q x}{L} +\frac{2\pi q \sigma(t-t_0)}{L},\\
    &x_{0;\sigma}(x,t) = \frac{L}{\pi q} \left[\cot^{-1}\left(\cot\frac{\pi qx}{L}+\frac{2\pi q\sigma(t-t_0)}{L}\right)\mod\pi\right]+\left\lfloor \frac{q x}{L}\right\rfloor\cdot \frac{L}{q},
\end{split}
\ee
where $x_{0,\sigma}(x,t_0)=x$. Substituting $x_{0;\sigma}(x,t)$ back into~\eqref{eq:QP-EE} and~\eqref{eq:QP-MI} with $t_0=0$, we obtain the green dashed curves in Fig.~\ref{fig:qSSD-Crosscap-Quench}. From~\eqref{SSDTrajectory}, if a quasiparticle is initially at $x$ ($x\in(X_m^f,X_{m+1}^f)$) it will converge to $X_m^f$ or $X_{m+1}^f$ in the limit $t\to\infty$ depending on whether it is a left- or right-mover.

We stress that the asymptotic expressions~\eqref{eq:qSSD-Asym-MI} differ for even and odd $q$. For $q>4$, this discrepancy originates from distinct quasiparticle graph-like patterns that encode non-local entanglement in the long-time limit, because all quasiparticles end up at fixed points. These patterns arise from the interplay between the fixed points and the crosscap initial state, as we will discuss in Sec.~\ref{sec:QP-EP}. However, for $q\leq 4$, the late-time quasiparticle graph patterns in the crosscap quench either coincide with those obtained from an initial short-ranged entangled state (e.g., a boundary state), or the quasiparticle picture fails to estimate the correct mutual information (if $q=1,2$) due to the $\log(t)$ term.

\subsection{$q$-Displacement (Heating) Crosscap Quench}
As a last case, we consider $H_1$ with the deformation profile $f(x) = \sin\left(\frac{2\pi q x}{L}\right)$. In this case, the Hamiltonian corresponds to the so-called $q$-Displacement Hamiltonian, which is the Hermitian generator of coherent states~\cite{Caputa:2021sib,Lapierre:2025zsg}. This Hamiltonian corresponds to taking $\sigma^0=\sigma^+=0$, $\sigma^-=1$ in~\eqref{eq:SL2-Hamiltonian}. For any fixed $q\in\mathbb{N}^+$, $H_1$ has $2q$ fixed points at $x=X_{\frac{m}{2}}^{f}=\frac{mL}{2q}$ with $m=0,1,\cdots,2q-1$. In this case, the post-quench coordinates $(w^{\text{new}},\bar{w}^{\text{new}})$ in~\eqref{eq:Post-Quench-Coordinate-Form} are given by
\be
\begin{split}
    &A(t)=\bar{A}(t)=D(t)=\bar{D}(t)=1,\quad B(t)=-\bar{B}(t)=C(t)=-\bar{C}(t)=\tanh\left(\frac{q\pi t}{L}\right).\\
\end{split}
\ee
In analogy with the $q$-SSD evolution, the entanglement entropy and mutual information generated by the heating Hamiltonian exhibit identical dynamical features in both holographic and free-fermionic theories, and do not result in thermalization. Furthermore, the late-time asymptotic values of $S_A(t)$, $S_A^{(2)}(t)$, $I_{A,B}(t)$, and $I_{A,B}^{(2)}(t)$ can be derived analytically from both holographic CFT and the quasiparticle picture.

\begin{figure}[htbp]
  \centering
  \begin{subfigure}[b]{0.97\textwidth}
    \centering  
    \includegraphics[width=\textwidth]{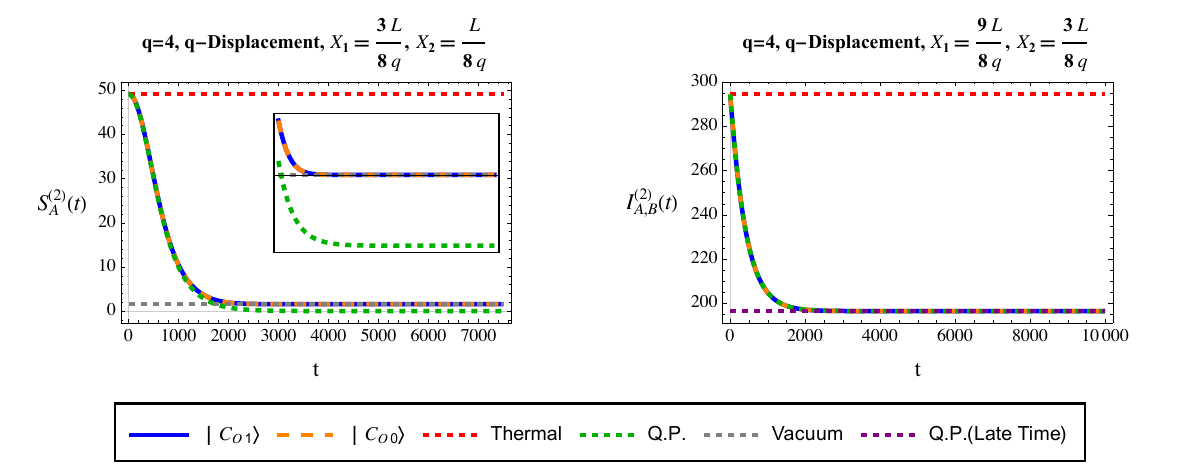}
    \caption{Free Dirac Fermion CFT.}
    \label{fig:FFCFT-qDis-EE}
  \end{subfigure}

  \vspace{0.5cm}

  \begin{subfigure}[b]{0.97\textwidth}
    \centering  
    \includegraphics[width=\textwidth]{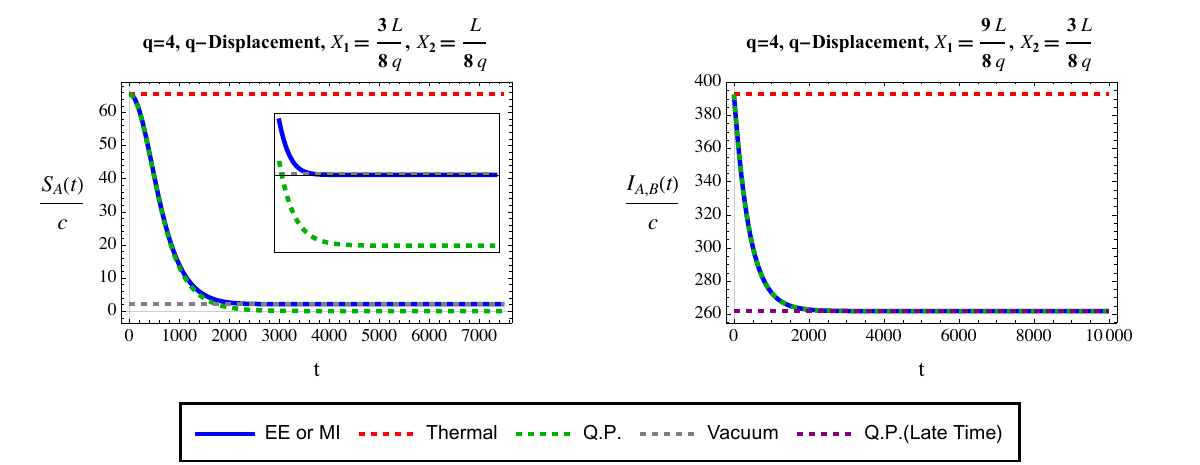}
    \caption{Holographic CFT.}
    \label{fig:HCFT-qDis-EE}
  \end{subfigure}

  \caption{
  EE and MI time evolution for $q$-Displacement quench. Upper panels: EE and MI in free Dirac fermion CFT; lower panels: EE and MI in holographic CFT; left panels: conformal cooling drives $S_{A}^{(2)}(t)$ to the vacuum entanglement value (gray dashed line) given in~\eqref{eq:Conformal-Cooling-Vacuum} when $A$ contains no fixed points, in both free-fermion and holographic CFTs. The insets detail how the entanglement entropy (EE) approaches the vacuum value; right panels: in both free-fermion and holographic CFTs, the late-time mutual information matches the graph-like pattern prediction (purple dashed line),~\eqref{eq:qDis-Asym-MI}.}
  \label{fig:qDis-Crosscap-Quench}
\end{figure}

If the subsystem $A$ contains no fixed point, the entanglement entropy and the second Rényi entropy reduce to their vacuum values,~\eqref{eq:Conformal-Cooling-Vacuum}, at late time due to the conformal cooling effect~\cite{Wen:2022pyj,Goto:2021sqx,Jiang:2024hgt,Nozaki:2023fkx, cn3z-vfgr}. On the contrary, subsystems $A$ that contain fixed points exhibit late-time linear growth of $S_A(t)$ and $S_A^{(2)}(t)$~\cite{Wen:2022pyj}. This linear growth, analogous to the logarithmic growth in the $q$-SSD evolution, originates from Jacobian factors in the conformal transformation rather than from thermalization.
In fact, under the $q$-Displacement evolution, the regularized crosscap state does not thermalize, with generically nonzero mutual information $I_{A,B}(t)$ and $I^{(2)}_{A,B}(t)$ at late time. Thus, the system avoids thermalization and scrambling in both holographic and free-fermionic theories.

Within the heating phase, we proceed case by case for $q>2$ and $q\leq2$. When $q>2$, the late time asymptotic values of $I_{A,B}(t)$ and $I_{A,B}^{(2)}(t)$ are approximately given by
\be\label{eq:qDis-Asym-MI}
\begin{split}
    &I_{A, B}(t)\overset{t\gg L}{\approx}\frac{c\pi}{3\beta}\cdot\frac{L}{q}\cdot\begin{cases}
        p~&\text{for odd}~q~\text{and}~1< p<q,\\
        p-1~&\text{for even}~q~\text{and}~1< p<q,\\
        0~&\text{for even}~q,~p=1,0~\text{and}~\text{odd }q,~p=0,\\
    \end{cases}\\
    &I_{A, B}^{(2)}(t)\overset{t\gg L}{\approx}\frac{\pi}{4\beta}\cdot\frac{L}{q}\cdot\begin{cases}
        p~&\text{for odd}~q~\text{and}~1< p<q,\\
        p-1~&\text{for even}~q~\text{and}~1< p<q,\\
        0~&\text{for even}~q,~p=1,0~\text{and}~\text{odd }q,~p=0,\\
    \end{cases}\\
\end{split}
\ee
where the subsystem $A$ and its antipodal partner $B=\overline{A}$ both contain $p$ fixed points with $p>1$. $p=1$ is excluded, as it renders the holographic CFT computation divergent; the details are provided in Appendix~\ref{app:EEHCFT}. The distinction between even and odd $q$ can be understood from the late-time quasiparticle graph-like patterns, as discussed in Sec.~\ref{sec:QP-EP}. If $q \le 2$, the long-time $q$-Displacement dynamics of the crosscap state cannot be described by quasiparticle picture because of a linear-in-time divergence; additionally, their graph-like patterns are indistinguishable from the dynamics of an initially short-range-entangled state.

Lastly, we derive the quasiparticle trajectory for the $q$-Displacement evolution by integrating
    $\frac{dx}{dt}=\sigma\sin\left(\frac{2q\pi x}{L}\right)$,
such that
\be\label{DisTrajectory}
\begin{split}
    &x_{0;\sigma}(x,t) \\
    &=\begin{cases}
    \frac{L}{2\pi q}  \cos^{-1}\left\{\tanh{\left[\tanh^{-1}\left(\cos{\frac{2\pi qx}{L}}\right)+\frac{2\pi q\sigma(t-t_0)}{L}\right]}\right\} ,& 0\leq\frac{2\pi qx}{L}\mod 2\pi<\pi\\ \nonumber
    \quad +\frac{L}{q}\left\lfloor\frac{qx}{L} \right\rfloor& \\
    -\frac{L}{2\pi q}  \cos^{-1}\left\{\tanh{\left[\tanh^{-1}\left(\cos{\frac{2\pi qx}{L}}\right)+\frac{2\pi q\sigma(t-t_0)}{L}\right]}\right\},& \pi\leq\frac{2\pi qx}{L}\mod 2\pi <2\pi\\
    \quad +\frac{L}{q}
    \left(    \left\lfloor\frac{qx}{L} \right\rfloor+1 \right)& \\
\end{cases},
\end{split}
\ee
where $x_{0;\sigma}(x,t_0)=x$. Substituting $x_{0;\sigma}(x,t)$ back into~\eqref{eq:QP-EE} and~\eqref{eq:QP-MI} with $t_0=0$, we obtain the quasiparticle curves shown in Fig.~\ref{fig:qDis-Crosscap-Quench}. 

We conclude this section by presenting cases where the quasiparticle description of mutual information fails in both free fermionic and holographic CFTs. This arises in our critical/heating crosscap quenches and in the boundary-state quenches of \cite{Nozaki:2023fkx}. In most cases, $I_{A,B}^{(n)}=S_{A}^{(n)}+S_{B}^{(n)}-S_{A\cup B}^{(n)}$ indicates that the conformal Jacobian factors producing divergences in $S_{A}^{(n)}+S_{B}^{(n)}$ are canceled by those in $S_{A\cup B}^{(n)}$. However, when $A\cup B$ contains only one fixed point, or when it contains all fixed points, the Jacobian factors contributing to $S_{A\cup B}^{(n)}$ are subject to conformal cooling and drive $S_{A\cup B}^{(n)}$ toward its vacuum value. Consequently, the divergent terms in $S_{A}^{(n)}+S_{B}^{(n)}$ fail to cancel, so the mutual information grows logarithmically in time under critical quenches and linearly in time under heating-phase quenches. This behavior is beyond the scope of the standard quasiparticle picture and is observed even in free-fermionic systems. We leave a systematic analysis of this issue to future work.

\section{Quasiparticle Pictures and Graph-like Entanglement Patterns\label{sec:QP-EP}}
\begin{figure}
    \centering
    \includegraphics[width=0.75\linewidth]{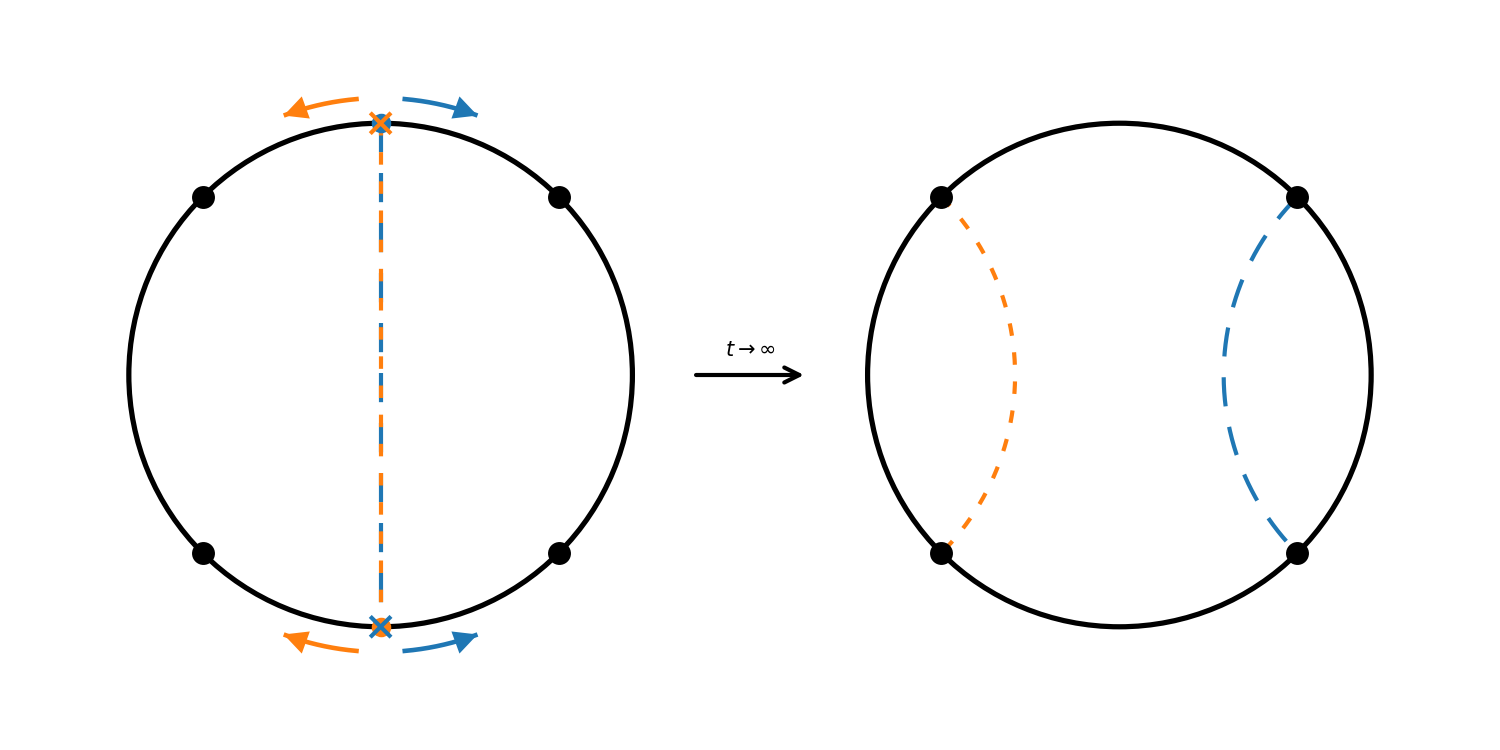}
    \caption{Motions of quasiparticle pairs for a critical point evolution with $q$ fixed points ($q=4$). Left panel: initial time $t=0$; right panel: late time $t\gg L$. Fixed points $x=X_m^f$ are denoted by the black dots.}
    \label{fig:Schematic QP3}
\end{figure}

In this section, we present an intriguing late-time, graph-like entanglement pattern predicted within the quasiparticle picture for the critical and heating-phase quench dynamics of the regularized crosscap state.
Here “graph-like” indicates a late-time quasiparticle representation in which fixed points are vertices and EPR links constitute edges, producing a graph structure.
These patterns, which capture late-time mutual information (cf.~\eqref{eq:qSSD-Asym-MI},~\eqref{eq:qDis-Asym-MI}), are generic across both critical-point and heating-phase Hamiltonian classes\footnote{In the non-heating phase, time evolution does not lead to any fixed point, and as a consequence, graph-like patterns do not emerge.}, and constitute a central result of this work.

\begin{figure}
    \centering
    \includegraphics[width=0.75\linewidth]{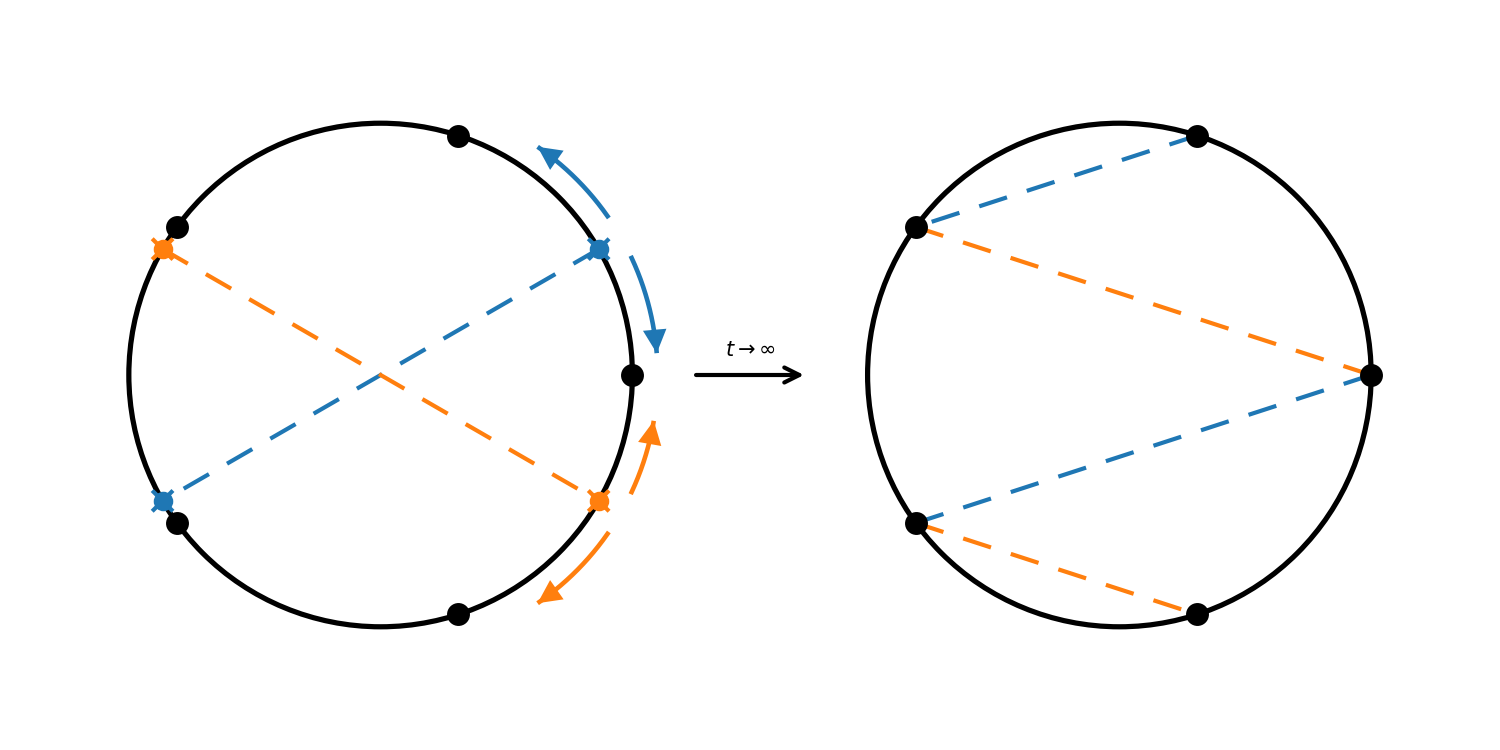}
    \caption{Motions of quasiparticle pairs for a critical point evolution with $q$ fixed points ($q=5$). Left panel: initial time $t=0$; right panel: late time $t\gg L$.}
    \label{fig:Schematic QP4}
\end{figure}

To start with, we analyze how an EPR pair, initially prepared at antipodal locations $x$ and $x+\frac{L}{2}$, evolves during the critical and heating phase dynamics. We recall that the system has PBC, thus quasiparticles move on a circle. We assume that right-moving modes propagate counterclockwise, while left-moving modes propagate clockwise. If the quenching Hamiltonian $H_1$ is the $q$-SSD, the quasiparticle velocity is given by $v(x)=2\sigma\sin^2\left(\frac{2q\pi x}{L}\right)$. Since $v(X_m^f)=0$, fixed points act as asymptotic barriers for quasiparticles. For even $q$, suppose a right-mover is within the interval $x\in[X_m^f,X_{m+1}^f]$ initially, so its left-moving partner is within {\footnotesize $\left[X_{m+\frac{q}{2}}^f,X_{m+1+\frac{q}{2}}^f\right]\mod L$}. When $t>0$ the right-moving quasiparticle moves to $X_{m+1}^f$, while the left-moving one moves to {\footnotesize $X_{m+\frac{q}{2}}^f\mod L$}, respectively, as shown in Fig.~\ref{fig:Schematic QP3}. Instead, if $q\in 2\mathbb{Z}+1$, and a right-mover is traveling within $x\in (X_m^f,X_{m+1}^f)$, then the left-moving mode has two possibilities: it is either initially in {\footnotesize $\left[X_{m+\frac{1}{2}+\lfloor\frac{q}{2}\rfloor}^f,X_{m+\lceil\frac{q}{2}\rceil}^f\right]\mod L$}, in which case it moves within {\footnotesize $\left[X_{m+\lfloor\frac{q}{2}\rfloor}^f,X_{m+\lceil\frac{q}{2}\rceil}^f\right]$}, or it is initially in {\footnotesize$\left[X_{m+\lceil\frac{q}{2}\rceil}^f,X_{m+\frac{1}{2}+\lceil\frac{q}{2}\rceil}^f\right]\mod L$}, in which case it moves within {\footnotesize$ \left[X_{m+\lceil\frac{q}{2}\rceil}^f,X_{m+1+\lceil\frac{q}{2}\rceil}^f\right]\mod L$}. Thus, when the right-mover reaches $X_{m+1}^f$, the corresponding left-mover arrives at {\footnotesize$ X_{m+\lfloor\frac{q}{2}\rfloor}^f\mod L$} in case one, or at {\footnotesize $X_{m+\lceil\frac{q}{2}\rceil}^f\mod L$} in case two; see Fig.~\ref{fig:Schematic QP4}. 
Under the $q$-SSD time evolution, entangled quasiparticle pairs drift toward one another and take an infinite time to reach their nearest fixed points (clockwise for left-movers (dots), counterclockwise for right-movers (crosses)). Since quasiparticles are effectively frozen at the fixed points, the details of the velocity profile in between the fixed points are irrelevant. Accordingly, the quasiparticle patterns extend beyond $q$-SSD to any Hamiltonian in the critical point class with $q$ fixed points (e.g., critical Floquet Hamiltonians~\cite{Fan:2020orx}).

\begin{figure}
    \centering
    \includegraphics[width=0.75\linewidth]{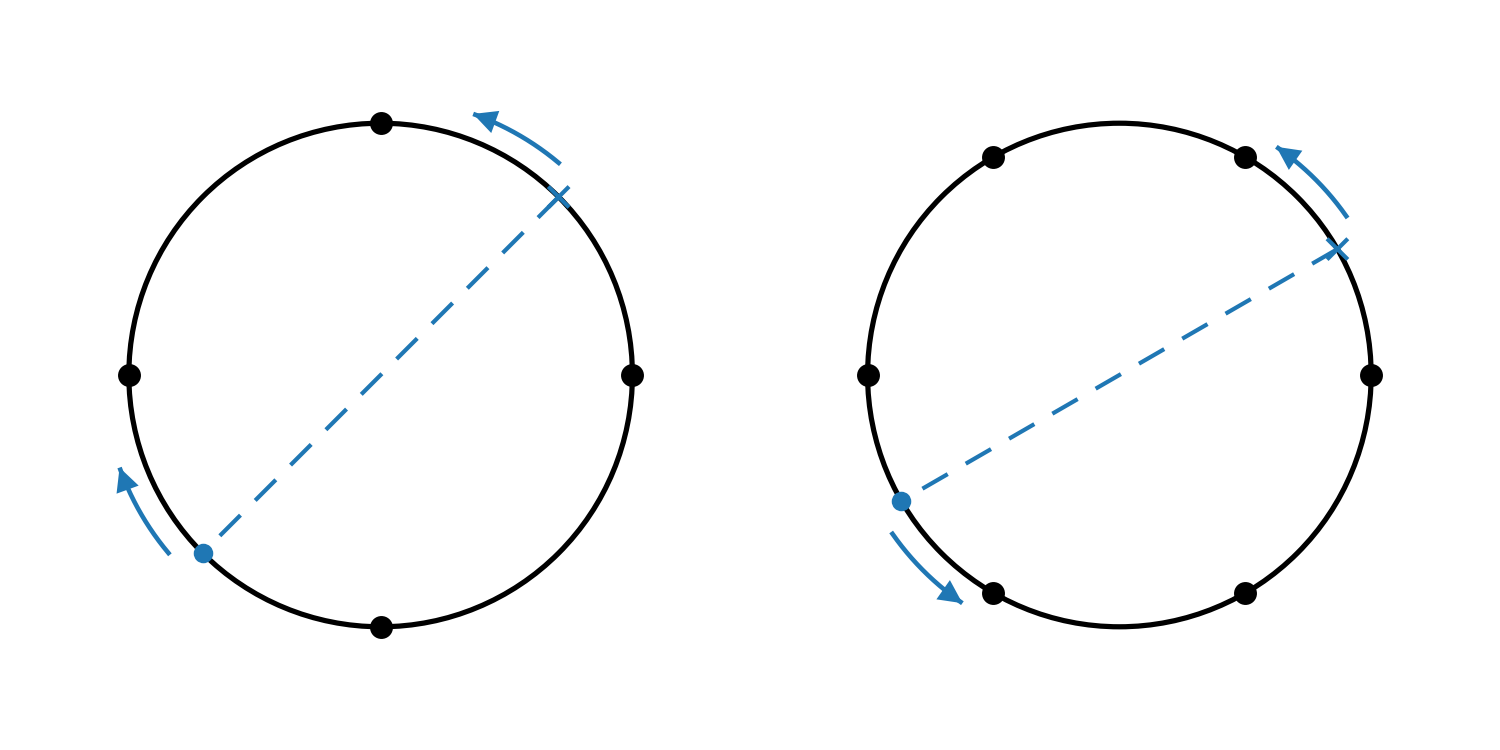}
    \caption{Motions of quasiparticle pairs for a heating phase evolution with $2q$ fixed points. Left panel: $q=2$; right panel: $q=3$. Fixed points $x=X_{\frac{m}{2}}^f$ are denoted by the black dots. Any two adjacent intervals bounded by fixed points have opposite quasiparticle velocity signs.}
    \label{fig:Schematic QP5}
\end{figure}

When the quasiparticle velocity is given by $v(x)=\sigma f(x)=\sigma\sin\left(\frac{2q\pi x}{L}\right)$, $H_1$ corresponds to the $q$-Displacement Hamiltonian. In this case, we note that
\be
    v(x+L)=(-1)^q\sigma\sin\left(\frac{2q\pi x}{L}\right)=\begin{cases}
        v(x)~&\text{for even}~q\\
        -v(x)~&\text{for odd}~q\\
    \end{cases},
\ee
where, for odd $q$, the quasiparticle velocity reverses sign between the antipodal intervals {\footnotesize $\left(X_{\frac{m}{2}}^f,X_{\frac{m+1}{2}}^f\right)$} and {\footnotesize $\left(X_{\frac{m+q}{2}}^f,X_{\frac{m+1+q}{2}}^f\right)\mod L$}. This implies that entangled quasiparticle pairs move toward each other when $q$ is even, and away from each other when $q$ is odd. Accordingly, if the right-mover begins in {\footnotesize $\left(X_{\frac{m}{2}}^f,X_{\frac{m+1}{2}}^f\right)$ and ends at $x=X_{\frac{m+1}{2}}^f$}, then its partner halts at {\footnotesize$x= X_{\frac{m+q}{2}}^f\mod L$} (moving counterclockwise) for even $q$, but at the antipodal fixed point {\footnotesize $x=X_{\frac{m+1+q}{2}}^f\mod L$} to the right-mover’s endpoint for odd $q$; see Fig.~\ref{fig:Schematic QP5}. In the same spirit as $q$-SSD, these patterns are generic to the heating phase and persist for any Hamiltonian with $2q$ fixed points, not only the specific example considered here.

\begin{figure}
    \centering
    \includegraphics[width=0.97\linewidth]{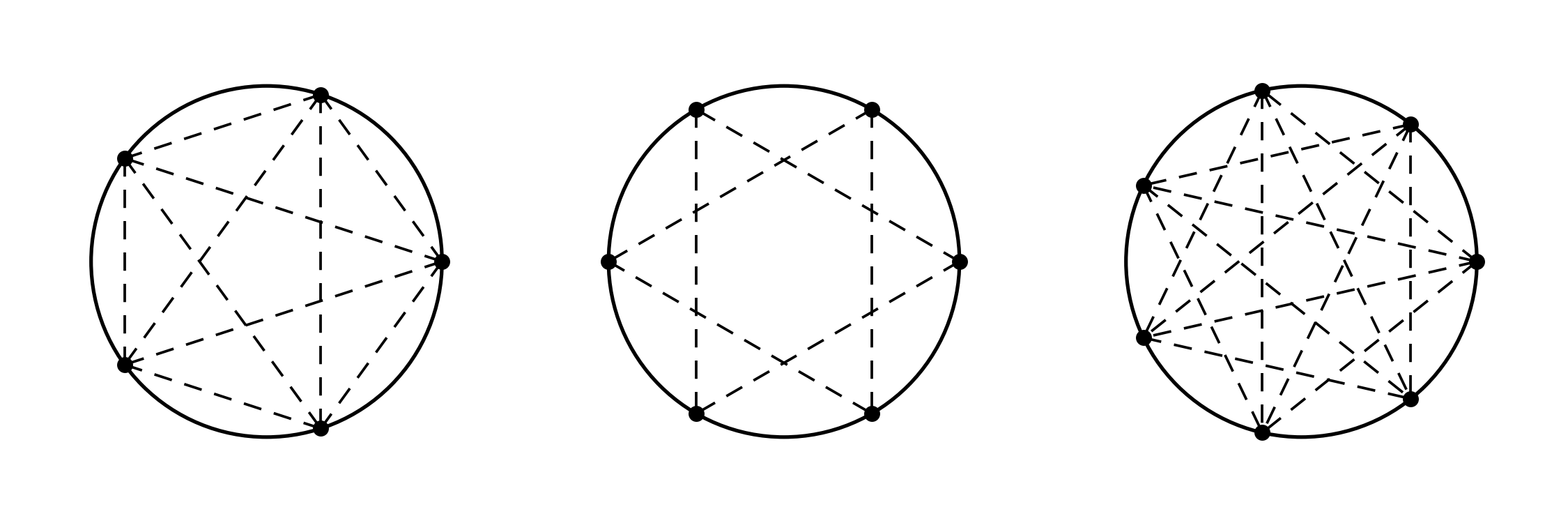}
    \caption{Late time graph-like entanglement patterns under critical point dynamics with $q$ fixed points. Left panel: $q=5$, quasiparticles form a $K_5$ complete graph at late times; middle panel: $q=6$, quasiparticles form a circulant graph $C(6;4,2)$; right panel: $q=7$, quasiparticles form a circulant graph $C(7;3,2)$}
    \label{fig:Schematic QP6}
\end{figure}

\begin{figure}
    \centering
    \includegraphics[width=0.7\linewidth]{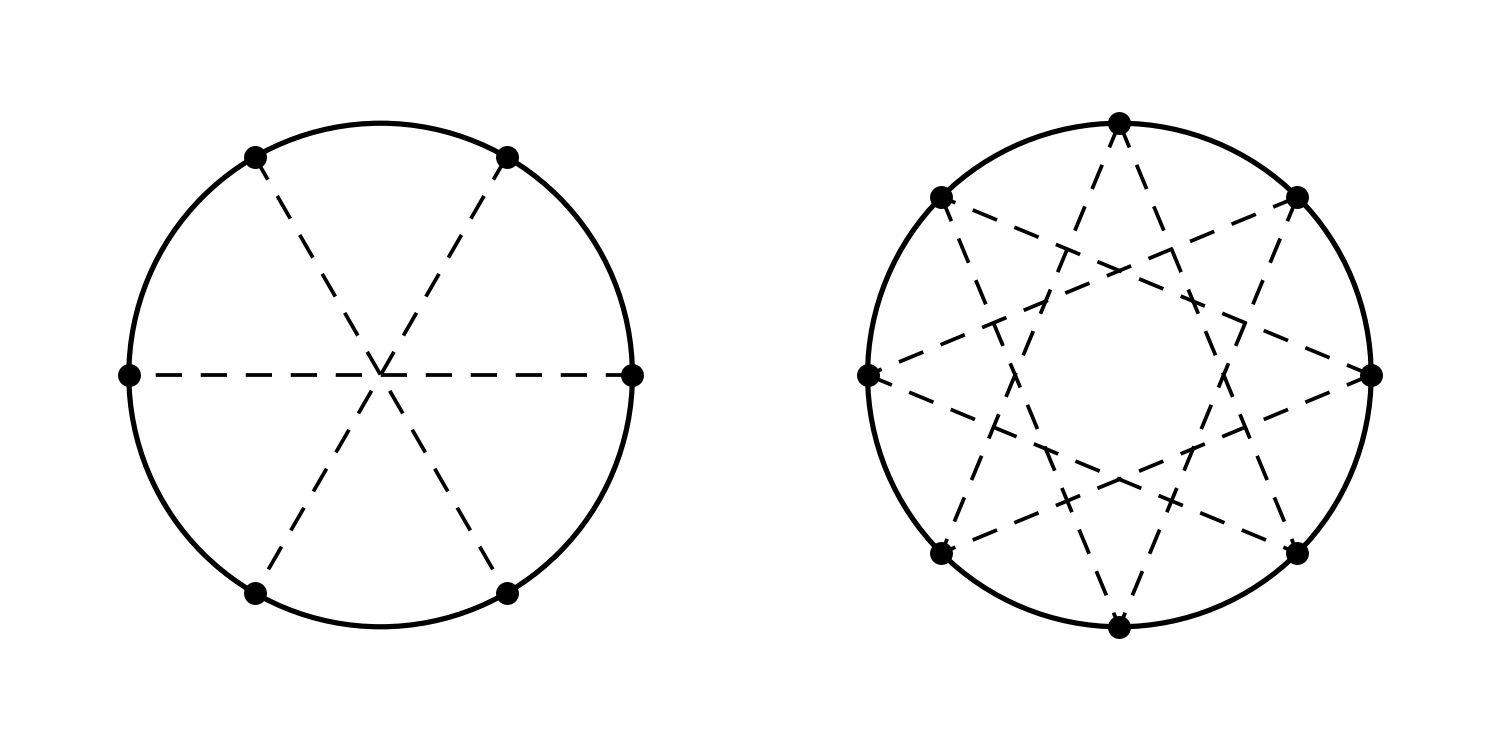}
    \caption{Late time graph-like entanglement patterns under heating phase dynamics with $2q$ fixed points. Left panel: $q=3$, quasiparticles form a $C(6;3)$ complete graph at late times; right panel: $q=4$, quasiparticles form a circulant graph $C(8;5,3)$. In particular, a heating-phase quench featuring $2q$ fixed points with $q\in2\mathbb{Z}+1$ exhibits an antipodal entanglement structure.}
    \label{fig:Schematic QP7}
\end{figure}

We now derive graph-like entanglement patterns within the quasiparticle picture, which capture the entanglement features of order $\beta^{-1}$. We treat vertices as quasiparticles (hence as fixed points at late times) and edges as EPR interactions, thereby forming graph structures in these figures, while ignoring the spatial circle since its entanglement contribution is subleading of order $\beta^{-1}$. As noted in Sec.~\ref{sec:QP}, at $t=0$ the quantum quench uniformly creates antipodal entangled pairs of quasiparticles across the entire system. For $t>0$, quasiparticles propagate with the position-dependent velocity $v(x)=\sigma f(x)$ and are eventually pinned by fixed points at late times. All pairs follow exactly the same rule outlined for a single pair previously. As a result, the quasiparticle picture yields different graph-like entanglement patterns for distinct choices of $H_1$ and $q$; see Fig.~\ref{fig:Schematic QP6} and~\ref{fig:Schematic QP7}. Following standard graph-theoretic terminology~\cite{GodsilRoyle2001AGT,Diestel2025GraphTheory}, we classify the late-time quasiparticle graph-like entanglement patterns and compare them with those produced by quenches from an initially uniformly and locally entangled state $\ket{\psi}_{\text{local}}$ (e.g., a boundary state); see Table~\ref{tab:Classification-QP-Patterns}.

It is easy to check that these graph-like patterns can capture the correct mutual information asymptotics,~\eqref{eq:qSSD-Asym-MI} and~\eqref{eq:qDis-Asym-MI}. In doing so, we have assumed the total number of quasiparticles is conserved throughout the inhomogeneous time evolution,
\be\label{eq:QP-Number}
    N_{\text{QP}}^{(n)}=\int_0^Ldx~2\rho_0^{(n)}=\int_0^Ldx~\left(\rho_{0;L}^{(n)}+\rho_{0;R}^{(n)}\right)=\frac{c}{6}\frac{n+1}{n}\frac{\pi L}{\beta}, \quad n\in\mathbb{N}_+.
\ee
Additionally, quasiparticles are uniformly distributed at $t=0$, and can never cross the fixed points. As a result, at late times, if there are in total $g$ edges for the graph-like pattern, each edge carries $g^{-1}N_{\text{QP}}$ units of entanglement. Then, by counting the number of edges connecting region $A$ to its antipodal partner $B=\overline{A}$, one obtains $I_{A,B}^{(n)}(t)$ for $t\gg L$. The results show perfect agreement with~\eqref{eq:qSSD-Asym-MI} and~\eqref{eq:qDis-Asym-MI}. However, the graph-like patterns do not determine the late-time entanglement entropy, which can contain divergent contributions\footnote{These divergent-in-time terms are solely contained in the conformal Jacobian factors, and do not rely on twist field correlators.} (linear in $t$ or $\log(t)$) that lie beyond the quasiparticle description. By contrast, for the mutual information, these divergences cancel; therefore, the graph-theoretic analysis applies.

These graph-like entanglement patterns emerge only when critical or heating-phase dynamics are combined with the crosscap initial state; if either ingredient is absent, no non-trivial graphs develop. In fact, without critical/heating Hamiltonian $H_1$, fixed points are absent; this is the case if we consider, e.g., the uniform (non-heating) crosscap quench. As discussed previously, in this case, holographic theories thermalize and thus $I_{A,B}(t)=0$ at $t\gg L$, while the free-fermionic model shows persistent revivals arising from quasiparticles’ periodic motion. 
On the other hand, we can consider the case where $H_1$ is a critical/heating Hamiltonian, but the initial state $\ket{\psi}_{\text{local}}$ contains only uniform, local entanglement, e.g.,
\be\label{eq:Regularized-Bdy-State}
    \ket{\psi}_{\text{local}}=\mathcal{N}_Be^{-\frac{\beta }{4}H_0}\ket{B}
\ee
(where $\mathcal{N}_B$ denotes a normalization constant).
In this example, because each quasiparticle pair is created at the same position $x$, and neither the left- nor right-movers can cross fixed points, the quasiparticle picture predicts late-time graphs of $C(q;1)$ for $q$-SSD and $C(2q;1)$ for $q$-Displacement evolution, where $q$ can be any positive integer. Evolving $\ket{\psi}_{\text{local}}$ with the $q$-SSD Hamiltonian, we find that the graph-like entanglement patterns of $\ket{\psi}_{\text{local}}$ are indistinguishable from those of the crosscap state for $q\leq 4$; moreover, for $q=1,2$, or when $A\cup B $ contains only one fixed point or all fixed points, the quasiparticle picture fails to capture the mutual information for subsystems $A,~ B$ that include fixed points, because both $I_{A,B}(t)$ and $I_{A,B}^{(2)}(t)$ are proportional to $\log(t)$ at late times. The first deviation appears at $q=5$. Under $q$-Displacement evolution, the patterns coincide for $q=1,2$ and begin to differ at $q=3$; in addition, for $q=1$, or when $A\cup B $ contains all fixed points, the patterns cannot be used to estimate the mutual information when the subsystems contain fixed points, due to a term that diverges linearly in $t$, which is beyond the quasiparticle picture.

We conclude this section with a brief discussion of relabeling (i.e., vertex-permutation) symmetries of the graph-like patterns—here, vertices are late-time fixed points and edges are EPR links\footnote{For graph-theoretic background, see~\cite{Diestel2025GraphTheory,GodsilRoyle2001AGT}.}. For concreteness, we analyze one representative example; the remaining graphs have the same construction but different relabeling groups (i.e., different sets of vertex permutations that leave the EPR-link pattern unchanged). Consider the $q$-SSD evolution, for which the patterns reduce to the complete graphs $K_5$ when $q=5$. In this case, every pair of distinct vertices is connected by a single edge, and the automorphism group is the full symmetric group $S_5$ (any permutation of vertex labels preserves the ``every-pair'' connectivity). Different dynamical protocols produce graph-like patterns that are distinguished by their relabeling symmetries. The concrete utility of the graph-like pattern in (1+1)d CFT/critical systems remains to be pinned down. Nevertheless, within the quasiparticle picture, these patterns provide a promising, symmetry-compatible descriptor of late-time mutual information—robust across both holographic and free-fermion models—even though the specific role of the relabeling (vertex-permutation) symmetries is still unclear. 

\begin{table}
    \centering
    \begin{tabular}{|c|c|c|c|}
        \hline
         \textbf{Hamiltonian Class and State}& $q\in 2\mathbb{Z}$ & $q\in 2\mathbb{Z}+1$\\
        \hline
        $t=0$ & $\sqcup_{m\in\mathbb{N}}K_2$ & $\sqcup_{m\in\mathbb{N}}K_2$\\
        \hline
        Non-heating phase and $\ket{C}$ $(t\gg L)$ & None & None\\
         \hline
        Critical point and $\ket{C}$ $(t\gg L)$ & $C\left(q;\frac{q}{2}+1,\frac{q}{2}-1\right)$ & $C\left(q;\frac{q-1}{2},\frac{q-3}{2}\right)$\\
         \hline
        Heating phase and $\ket{C}$ $(t\gg L)$ & $C\left(2q;q+1,q-1\right)$  & $C\left(2q;q\right)$\\
         \hline
         Non-heating phase and $\ket{\psi}_{\text{local}}$ $(t\gg L)$ & None  & None\\
         \hline
        Critical point and $\ket{\psi}_{\text{local}}$ $(t\gg L)$ & $C\left(q;1\right)$  & $C\left(q;1\right)$\\
         \hline
        Heating phase and $\ket{\psi}_{\text{local}}$ $(t\gg L)$ & $C\left(2q;1\right)$  & $C\left(2q;1\right)$\\
         \hline
    \end{tabular}
    \caption{Classification of the late time quasiparticle graph-like entanglement patterns. The integer $q$ refers to Virasoro modes. A circulant graph $C(q;a,b)$ is a 1D ring of $q$ vertices where each vertex $m\in \mathbb{Z}_{m}$ is linked to its fixed-offset neighbors $m\pm a$ and $m\pm b~(\text{mod}~q)$. Here, $a,b\in\{0,1,\cdots,\lfloor\frac{q}{2}\rfloor\}$ are the step (range) distances—analogous to tight-binding couplings at lattice spacings $a$ and $b$. Duplicate offsets collapse (e.g., $a=b$ such that $C(q;a,b)=C(q;a)$). In our diagrams, edges denote EPR links between the corresponding quasiparticle aggregates (vertices). At $t=0$, the entanglement graph is a disjoint union of 2-vertex complete graphs, $\sqcup_{m\in\mathbb{N}}K_2=\lim_{q\to\infty}C(2q;q)$, (infinite independent EPR pairs, i.e., a perfect matching), where each $K_2$ labels a quasiparticle pair. Graphs $C\left(q;\frac{q}{2}+1,\frac{q}{2}-1\right)$, $C\left(q;\frac{q-1}{2},\frac{q-3}{2}\right)$, $C\left(2q;q+1,q-1\right)$, $C\left(2q;q\right)$ and $C(q;1)$ have $q$, $2q$, $2q$, $q$, $q$ edges, respectively. Non-heating phase Hamiltonians have no fixed points, thus no graph-like pattern exists at late times.}
    \label{tab:Classification-QP-Patterns}
\end{table}

\section{Holographic Dual for Inhomogeneous Crosscap Quenches\label{sec:Holo-Dual}}

\begin{figure}
    \centering
    \includegraphics[width=0.9\linewidth]{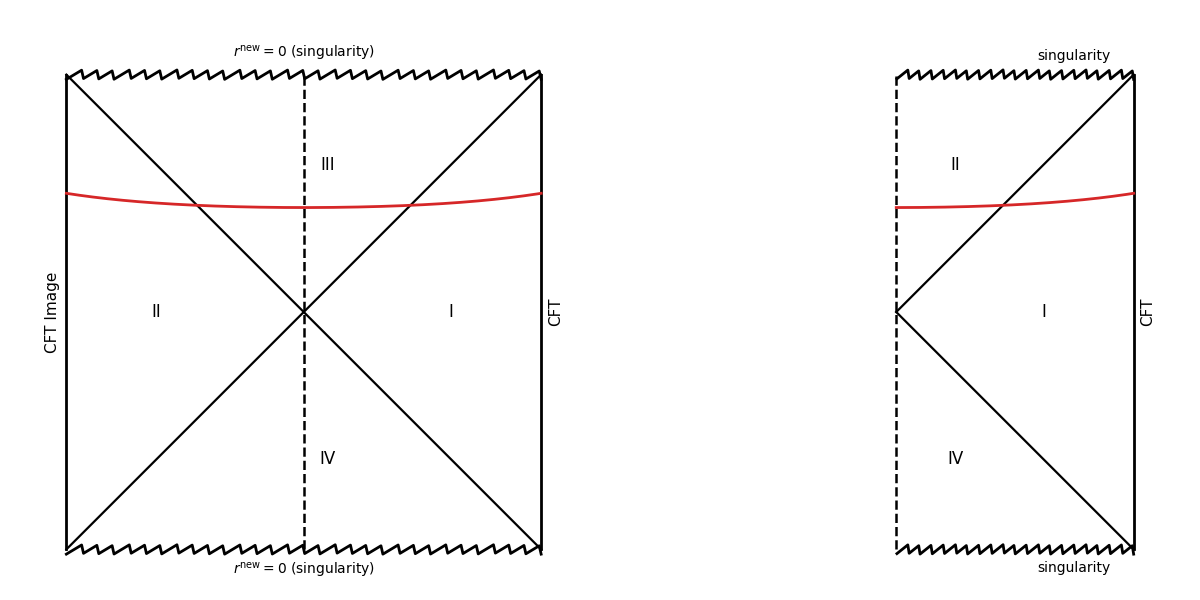}
    \caption{Penrose diagram for $\AdS_3$ geon. Left panel: double cover of geon, i.e., two-sided $\AdS_3$ eternal black hole; the geon is constructed by identifying the two sides along the dashed line. Regions I-IV are divided by the black hole horizon. Region I/II corresponds to the right/left exterior, the holographic CFT and its imaged theory are located at their asymptotic boundaries (thick vertical lines) with $r^{\text{new}}\to\infty$. Regions III and IV correspond to the black hole interior. The red lines represent disconnected geodesics or non-traversable wormholes. The crosscap lives on the dashed line.}
    \label{fig:Penrose Diagram}
\end{figure}

In this section, we apply holographic methods to study inhomogeneous crosscap quenches in Lorentzian spacetime using the RT/HRT formula~\cite{Ryu:2006bv,Hubeny:2007xt}. The holographic dual of the uniform crosscap quench is the $\AdS_3$ geon geometry~\cite{Wei:2024kkp}, which is a single-sided spacetime with a black hole~\cite{Louko:1998hc}. In our case, even though the time evolution is inhomogeneous, the Lorentzian geometry is still obtained by the BTZ black hole metric with a $\mathbb{Z}_2$ identification
\be\label{eq:Geon-Schwarzschild-Form}
\begin{split}
    &ds^2= -\left({r^{\text{new}}}^2-{r^{\text{new}}_h}^2\right)d{t^{\text{new}}}^2+\frac{d{r^{\text{new}}}^2}{{r^{\text{new}}}^2-{r^{\text{new}}_h}^2}+{r^{\text{new}}}^2d{x^{\text{new}}}^2,
    \quad r^{\text{new}}\geq r^{\text{new}}_h=\frac{2\pi}{\beta},\\
    &\mathbb{Z}_2\text{ quotient:}~(t^{\text{new}},r^{\text{new}},x^{\text{new}})_{\text{right exterior}}\sim \left(-t^{\text{new}},r^{\text{new}},x^{\text{new}}+\frac{L}{2}\right)_{\text{left exterior}},
\end{split}
\ee
which is a smooth\footnote{By ``smooth'' we precisely mean that the quotient does not introduce additional singularities in the spacetime.} $\mathbb{Z}_2$ quotient of a two-sided eternal $\AdS_3$ black hole, as depicted in Fig.~\ref{fig:Penrose Diagram}
~\cite{Maldacena:2001kr}; the subscripts ``right/left exterior'' indicate the corresponding asymptotic regions. The black hole horizon has a radius given by $r^{\text{new}}_h=\frac{2\pi}{\beta}$, and the $\mathbb{Z}_2$ quotient is the (2+1)d bulk extension of the crosscap identification $x\sim x+L/2$~\cite{Louko:1998hc,Maloney:2016gsg}. In the Heisenberg picture,~\eqref{eq:Post-Quench-Coordinate}, the metric is fixed in time, and the time dependence is implemented by a reparametrization (diffeomorphism) of the spacetime coordinates:
\be\label{app-eq:Post-Quench-Space-Time-Coordinates}
    t^{\text{new}}=\frac{w^{\text{new}}+\bar{w}^{\text{new}}}{2i},
    \quad x^{\text{new}}=\frac{w^{\text{new}}-\bar{w}^{\text{new}}}{2i}.
\ee

In the $\AdS_{d+1}/\CFT_d$ correspondence~\cite{Maldacena:1997re}, a time slice of the $\CFT_d$ describes a quantum state of a $(d-1)$-dimensional quantum system. Within a time slice $t$, one can separate the system into region $\mathcal{V}$ and its complement $\mathcal{V}^c$. Then, the entanglement entropy $S_{\mathcal{V}}(t)$ can be computed in the $\AdS_{d+1}$ spacetime using the RT/HRT formula, 
\be\label{eq:HRT-Formula}
    S_{\mathcal{V}}(t)=\text{Min}\left[\underset{\gamma_{\mathcal{V}}\sim\mathcal{V}}{\text{Ext}}\frac{\text{Area}\left(\gamma_{\mathcal{V}}\right)}{4G_N^{(d+1)}}\right],
\ee
where $G_N^{(d+1)}$ is the Newton constant of $\AdS_{d+1}$, $\text{Area}\left(\gamma_{\mathcal{V}}\right)$ denotes the area of the surface $\gamma_{\mathcal{V}}$. The latter is a space-like co-dimensional two extremized surface that shares the boundary with the subsystem $\mathcal{V}$:
\be
    \gamma_{\mathcal{V}}\sim\mathcal{V}:~\partial\gamma_{\mathcal{V}}=\partial \mathcal{V}~\&~\exists~d\text{-dimensional hypersurface}~\mathcal{R}_{\mathcal{V}}~\text{s.t.}~\partial\mathcal{R}_{\mathcal{V}}=\gamma_{\mathcal{V}}\cup\mathcal{V},
\ee
where $\partial\mathcal{V}$ denotes the boundary of $\mathcal{V}$. By minimizing over all $\gamma_{\mathcal{V}}$ surfaces, we obtain~\eqref{eq:HRT-Formula}. In our case with geon geometry,~\eqref{eq:HRT-Formula} is given by
\be
    S_{\mathcal{V}}(t)=\frac{\text{Min}\left[D_{\text{dis.}},D_{\text{con.}}\right]}{4G_N^{(3)}},
\ee
where $D_{\text{dis.}}$ (the ``disconnected'' phase or ``wormhole'' phase) denotes the family of extremal spacelike geodesics that connect the black-hole interior to the boundary of $\mathcal{V}$, and $D_{\text{con.}}$ (the ``connected'' phase or ``thermal'' phase) is the set of extremized spacelike geodesics joining the boundary endpoints of $\mathcal{V}$. The details of $D_{\text{dis.}},D_{\text{con.}}$ can be found in Appendix~\ref{app:RTHRT}. Minimizing over the two phases provides a thermalization diagnostic~\cite{Hartman:2013qma,Asplund:2015eha,Wei:2024kkp}: for any subsystem $\mathcal{V}$ of size larger than $\beta$, late-time dominance of $D_{\text{con.}}$ indicates thermalization. By contrast, dominance of the disconnected phase for some subsystems signals the absence of thermalization/scrambling and a persistent, nonzero mutual information for suitable partitions.

Across all the scenarios considered in this work, at late times, $S_{A}(t)$ is controlled by the connected (thermal) phase; on the other hand, the antipodal entanglement entropy $S_{A\cup B}(t)$ depends on the specific dynamics. For uniform and $q$-M\"obius evolutions, the connected phase $D_{\text{con.}}$ controls $S_{A\cup B}(t)$, implying thermalization and $I_{A,B}(t)=0$ at late times. For $q$-SSD and $q$-Displacement evolutions, the antipodal entropy selects the wormhole (disconnected) phase, yielding a persistent nonzero mutual information, as demonstrated analytically in Sec.~\ref{sec:Inh-Quench} and visually in Sec.~\ref{sec:QP-EP}. We summarize the late-time geodesic dominance in Table~\ref{tab:Geodeisc-Dominance} and relegate the RT/HRT derivations to Appendix~\ref{app:RTHRT}.
Finally, while the quasiparticle picture on the CFT side yields sharp graph-like patterns (see Sec.~\ref{sec:QP-EP}), an obvious counterpart is absent in the $(2+1)$d geon spacetime. We nevertheless observe an intriguing, possibly related phenomenon on the gravity side: a mismatch between geodesics inside the black hole. For a fixed subsystem $\mathcal{V}$, the wormhole phase involves two endpoint-anchored geodesics: one that realizes the wormhole extremal surface, and one that connects an endpoint to its mirror image on the left-exterior asymptotic boundary. In standard cases (boundary-state quenches; uniform crosscap quench), the interior pieces coincide in length. However, for inhomogeneous crosscap quenches that violate antipodal symmetry (e.g., $q$-M\"obius with odd $q$), deviations occur, as we explain with details in Appendix~\ref{app:GM}.

\begin{table}
    \centering
    \begin{tabular}{|c|c|c|}
    \hline
    Late-time, $S_{A\cup B}(t)$ &  Late-time ($t\gg L$)\\
    \hline
    Uniform &  Thermal Phase\\
    \hline
    $q$-M\"obius &  Thermal Phase \\
    \hline
    $q$-SSD &  Wormhole Phase\\
    \hline
    $q$-Displacement & Wormhole Phase\\
        \hline
    \end{tabular}
    \caption{Late-time selection of the dominant phase (thermal vs. wormhole) in $S_{A\cup B}(t)$ for the crosscap state under different dynamics.}
    \label{tab:Geodeisc-Dominance}
\end{table}

\section{Conclusion and Discussion\label{sec:Conclusion}}

In this work, we have investigated the dynamics of the crosscap state after inhomogeneous quenches in $(1+1)$d critical systems described by (1+1)d (holographic or free Dirac fermion) CFT. Specifically, the dynamics are generated by $\text{SL}(2,\mathbb{R})$ deformed Hamiltonians, which can be classified into three distinct classes. Under both the critical and heating phase time evolution, the crosscap state showcases a clear absence of thermalization at late times, even in the holographic case. Instead, the late-time entanglement self-organizes into robust, graph-like patterns that resist thermalization and accurately capture the mutual information.
These graph-like entanglement patterns are constructed directly from the quasiparticle picture, and have been confirmed using analytical calculations. In particular, they coincide in both holographic and free Dirac fermion CFTs, remain insensitive to the specific Hamiltonian choice, and are naturally expressed in terms of circulant graphs.
We stress that this graph-theoretic description arises from the interplay between the non-local properties of the crosscap state and the spatially inhomogeneous Hamiltonian dynamics with emergent fixed points. In contrast, a non-heating quench leads to thermalization in holographic CFT but periodic revivals in free fermion theory, and no graph-like entanglement pattern forms at late times.
As a special case, the uniform crosscap quench studied previously~\cite{Wei:2024kkp} is an example of non-heating phase evolution. 
Our results thus provide a platform to design either scrambling or non-thermalizing dynamics for crosscap states depending on the Hamiltonian dynamics.

Given the AdS/CFT duality between (1+1)d holographic CFT and 3D gravity, the bulk counterpart of the crosscap quench is a geon spacetime. First, we showed that applying the RT/HRT prescription in the geon correctly reproduces the entanglement entropy and mutual information calculations. Second, we observed a discrepancy between the interior portions of two geodesics that appears only when the geon geometry is combined with inhomogeneous time evolution. This is analogous to the graph-like structures seen in the $(1+1)$d critical quench. 
However, the relationship between the two phenomena remains unclear; we leave a broader understanding of this potential connection for future work.

\noindent We close by highlighting several directions for future research:
\begin{itemize}
\renewcommand{\labelitemi}{--}
  \item  \textit{Relation between crosscap, EAP states, and lattice models}: This work focused on crosscap dynamics in CFT and naturally calls for EAP-state calculations on lattices; yet only two TFIM EAP states are known to flow to the Ising field theory crosscaps at criticality~\cite{Zhang:2024rnh,Harada:2025uhh}, and the broader crosscap–EAP link is still unknown. 
  \item \textit{Probes of multipartite entanglement}: As suggested by the graph-like structure of entanglement at late times after the quench, we anticipate robust multipartite entanglement patterns. It will be particularly interesting to use information-theoretic quantities beyond entanglement entropy and mutual information to better characterize such entanglement features. 
  \item \textit{Dissipative preparation of crosscap states}:  Previous works have shown that certain pure states with volume law entanglement, such as the so-called rainbow state, could be engineered using Lindbladians with local 
  dissipation~\cite{PhysRevB.105.L140301}. It would be fruitful to understand whether the crosscap state can similarly be prepared in many-body lattice systems. Such dissipative protocols could be used to engineer crosscap states and experimentally study their nonequilibrium dynamics.
  \item \textit{Beyond }$\mathrm{SL}^{(q)}(2,\mathbb{R})$: For $\mathrm{SL}^{(q)}(2,\mathbb{R})$ deformations, a conformal map uniformizes the post-quench coordinates to the complex plane, simplifying calculations. This works because all such deformations lie on Virasoro coadjoint orbits of type $S^1$ or $\text{PSL}(2,\mathbb{R})$~\cite{Witten:1987ty}. Two additional orbits, $T_{n,\Delta}$ and $\tilde{T}_{n,\pm}$~\cite{Witten:1987ty}, can lead to quenches that are much harder to treat directly~\cite{Li:2025rzl}. Nevertheless, since the graph-like patterns are insensitive to Hamiltonian details, their thermalization features can still be assessed via the quasiparticle picture.
\end{itemize}

\noindent\textit{Note added:} Upon completion of this manuscript, we became aware of the recent preprint~\cite{Chen:2025wzd}, the results of which partially overlap with our Section~\ref{sec:uniformcrosscap}, in particular with the formula \eqref{eq:Uniform-Crosscap-Quench-Fermion-EE} for the single interval entanglement entropy.

\section*{Acknowledgements}

We thank Masahiro Nozaki, Akihiro Miyata, Zixia Wei, Hong-Hao Tu, Xueda Wen, Weibo Mao, Jia Tian, Bo-Ting Chen, Dongyeob Kim, Wucheng Zhang, Nathan Seiberg, Liang-Hong Mo, Bing-Xin Lao, Minxuan Wang and Zihan Zhou for inspiring discussions.
B.L. acknowledges financial support from the Swiss National Science Foundation (Postdoc.Mobility Grant No. 214461). S.R. is supported by a Simons
Investigator Grant from the Simons Foundation (Award No. 566116). M.T.T. acknowledges funding from the European Research
Council (ERC) through the Advanced grant QUEST (Grant Agreement No. 101096208).
The author(s) would like to thank the Isaac Newton Institute for Mathematical Sciences, Cambridge, for support and hospitality during the programme Quantum field theory with boundaries, impurities, and defects, where work on this paper was undertaken. This work was supported by EPSRC grant EP/Z000580/1.

\appendix

\section{Entanglement Entropy and Mutual Information for Dirac Fermions\label{app:EEFDF}}
In this appendix, we derive the expressions for $S_A^{(n)}(t)$, $S_{A\cup B}^{(n)}(t)$ and $I_{A,B}^{(n)}(t)$ after crosscap quenches for free Dirac fermions. 
For simplicity and notation consistency with~\cite{Takayanagi:2010wp}, we set $L=2\pi$, $w\to y=\tau+i\sigma,\bar{w}\to\bar{y}=\tau-i\sigma$, $\beta\to 4\epsilon$ and $\tau_{\text{mod.}}=\frac{i2\epsilon}{\pi}$. We recall that $w$ does not represent the coordinate but winding number. 

The (1+1)d free massless Dirac fermion $(\psi,\bar{\psi})$ can be bosonized into a compact boson $X(y,\bar{y})=X_L(y)+X_R(\bar{y})$ following the standard Bosonization procedure~\cite{Senechal:1999us}. Specifically, we have
\be
\begin{split}
    &\psi_L(y)=e^{iX_L(y)},
    \quad
    \bar{\psi}_L(y)=e^{-iX_L(y)},
    \quad \psi_R(\bar{y})=e^{iX_R(\bar{y})},
    \quad
    \bar{\psi}_R(\bar{y})=e^{-iX_R(\bar{y})},
\end{split}
\ee
where our normalization of the (1+1)d CFT is $\alpha'=l_s=2$ in the string theory worldsheet~\cite{Polchinski:1998rq}, or equivalently $4\pi g=1$ as is standard in CFT. The compact boson satisfies 
\be\label{eq:PBC-Compact-Boson}
    X(\tau,\sigma+2\pi)=X(y+2\pi i,\bar{y}-2\pi i)=X(y,\bar{y})-2s\pi w R,
\ee
where $s=\pm1$ (and we choose $s=-1$ in this appendix\footnote{Note that to reproduce the correct expression of entanglement entropy for global quench in~\cite{Takayanagi:2010wp}, $s$ must be $-1$ instead of $+1$ chosen in~\cite{Takayanagi:2010wp}.}) and $w\in\mathbb{Z},R$ are winding number and the circle of radius for the boson, respectively.
Using~\eqref{eq:PBC-Compact-Boson}, the mode expansions for the compact boson are given by
($m\in\mathbb{Z}$)
\be\label{eq:Compact-Boson-Mode-Expansion}
\begin{split}
    &X_L(y)=x_L-isp_Ly+i\sum_{m\neq 0}\frac{e^{-my}}{m}\alpha_m,
    \quad 
    X_R(\bar{y})=x_R-isp_R\bar{y}+i\sum_{m\neq 0}\frac{e^{-m\bar{y}}}{m}\tilde{\alpha}_m,
\end{split}
\ee
where we also introduce a simplified expression $\phi_0=x_L+x_R$. After applying canonical quantization, the ladder operators and zero modes part satisfy
\be\label{eq:Compact-Boson-Canonical-Commutator}
\begin{split}
    &(\alpha_m)^{\dagger}=\alpha_{-m},
    \quad(\tilde{\alpha}_m)^{\dagger}=\tilde{\alpha}_{-m},
    \quad 
    [\alpha_m,\alpha_n]=[\tilde{\alpha}_m,\tilde{\alpha}_n]=m\delta_{m+n,0},
    \nonumber \\
    &
    [x_{i=L,R},p_{j=L,R}]=i\delta_{ij}.
\end{split}
\ee

Besides, the CFT Hamiltonian of the compact boson can be written as
\be
\begin{split}
    &L_0=\frac{p_L^2}{2}+\sum_{m\in \mathbb{Z}^+}\alpha_{-m}\alpha_m=\frac{p_L^2}{2}+N_L,
    \quad [N_L,\alpha_m]=-2m\alpha_m,\\
    &\bar{L}_0=\frac{p_R^2}{2}+\sum_{m\in \mathbb{Z}^+}\tilde{\alpha}_{-m}\tilde{\alpha}_m=\frac{p_R^2}{2}+N_R,
    \quad
[N_R,\tilde{\alpha}_m]=-2m\tilde{\alpha}_m,\\
    &H=\frac{p_L^2+p_R^2}{2}+\sum_{m\in \mathbb{Z}^+}(\alpha_{-m}\alpha_m+\tilde{\alpha}_{-m}\tilde{\alpha}_m)-\frac{c}{12},
\end{split}
\ee
where the zero mode parts $p_L,p_R$ are momenta operators. The eigenstates $\ket{n,w}$ have the corresponding conformal dimensions
\be
    h_{n,w}=\frac{1}{2}\left(\frac{n}{R}+\frac{wR}{2}\right)^2,
    \quad \bar{h}_{n,w}=\frac{1}{2}\left(\frac{n}{R}-\frac{wR}{2}\right)^2,
    \quad 
    n,w\in\mathbb{Z}.
\ee
Note that if one wants to consider open boundary conditions, e.g., Neumann or Dirichlet boundary conditions, the corresponding boundary states are comprised of $\ket{w}=\ket{n=0,w}$ or $\ket{n}=\ket{n,w=0}$, respectively. Specifically, we have
\be\label{eq:Momenta-Eigenstate}
    p_L\ket{n,w}=\left(\frac{n}{R}+\frac{wR}{2}\right)\ket{n,w},
    \quad p_R\ket{n,w}=\left(\frac{n}{R}-\frac{wR}{2}\right)\ket{n,w}.
\ee
Furthermore, as $\ket{n,w}=\ket{h_{n,w}}$ is a primary state, it will be annihilated by $\alpha_m,\tilde{\alpha}_m$ for $m>0$. For later use, we define some notations and present some useful identities. First, we define
\be\label{eq:Useful-Notations-1}
    \alpha=\frac{\alpha_m}{\sqrt{m}},
    \quad \alpha^{\dagger}=\frac{\alpha_{-m}}{\sqrt{m}},
    \quad \beta=\frac{\tilde{\alpha}_m}{\sqrt{m}},
    \quad \beta^{\dagger}=\frac{\tilde{\alpha}_{-m}}{\sqrt{m}},\quad \text{s.t.}\quad [\alpha,\alpha^{\dagger}]=[\beta,\beta^{\dagger}]=1,
\ee
which directly lead to commutators
\be
\begin{split}
    &[\alpha,(\alpha^{\dagger})^k]=k(\alpha^{\dagger})^{k-1},
    \quad [\beta,(\beta^{\dagger})^k]=k(\beta^{\dagger})^{k-1},\quad [\mu\alpha^{\dagger}+\nu\beta^{\dagger},\gamma\alpha+\delta\beta]=-\mu\gamma-\nu\delta.
\end{split}
\ee
The crosscap states are solutions of $\left(L_q-(-1)^q\bar{L}_{-q}\right)\ket{C}=0$~\cite{Ishibashi:1988kg}. In particular, two typical solutions are
\be\label{eq:Compact-Boson-Crosscap-States}
\begin{split}
        \ket{C_{\text{O0}}}&=\mathcal{N}_{\text{O0}}e^{\sum_{m\in\mathbb{Z}^+}\frac{(-1)^m}{m}\alpha_{-m}\tilde{\alpha}_{-m}}\sum_{n\in\mathbb{Z}}\ket{2n},\\
        \ket{C_{\text{O1}}}&=\mathcal{N}_{\text{O1}}e^{-\sum_{m\in\mathbb{Z}^+}\frac{(-1)^m}{m}\alpha_{-m}\tilde{\alpha}_{-m}}\sum_{w\in\mathbb{Z}}\ket{2w},\\
\end{split}
\ee
where  $\mathcal{N}_{\text{O0}},\mathcal{N}_{\text{O1}}$ depend on $R$ and are fixed by the crosscap Cardy condition~\cite{Blumenhagen:2009zz,Tan:2024dcd}.
We stress that the following calculations for $\ket{C_{O0}},\ket{C_{O1}}$ can be performed independently, and in the high-temperature regime entanglement entropy and mutual information are insensitive to their distinction. Moreover, by setting $R=\sqrt{2}$, they share the same expressions of entanglement entropy. Hence, we only consider $\ket{C_{O0}}$ here as an illustrative example and present the computational details below.

\subsection{Two-Point function}
Here, we compute
\be\label{eq:Two-Pt-Fn-O0-Crosscap}
    \frac{\bra{C_{\text{O0}}}e^{-2\epsilon H}V_{k_L^{(1)},k_R^{(1)}}(y_1,\bar{y}_1)V_{k_L^{(2)},k_R^{(2)}}(y_2,\bar{y}_2)\ket{C_{\text{O0}}}}{\bra{C_{\text{O0}}}e^{-2\epsilon H}\ket{C_{\text{O0}}}},
\ee
where we set $k=k_L^{(1)}=k_R^{(1)}=-k_L^{(2)}=-k_R^{(2)}$ for $\ket{C_{\text{O0}}}$ state.\footnote{For state $\ket{C_{O1}}$, we choose $k=k_L^{(1)}=-k_R^{(1)}=k_L^{(2)}=-k_R^{(2)}$ instead. The single-valuedness of the vertex-operator OPE requires ${k_L^{(j)}}^2={k_R^{(j)}}^2$, which admits the two branches $k_L^{(j)}=\pm k_R^{(j)}$, \cite{Polchinski:1998rq}.} Using the BCH formula $e^{X}e^{Y}=e^{X+Y}e^{\frac{[X,Y]}{2}},~\text{iff }[X,[X,Y]]=[Y,[X,Y]]=0$, the zero mode part is given by
{\small \be
\begin{split}
    &e^{-\frac{\epsilon}{6}}\sum_{n\in\mathbb{Z}}\bra{2n}e^{-\epsilon(p_L^2+p_R^2)}e^{i[k_L^{(1)}(x_L-isy_1p_L)+k_R^{(1)}(x_R-is\bar{y}_1p_R)]}e^{i[k_L^{(2)}(x_L-isy_2p_L)+k_R^{(2)}(x_R-is\bar{y}_2p_R)]}\ket{2n}\\
    &=e^{-\frac{\epsilon}{6}}e^{\frac{k^2}{2}\sum_{j=1}^2(-1)^j\left[y_j+\bar{y}_j\right]}\sum_{n\in\mathbb{Z}}e^{-\frac{8\epsilon n^2}{R^2}}e^{\frac{2[\sum_{j=1}^2(-1)^j(y_j+\bar{y}_j)]n}{R}}
\end{split}
\ee}
where $p_L\ket{2n}=p_R\ket{2n}=\frac{2n}{R}\ket{2n}$. The zero mode part of the partition function is obtained as
\be
    e^{-\frac{\epsilon}{6}}\sum_{n\in\mathbb{Z}}\bra{2n}e^{-\epsilon(p_L^2+p_R^2)}\ket{2n}=e^{-\frac{\epsilon}{6}}\sum_{n\in\mathbb{Z}}e^{-\frac{8\epsilon n^2}{R^2}}.
\ee
In summary, the zero mode part of~\eqref{eq:Two-Pt-Fn-O0-Crosscap} reads
\be\label{eq:O0-Two-Pt-Zero-Mode-Part-Result}
    \left[\frac{\sum_{n\in\mathbb{Z}}e^{-\frac{8\epsilon n^2}{R^2}}e^{\frac{2[\sum_{j=1}^2(-1)^j(y_j+\bar{y}_j)]n}{R}}}{\sum_{n\in\mathbb{Z}}e^{-\frac{8\epsilon n^2}{R^2}}}\right]\cdot e^{\frac{k^2}{2}\sum_{j=1}^2(-1)^j\left[y_j+\bar{y}_j\right]}.
\ee
Next, applying the normal ordering, the oscillating mode part for the vertex operators is given by
\be
\begin{split}
    &\prod_{j=1}^2e^{\sum_{m>0}(\frac{k_L^{(j)}e^{my_j}}{\sqrt{m}}\alpha^{\dagger}+\frac{k_R^{(j)}e^{m\bar{y}_j}}{\sqrt{m}}\beta^{\dagger})}e^{-\sum_{m>0}(\frac{k_L^{(j)}e^{-my_j}}{\sqrt{m}}\alpha+\frac{k_R^{(j)}e^{-m\bar{y}_j}}{\sqrt{m}}\beta)}\\
    =&\prod_{m=1}^{\infty}e^{a_L\alpha+a_R\beta}e^{b_L\alpha^{\dagger}+b_R\beta^{\dagger}}e^{\frac{\left[k_L^{(1)}k_L^{(2)}e^{m(y_1-y_2)}+k_R^{(1)}k_R^{(2)}e^{m(\bar{y}_1-\bar{y}_2)}\right]}{m}}e^{\frac{\sum_{j=1}^2[k_L^{(j)}+k_R^{(j)}]}{m}},\\
    a_L&=-\sum_{j=1}^2\left(\frac{k_{L}^{(j)}e^{-my_j}}{\sqrt{m}}\right)=-\frac{k}{\sqrt{m}}\left(e^{-my_1}-e^{-my_2}\right), \\ a_R&=-\sum_{j=1}^2\left(\frac{k_{R}^{(j)}e^{-m\bar{y}_j}}{\sqrt{m}}\right)=-\frac{k}{\sqrt{m}}\left(e^{-m\bar{y}_1}-e^{-m\bar{y}_2}\right),\\
    b_L&=\sum_{j=1}^2\left(\frac{k_{L}^{(j)}e^{my_j}}{\sqrt{m}}\right)=\frac{k}{\sqrt{m}}\left(e^{my_1}-e^{my_2}\right),~b_R=\sum_{j=1}^2\left(\frac{k_{R}^{(j)}e^{m\bar{y}_j}}{\sqrt{m}}\right)=\frac{k}{\sqrt{m}}\left(e^{m\bar{y}_1}-e^{m\bar{y}_2}\right),
\end{split}
\ee
which leads to the following expression of the oscillating mode part of~\eqref{eq:Two-Pt-Fn-O0-Crosscap}:
{\footnotesize
\be\label{eq:O0-Two-Pt-Oscillating-Mode-Part}
\begin{split}
    &\frac{\prod_{m=1}^{\infty}e^{\frac{\left[k_L^{(1)}k_L^{(2)}e^{m(y_1-y_2)}+k_R^{(1)}k_R^{(2)}e^{m(\bar{y}_1-\bar{y}_2)}\right]}{m}}\bra{2n}e^{(-1)^m\alpha\beta}e^{-2m\epsilon(\alpha^{\dagger}\alpha+\beta^{\dagger}\beta)}e^{a_L\alpha+a_R\beta}e^{b_L\alpha^{\dagger}+b_R\beta^{\dagger}}e^{(-1)^m\alpha^{\dagger}\beta^{\dagger}}\ket{2n}}{\prod_{m=1}^{\infty}\bra{2n}e^{(-1)^m\alpha\beta}e^{-2m\epsilon(\alpha^{\dagger}\alpha+\beta^{\dagger}\beta)}e^{(-1)^m\alpha^{\dagger}\beta^{\dagger}}\ket{2n}\left[\prod_{m=1}^{\infty}e^{\frac{\sum_{j=1}^2[k_L^{(j)}+k_R^{(j)}]}{m}}\right]^{-1}}.
\end{split}
\ee
}
We first look at the numerator of the above expression. Notice that
\be
\bra{2n}e^{(-1)^m\alpha\beta}e^{-2m\epsilon(\alpha^{\dagger}\alpha+\beta^{\dagger}\beta)}=\bra{2n}\exp\left(e^{-4m\epsilon}[(-1)^m\alpha\beta]\right).
\ee
Then, the matrix element of the numerator is simplified using the identity~\cite{Takayanagi:2010wp}
\be\label{eq:Important-Identity-1}
    \bra{n,w}e^{d\alpha\beta}e^{a_L\alpha+a_R\beta}e^{a_L\alpha^{\dagger}+b_L\beta^{\dagger}}e^{c\alpha^{\dagger}\beta^{\dagger}}\ket{n,w}=\frac{\exp\left[\frac{a_Lb_L+a_Rb_R+ca_La_R+db_Lb_R}{1-cd}\right]}{1-cd},
    \quad |cd|<1,
\ee
which is $w$- and $n$-independent and only depends on $m$, as
\be\label{eq:O0-Two-Pt-Oscillating-Mode-Part-Numerator-Lego}
\begin{split}
    &\bra{2n}e^{(-1)^m\alpha\beta}e^{-2m\epsilon(\alpha^{\dagger}\alpha+\beta^{\dagger}\beta)}e^{a_L\alpha+a_R\beta}e^{b_L\alpha^{\dagger}+b_R\beta^{\dagger}}e^{(-1)^m\alpha^{\dagger}\beta^{\dagger}}\ket{2n}\\
    =& \frac{1}{1-z}\exp\left(\frac{a_Lb_L+a_Rb_R+(-1)^m[a_La_R+b_Lb_R]}{1-z}-(-1)^mb_Lb_R\right)\\
    c&=(-1)^m,~d=(-1)^me^{-4m\epsilon}=(-1)^mz=(-1)^mq^m,~z=q^m=e^{-4m\epsilon}.
\end{split}
\ee
On the other hand, the denominator reads
\be
    \prod_{m=1}^{\infty}\bra{2n}e^{(-1)^m\alpha\beta}e^{-2m\epsilon(\alpha^{\dagger}\alpha+\beta^{\dagger}\beta)}e^{(-1)^m\alpha^{\dagger}\beta^{\dagger}}\ket{2n}=\prod_{m=1}^{\infty}\frac{1}{1-q^m}=\frac{q^{\frac{1}{24}}}{\eta(\tau_{\text{mod.}})},
    \quad \tau_{\text{mod.}}=i\frac{2\epsilon}{\pi}.
\ee
Combining~\eqref{eq:O0-Two-Pt-Oscillating-Mode-Part},~\eqref{eq:O0-Two-Pt-Oscillating-Mode-Part-Numerator-Lego},~\eqref{eq:O0-Two-Pt-Oscillating-Mode-Part-Denominator} and~\eqref{eq:Dedekind-eta-Fn}, we therefore obtain the oscillating mode part of~\eqref{eq:Two-Pt-Fn-O0-Crosscap}
\be\label{eq:O0-Two-Pt-Oscillating-Mode-Part-Denominator}
\begin{split}
    \prod_{m=1}^{\infty}\left[e^{\frac{4k^2}{m}}e^{-\frac{k^2}{m}[e^{m(y_1-y_2)}+e^{m(\bar{y}_1-\bar{y}_2)}]}\cdot\exp\left(\frac{a_Lb_L+a_Rb_R+(-1)^m[a_La_R+b_Lb_R]}{1-z}-(-1)^mb_Lb_R\right)\right],
\end{split}
\ee
where 
\be
\begin{split}
    &\frac{a_Lb_L+a_Rb_R+(-1)^m[a_La_R+b_Lb_R]}{1-z}-(-1)^mb_Lb_R\\
    =&\frac{k^2}{m(1-z)}\bigg[-4+e^{m(y_1-y_2)}+e^{-m(y_1-y_2)}+e^{m(\bar{y}_1-\bar{y}_2)}+e^{-m(\bar{y}_1-\bar{y}_2)}\bigg]\\
    +&\frac{(-1)^mk^2}{m(1-z)}\bigg[\sum_{j=1}^2\left(e^{m(y_j+\bar{y}_j)}+e^{-m(y_j+\bar{y}_j)}\right)-e^{-m(y_1+\bar{y}_2)}-e^{m(y_1+\bar{y}_2)}-e^{-m(\bar{y}_1+y_2)}-e^{m(\bar{y}_1+y_2)}\bigg]\\
    -&\frac{(-1)^mk^2}{m}\bigg[\sum_{j=1}^2e^{m(y_j+\bar{y}_j)}-e^{m(y_1+\bar{y}_2)}-e^{m(\bar{y}_1+y_2)}\bigg].
\end{split}
\ee

The following identities are useful when summing over the modes:
\be
\begin{split}
    &\prod_{m\in\mathbb{Z}^+}e^{-\frac{x^m}{m}}=1-x,~\prod_{m\in\mathbb{Z}^+}e^{-\frac{(-1)^mx^m}{m}}=1+x,~\prod_{m\in\mathbb{Z}^+}\frac{1}{1-z}=\frac{q^{\frac{1}{24}}}{\eta(\tau_{\text{mod.}})},\\
    &\prod_{m\in\mathbb{Z}^+}e^{\frac{x^m}{m}}=\frac{1}{1-x},~\prod_{m\in\mathbb{Z}^+}e^{\frac{(-1)^mx^m}{m}}=\frac{1}{1+x},\\
    &\prod_{m\in\mathbb{Z}^+}e^{\frac{x^m}{m(1-z)}}=\prod_{m'\in\mathbb{Z}}\frac{1}{1-xq^{m'}},~\prod_{m\in\mathbb{Z}^+}e^{\frac{(-1)^mx^m}{m(1-z)}}=\prod_{m'\in\mathbb{Z}}\frac{1}{1+xq^{m'}},\\
    &\prod_{m\in\mathbb{Z}^+}e^{\frac{zx^m}{m(1-z)}}=\prod_{m'\in\mathbb{Z}}\frac{1}{1-xq^{m'+1}},~\prod_{m\in\mathbb{Z}^+}e^{\frac{z(-1)^mx^m}{m(1-z)}}=\prod_{m'\in\mathbb{Z}}\frac{1}{1+xq^{m'+1}},
\end{split}
\ee
where the modular parameter is $\tau_{\text{mod.}}=i\frac{2\epsilon}{\pi}$ and $z=q^m,q=e^{-4\epsilon}$. Additionally, the Dedekind eta function and Jacobi theta functions are defined as
\be\label{eq:Dedekind-eta-Fn}
\begin{split}
    &\eta(\tau_{\text{mod.}})=q^{\frac{1}{24}}\prod_{m\in\mathbb{Z}^+}(1-q^m),~x=e^{2\pi i z},~q=e^{2\pi i\tau_{\text{mod.}}},\\
    &\theta_1(z|\tau_{\text{mod.}})=-ix^{\frac{1}{2}}q^{\frac{1}{8}}\prod_{m\in\mathbb{Z}^+}(1-q^m)\prod_{m\in\mathbb{N}}(1-xq^{m+1})(1-x^{-1}q^m),\\
    &\theta_2(z|\tau_{\text{mod.}})=x^{\frac{1}{2}}q^{\frac{1}{8}}\prod_{m\in\mathbb{Z}^+}(1-q^m)\prod_{m\in\mathbb{N}}(1+xq^{m+1})(1+x^{-1}q^m).
\end{split}
\ee
Using the above identities,~\eqref{eq:O0-Two-Pt-Oscillating-Mode-Part-Denominator} can be calculated as follows:
{ \be\label{eq:O0-Two-Pt-Oscillating-Mode-Part-Result}
\begin{split}
    &\prod_{m=1}^{\infty}\left[e^{-\frac{k^2}{m}[e^{m(y_1-y_2)}+e^{m(\bar{y}_1-\bar{y}_2)}]}\cdot\exp\left(\frac{a_Lb_L+a_Rb_R+(-1)^m[a_La_R+b_Lb_R]}{1-z}-(-1)^mb_Lb_R\right)\right]~\\
    =&\left[-\frac{\eta^6(\tau_{\text{mod.}})e^{-\frac{1}{2}\sum_{j=1}^2(-1)^j[y_j+\bar{y}_j]}\theta_2\left(\frac{y_1+\bar{y}_2}{2\pi i}|\tau_{\text{mod.}}\right)\theta_2\left(\frac{\bar{y}_1+y_2}{2\pi i}|\tau_{\text{mod.}}\right)}{\theta_1\left(\frac{y_1-y_2}{2\pi i}|\tau_{\text{mod.}}\right)\theta_1\left(\frac{\bar{y}_1-\bar{y}_2}{2\pi i}|\tau_{\text{mod.}}\right)\prod_{j=1}^2\theta_2\left(\frac{y_j+\bar{y}_j}{2\pi i}|\tau_{\text{mod.}}\right)}\right]^{k^2}.
\end{split}
\ee}
Combining~\eqref{eq:O0-Two-Pt-Zero-Mode-Part-Result} and~\eqref{eq:O0-Two-Pt-Oscillating-Mode-Part-Result},~\eqref{eq:Two-Pt-Fn-O0-Crosscap} is explicitly given by
{\small \be\label{eq:Two-Pt-Fn-O0-Crosscap-Result}
\begin{split}
    &\frac{\bra{C_{\text{O0}}}e^{-2\epsilon H}V_{k_L^{(1)},k_R^{(1)}}(y_1,\bar{y}_1)V_{k_L^{(2)},k_R^{(2)}}(y_2,\bar{y}_2)\ket{C_{\text{O0}}}}{\bra{C_{\text{O0}}}e^{-2\epsilon H}\ket{C_{\text{O0}}}}\\
    =&\left[\frac{\sum_{n\in\mathbb{Z}}e^{-\frac{8\epsilon n^2}{R^2}}e^{\frac{2[\sum_{j=1}^2(-1)^j(y_j+\bar{y}_j)]n}{R}}}{\sum_{n\in\mathbb{Z}}e^{-\frac{8\epsilon n^2}{R^2}}}\right]\left[-\frac{\eta^6(\tau_{\text{mod.}})\theta_2\left(\frac{y_1+\bar{y}_2}{2\pi i}|\tau_{\text{mod.}}\right)\theta_2\left(\frac{\bar{y}_1+y_2}{2\pi i}|\tau_{\text{mod.}}\right)}{\theta_1\left(\frac{y_1-y_2}{2\pi i}|\tau_{\text{mod.}}\right)\theta_1\left(\frac{\bar{y}_1-\bar{y}_2}{2\pi i}|\tau_{\text{mod.}}\right)\prod_{j=1}^2\theta_2\left(\frac{y_j+\bar{y}_j}{2\pi i}|\tau_{\text{mod.}}\right)}\right]^{k^2}.
\end{split}
\ee}
By convention in this work, we set $R=\sqrt{2}$ \footnote{$R=\sqrt{2}$ gives rise to the same Klein bottle partition function for $\ket{C_{O0}}$ and $\ket{C_{O1}}$~\cite{Tan:2024dcd}.}, which simplifies the first bracket to
\be
    \left[\frac{\sum_{n\in\mathbb{Z}}e^{-4\epsilon n^2}e^{\sqrt{2}[\sum_{j=1}^2(-1)^j(y_j+\bar{y}_j)]n}}{\sum_{n\in\mathbb{Z}}e^{-4\epsilon n^2}}\right]=\frac{\theta_3\left(\frac{k\sum_{j=1}^2(-1)^j[y_j+\bar{y}_j]}{\sqrt{2}\pi i}|2\tau_{\text{mod.}}\right)}{\theta_3\left(0|2\tau_{\text{mod.}}\right)}.
\ee

\subsection{Four-Point Function}
The computation of the four point function is analogous to the two-point analysis; we, therefore, omit its derivation and simply state the final result:
{
\be\label{eq:Four-Pt-Fn-O0-Crosscap}
\begin{split}
    &\frac{\bra{C_{\text{O0}}}e^{-2\epsilon H}V_{k_L^{(1)},k_R^{(1)}}(y_1,\bar{y}_1)V_{k_L^{(2)},k_R^{(2)}}(y_2,\bar{y}_2)V_{k_L^{(3)},k_R^{(3)}}(y_3,\bar{y}_3)V_{k_L^{(4)},k_R^{(4)}}(y_4,\bar{y}_4)\ket{C_{\text{O0}}}}{\bra{C_{\text{O0}}}e^{-2\epsilon H}\ket{C_{\text{O0}}}}\\
    =&\left[\frac{\sum_{n\in\mathbb{Z}}e^{-8\epsilon n^2}e^{2nk\sum_{j=1}^4(-1)^j[y_j+\bar{y}_j]}}{\sum_{n\in\mathbb{Z}}e^{-8\epsilon n^2}}\right]\cdot e^{\frac{k^2}{2}\sum_{j=1}^4(-1)^j[y_j+\bar{y}_j]}\cdot\left[\prod_{m=1}^{\infty}e^{\frac{8k^2}{m}-\frac{8k^2}{m(1-z)}}\right]\\
    \times&\left[\prod_{j=1}^4\prod_{m=1}^{\infty}e^{-\frac{k^2(-1)^m}{m}e^{-m(y_j+\bar{y}_j)}}e^{\frac{k^2(-1)^m}{m(1-z)}\left(e^{m(y_j+\bar{y}_{j})}+e^{-m(y_j+\bar{y}_{j})}\right)}\right]\\
    \times&\left[\prod_{j=1}^3\prod_{m=1}^{\infty}e^{\frac{k^2(-1)^m}{m}e^{-m(y_j+\bar{y}_{j+1})}}e^{-\frac{k^2(-1)^m}{m(1-z)}\left(e^{m(y_j+\bar{y}_{j})}+e^{-m(y_j+\bar{y}_{j})}\right)}\right]\\
    \times&\left[\prod_{j=1}^3\prod_{m=1}^{\infty}e^{\frac{k^2(-1)^m}{m}e^{-m(\bar{y}_j+y_{j+1})}}e^{-\frac{k^2(-1)^m}{m(1-z)}\left(e^{m(\bar{y}_j+y_{j})}+e^{-m(y_j+\bar{y}_{j})}\right)}\right]\\
    \times&\left[\prod_{j=1}^3\prod_{m=1}^{\infty}e^{-\frac{k^2}{m}e^{m(y_j-y_{j+1}+\bar{y}_j-\bar{y}_{j+1})}}e^{\frac{k^2}{m(1-z)}\left(e^{m(y_j-y_{j+1})}+e^{-m(y_j-y_{j+1})}+e^{m(\bar{y}_j-\bar{y}_{j+1})}+e^{-m(\bar{y}_j-\bar{y}_{j+1})}\right)}\right]\\
    \times&\left[\prod_{j=1}^2\prod_{m=1}^{\infty}e^{\frac{k^2}{m}e^{m(y_j-y_{j+2}+\bar{y}_j-\bar{y}_{j+2})}}e^{-\frac{k^2}{m(1-z)}\left(e^{m(y_j-y_{j+2})}+e^{-m(y_j-y_{j+2})}+e^{m(\bar{y}_j-\bar{y}_{j+2})}+e^{-m(\bar{y}_j-\bar{y}_{j+2})}\right)}\right]\\
    \times&\left[\prod_{j=1}^2\prod_{m=1}^{\infty}e^{-\frac{k^2(-1)^m}{m}e^{-m(y_j+\bar{y}_{j+2})}}e^{\frac{k^2(-1)^m}{m(1-z)}\left(e^{m(y_j+\bar{y}_{j+2})}+e^{-m(y_j+\bar{y}_{j+2})}\right)}\right]\\
    \times&\left[\prod_{j=1}^2\prod_{m=1}^{\infty}e^{-\frac{k^2(-1)^m}{m}e^{-m(\bar{y}_j+y_{j+2})}}e^{\frac{k^2(-1)^m}{m(1-z)}\left(e^{m(\bar{y}_j+y_{j+2})}+e^{-m(\bar{y}_j+y_{j+2})}\right)}\right]\\
    \times&\left[\prod_{m=1}^{\infty}e^{-\frac{k^2}{m}e^{m(y_1-y_{4}+\bar{y}_1-\bar{y}_{4})}}e^{\frac{k^2}{m(1-z)}\left(e^{m(y_1-y_{4})}+e^{-m(y_1-y_{4})}+e^{m(\bar{y}_1-\bar{y}_{4})}+e^{-m(\bar{y}_1-\bar{y}_{4})}\right)} \right]\\
    \times&\left[\prod_{m=1}^{\infty}e^{\frac{k^2(-1)^m}{m}e^{-m(y_1+\bar{y}_{4}+\bar{y}_1+y_{4})}}e^{-\frac{k^2(-1)^m}{m(1-z)}\left(e^{m(y_1+\bar{y}_{4})}+e^{-m(y_1+\bar{y}_{4})}+e^{m(\bar{y}_1+y_{4})}+e^{-m(\bar{y}_1+y_{4})}\right)} \right],\\
\end{split}
\ee
}
which is readily simplified to
{
\be\label{eq:Four-Pt-Fn-O0-Crosscap-Result}
\begin{split}
    &\frac{\bra{C_{\text{O0}}}e^{-2\epsilon H}V_{k_L^{(1)},k_R^{(1)}}(y_1,\bar{y}_1)V_{k_L^{(2)},k_R^{(2)}}(y_2,\bar{y}_2)V_{k_L^{(3)},k_R^{(3)}}(y_3,\bar{y}_3)V_{k_L^{(4)},k_R^{(4)}}(y_4,\bar{y}_4)\ket{C_{\text{O0}}}}{\bra{C_{\text{O0}}}e^{-2\epsilon H}\ket{C_{\text{O0}}}}\\
    =&\left[\frac{\sum_{n\in\mathbb{Z}}e^{-\frac{8\epsilon n^2}{R^2}}e^{\frac{2nk\sum_{j=1}^4(-1)^j[y_j+\bar{y}_j]}{R}}}{\sum_{n\in\mathbb{Z}}e^{-\frac{8\epsilon n^2}{R^2}}}\right]\\
    \times&\left\{\frac{\eta^{12}(\tau_{\text{mod.}}) \left[\prod_{j=1}^3\theta_2\left(-\frac{y_j+\bar{y}_{j+1}}{2\pi i}|\tau_{\text{mod.}}\right)\theta_2\left(-\frac{\bar{y}_j+y_{j+1}}{2\pi i}|\tau_{\text{mod.}}\right)\right]}{\left[\prod_{j=1}^4\theta_2\left(-\frac{y_j+\bar{y}_j}{2\pi i}|\tau_{\text{mod.}}\right)\right]\left[\prod_{j=1}^3\theta_1\left(\frac{y_j-y_{j+1}}{2\pi i}|\tau_{\text{mod.}}\right)\theta_1\left(\frac{\bar{y}_j-\bar{y}_{j+1}}{2\pi i}|\tau_{\text{mod.}}\right)\right] }\right.\\
    \times& \left.\frac{\theta_2\left(-\frac{y_1+\bar{y}_{4}}{2\pi i}|\tau_{\text{mod.}}\right)\theta_2\left(-\frac{\bar{y}_1+y_{4}}{2\pi i}|\tau_{\text{mod.}}\right)\left[\prod_{j=1}^2\theta_1\left(\frac{y_j-y_{j+2}}{2\pi i}|\tau_{\text{mod.}}\right)\theta_1\left(\frac{\bar{y}_j-\bar{y}_{j+2}}{2\pi i}|\tau_{\text{mod.}}\right) \right]}{\theta_1\left(\frac{y_1-y_{4}}{2\pi i}|\tau_{\text{mod.}}\right)\theta_1\left(\frac{\bar{y}_1-\bar{y}_{4}}{2\pi i}|\tau_{\text{mod.}}\right) \left[\prod_{j=1}^2\theta_2\left(-\frac{y_j+\bar{y}_{j+2}}{2\pi i}|\tau_{\text{mod.}}\right)\theta_2\left(-\frac{\bar{y}_j+y_{j+2}}{2\pi i}|\tau_{\text{mod.}}\right)\right]}\right\}^{k^2}.
\end{split}
\ee} Finally, we set $R=\sqrt{2}$ (the self-dual point) per our convention, and the first term simplifies to
\be
\begin{split}
    &\left[\frac{\sum_{n\in\mathbb{Z}}e^{-4\epsilon n^2}e^{\sqrt{2}nk\sum_{j=1}^4(-1)^j[y_j+\bar{y}_j]}}{\sum_{n\in\mathbb{Z}}e^{-4\epsilon n^2}}\right]=\frac{\theta_3\left(\frac{k\sum_{j=1}^4(-1)^j[y_j+\bar{y}_j]}{\sqrt{2}\pi i}|2\tau_{\text{mod.}}\right)}{\theta_3\left(0|2\tau_{\text{mod.}}\right)},
\end{split}
\ee
which is same for both $\ket{C_{O1}}$ and $\ket{C_{O0}}$ at the self-dual point.

\subsection{Entanglement Entropies}
Now, replacing $(y_j,\bar{y}_j)$ by $\left(w^{\text{new}}_j,\bar{w}^{\text{new}}_j\right)$ in~\eqref{eq:REE-Inhomo-Expression}, $L=2\pi$ by generic $L$ and $\epsilon$ by $\frac{\beta}{4}$, the $n$th R\'enyi entanglement entropies for free Dirac fermion are given by
{\small
\be\label{eq:REE-Inhomo-Expression-FFCFT}
\begin{split}
    &S_{A}^{(n)}(t)=\frac{1}{1-n}\sum_{a=-\frac{n-1}{2}}^{\frac{n-1}{2}}\log\left[\frac{\theta_3\left(\frac{2a\sum_{j=1}^2(-1)^j[w^{\text{new}}_j+\bar{w}^{\text{new}}_j]}{\sqrt{2}nL i}|2\tau_{\text{mod.}}\right)}{\theta_3\left(0|2\tau_{\text{mod.}}\right)}\right]\\
    &\quad +\frac{1}{1-n}\sum_{a=-\frac{n-1}{2}}^{\frac{n-1}{2}}\log\frac{\bra{C_{\text{O0}}}e^{-\frac{\beta}{2} H_0}V_{\frac{a}{n},\frac{a}{n}}\left(w^{\text{new}}_1,\bar{w}^{\text{new}}_1\right)V_{-\frac{a}{n},-\frac{a}{n}}\left(w^{\text{new}}_2,\bar{w}^{\text{new}}_2\right)\ket{C_{\text{O0}}}}{\bra{C_{\text{O0}}}e^{-2\frac{\beta}{2} H_0}\ket{C_{\text{O0}}}},\\
    &S_{A\cup B}^{(n)}(t)=\frac{1}{1-n}\sum_{a=-\frac{n-1}{2}}^{\frac{n-1}{2}}\log\left[\frac{\theta_3\left(\frac{2a\sum_{j=1}^4(-1)^j[w^{\text{new}}_j+\bar{w}^{\text{new}}_j]}{\sqrt{2}nL i}|2\tau_{\text{mod.}}\right)}{\theta_3\left(0|2\tau_{\text{mod.}}\right)}\right]+\frac{1}{1-n}\sum_{a=-\frac{n-1}{2}}^{\frac{n-1}{2}}\log\Bigg\{\\
    &\frac{\bra{C_{\text{O0}}}e^{-\frac{\beta}{2} H_0}V_{\frac{a}{n},\frac{a}{n}}\left(w^{\text{new}}_1,\bar{w}^{\text{new}}_1\right)V_{-\frac{a}{n},-\frac{a}{n}}\left(w^{\text{new}}_2,\bar{w}^{\text{new}}_2\right)V_{\frac{a}{n},\frac{a}{n}}\left(w^{\text{new}}_3,\bar{w}^{\text{new}}_3\right)V_{-\frac{a}{n},-\frac{a}{n}}\left(w^{\text{new}}_4,\bar{w}^{\text{new}}_4\right)\ket{C_{\text{O0}}}}{\bra{C_{\text{O0}}}e^{-\frac{\beta}{2} H_0}\ket{C_{\text{O0}}}}\Bigg\},\\
\end{split}
\ee}
where the two- and four-point functions are given by~\eqref{eq:Two-Pt-Fn-O0-Crosscap-Result} and~\eqref{eq:Four-Pt-Fn-O0-Crosscap-Result}, respectively. Particularly, if $H_1=H_0$, $\left(w^{\text{new}}_j,\bar{w}^{\text{new}}_j\right)$ reduces to $(w_j,\bar{w}_j)$ with $j=1,2,3,4$. Accordingly, the first $\log$-terms for $S_{A}^{(n)}(t)$ and $S_{A\cup B}^{(n)}(t)$ vanish. Then, taking von Neumann limit $n\to 1$, the single and double intervals entropies~\eqref{eq:Uniform-Crosscap-Quench-Fermion-EE} are obtained from~\eqref{eq:REE-Inhomo-Expression-FFCFT}. Furthermore, it is easy to derive the $n$th order R\'enyi mutual information as follows:
{\small
\be
\begin{split}
    &I_{A,B}^{(n)}(t)=\frac{1}{1-n}\sum_{a=-\frac{n-1}{2}}^{\frac{n-1}{2}}\log\left[\frac{\theta_3\left(\frac{2a\sum_{j=1}^2(-1)^j[w^{\text{new}}_j+\bar{w}^{\text{new}}_j]}{\sqrt{2}nL i}|2\tau_{\text{mod.}}\right)\theta_3\left(\frac{2a\sum_{j=3}^4(-1)^j[w^{\text{new}}_j+\bar{w}^{\text{new}}_j]}{\sqrt{2}nL i}|2\tau_{\text{mod.}}\right)}{\theta_3\left(0|2\tau_{\text{mod.}}\right)\theta_3\left(\frac{2a\sum_{j=1}^4(-1)^j[w^{\text{new}}_j+\bar{w}^{\text{new}}_j]}{\sqrt{2}nL i}|2\tau_{\text{mod.}}\right)}\right]\\
    &+\frac{c(n+1)}{12n}\times\\
    &\log\left\{\left|\frac{\theta_1\left(i\frac{w^{\text{new}}_3-w^{\text{new}}_1}{L}|\tau_{\text{mod.}}\right)\theta_1\left(i\frac{\bar{w}^{\text{new}}_3-\bar{w}^{\text{new}}_1}{L}|\tau_{\text{mod.}}\right) \theta_1\left(i\frac{w^{\text{new}}_4-w^{\text{new}}_2}{L}|\tau_{\text{mod.}}\right)\theta_1\left(i\frac{\bar{w}^{\text{new}}_4-\bar{w}^{\text{new}}_2}{L}|\tau_{\text{mod.}}\right)}{\theta_1\left(i\frac{w^{\text{new}}_3-w^{\text{new}}_2}{L}|\tau_{\text{mod.}}\right)\theta_1\left(i\frac{\bar{w}^{\text{new}}_3-\bar{w}^{\text{new}}_2}{L}|\tau_{\text{mod.}}\right) \theta_1\left(i\frac{w^{\text{new}}_4-w^{\text{new}}_1}{L}|\tau_{\text{mod.}}\right)\theta_1\left(i\frac{\bar{w}^{\text{new}}_4-\bar{w}^{\text{new}}_1}{L}|\tau_{\text{mod.}}\right)} \right.\right.\\
    &\times \left.\left.\frac{\theta_3\left(i\frac{w^{\text{new}}_3+\bar{w}^{\text{new}}_2}{L}|\tau_{\text{mod.}}\right)\theta_3\left(i\frac{\bar{w}^{\text{new}}_3+{w}^{\text{new}}_2}{L}|\tau_{\text{mod.}}\right) \theta_3\left(i\frac{w^{\text{new}}_4+\bar{w}^{\text{new}}_1}{L}|\tau_{\text{mod.}}\right)\theta_3\left(i\frac{\bar{w}^{\text{new}}_4+{w}^{\text{new}}_1}{L}|\tau_{\text{mod.}}\right)}{\theta_3\left(i\frac{w^{\text{new}}_3+\bar{w}^{\text{new}}_1}{L}|\tau_{\text{mod.}}\right)\theta_3\left(i\frac{\bar{w}^{\text{new}}_3+{w}^{\text{new}}_1}{L}|\tau_{\text{mod.}}\right) \theta_3\left(i\frac{w^{\text{new}}_4+\bar{w}^{\text{new}}_2}{L}|\tau_{\text{mod.}}\right)\theta_3\left(i\frac{\bar{w}^{\text{new}}_4+{w}^{\text{new}}_2}{L}|\tau_{\text{mod.}}\right)}\right|\right\},
\end{split}
\ee
}
where we use the relation $\theta_2(z+\tau/2|\tau)=e^{-i\pi z}q^{-\frac{1}{8}}\theta_3(z|\tau)$.

\section{Entanglement Entropy in Holographic CFT\label{app:EEHCFT}}
In this appendix, we derive analytic expressions for entanglement entropies in two-dimensional holographic CFTs using twist-field techniques. Moreover, we derive~\eqref{eq:qSSD-Asym-MI} from these results.

In the twist-field formalism, we first compute correlation functions in the Euclidean signature, and then rotate to the Lorentzian one to obtain the entanglement entropies. The single-interval $n$th order Euclidean R\'enyi entanglement entropy can be expressed as
\be\label{app-eq:Single-Renyi-EE}
\begin{split}
    S_{A;\text{E}}^{(n)}&=\frac{1}{1-n}\log\left\langle\Psi(\tau)\left|\sigma_n(0,X_2)\bar{\sigma}_n(0,X_1)\right|\Psi(\tau)\right\rangle\\
    &=\frac{h_n}{1-n}\log\left(\prod_{j=1}^2\left[\frac{dw^{\text{new}}_{j}}{dw_{j}}\frac{d\bar{w}^{\text{new}}_{j}}{d\bar{w}_{j}}\right]\right)\\
    &\quad +\frac{1}{1-n}\log\left\langle \sigma_n\left(\tau^{\text{new}}_{X_2}+\frac{\beta}{4},x^{\text{new}}_{X_2}\right)\bar{\sigma}_n\left(\tau^{\text{new}}_{X_1}+\frac{\beta}{4},x^{\text{new}}_{X_1}\right)\right\rangle_{\mathbb{K}^2},
\end{split}
\ee
where $\tau=it$ and the post-quench Euclidean time coordinate is $\tau^{\text{new}}_{X_j}=\frac{w^{\text{new}}_{j}-\bar{w}^{\text{new}}_{j}}{2i}$, and the two-point function is calculated using the doubling trick \cite{Wei:2024kkp,Tsiares:2020ewp}
{\footnotesize
\be\label{eq:Method-Image-Single-Interval}
\begin{split}
    &\left\langle \sigma_n\left(\tau^{\text{new}}_{X_2}+\frac{\beta}{4},x^{\text{new}}_{X_2}\right)\bar{\sigma}_n\left(\tau^{\text{new}}_{X_1}+\frac{\beta}{4},x^{\text{new}}_{X_1}\right)\right\rangle_{\mathbb{K}^2}=\\
    &\sqrt{\left\langle \sigma_n\left(\tau^{\text{new}}_{X_2}+\frac{\beta}{4},x^{\text{new}}_{X_2}\right)\bar{\sigma}_n\left(\tau^{\text{new}}_{X_1}+\frac{\beta}{4},x^{\text{new}}_{X_1}\right)\bar{\sigma}_n\left(\frac{3\beta}{4}-\tau^{\text{new}}_{X_2},x^{\text{new}}_{X_2}+\frac{L}{2}\right)\sigma_n\left(\frac{3\beta}{4}-\tau^{\text{new}}_{X_1},x^{\text{new}}_{X_1}+\frac{L}{2}\right)\right\rangle_{\mathbb{T}^2}}.
\end{split}
\ee
}
Note that the torus two-point functions in the high temperature region are approximately given by the cylinder (parameterized by $x\in\mathbb{R},\tau\in\left[0,\beta\right)$) two-point functions
\be\label{eq:Cylinder-Two-Pt-Fn}
\begin{split}
    &\left\langle\mathcal{O}\left(\tau^{\text{new}}_1,x^{\text{new}}_1\right) \mathcal{O}\left(\tau^{\text{new}}_2,x^{\text{new}}_2\right)\right\rangle_{\mathbb{T}^2}\approx\left\langle\mathcal{O}\left(\tau^{\text{new}}_1,x^{\text{new}}_1\right) \mathcal{O}\left(\tau^{\text{new}}_2,x^{\text{new}}_2\right)\right\rangle_{\text{cylinder}}\\
    &=\left(\frac{\pi}{\beta}\right)^{4h_{\mathcal{O}}}\left(\frac{1}{2}\left[\cosh\left(\frac{2\pi(x^{\text{new}}_1-x^{\text{new}}_2)}{\beta}\right)-\cos\left(\frac{2\pi(\tau^{\text{new}}_1-\tau^{\text{new}}_2)}{\beta}\right)\right]\right)^{-2h_{\mathcal{O}}},
\end{split}
\ee
where we assume $\bar{h}_{\mathcal{O}}=h_{\mathcal{O}}$ and $\tau^{\text{new}}_x=it^{\text{new}}(t=-i\tau,x)$ under the analytic continuation.
For the antipodal double interval $A\cup B$, the Euclidean $n$th order R\'enyi entanglement entropy reads
{\be\label{eq:Antipodal-Renyi-EE}
\begin{split}
    &S_{A\cup B;\text{E}}^{(n)}=\frac{1}{1-n}\log\left\langle\Psi(\tau)\left|\sigma_n(0,X_2)\bar{\sigma}_n(0,X_1)\sigma_n\left(0,X_4\right)\bar{\sigma}_n\left(0,X_3\right)\right|\Psi(\tau)\right\rangle\\
    &=\frac{h_n}{1-n}\log\left(\prod_{j=1}^4\left[\frac{dw^{\text{new}}_{j}}{dw_{j}}\frac{d\bar{w}^{\text{new}}_{j}}{d\bar{w}_{j}}\right]\right)+\frac{1}{1-n}\times\\
    &\log\left\langle \sigma_n\left(\tau^{\text{new}}_{X_2}+\frac{\beta}{4},x^{\text{new}}_{X_2}\right)\bar{\sigma}_n\left(\tau^{\text{new}}_{X_1}+\frac{\beta}{4},x^{\text{new}}_{X_1}\right)\sigma_n\left(\tau^{\text{new}}_{X_4}+\frac{\beta}{4},x^{\text{new}}_{X_4}\right)\bar{\sigma}_n\left(\tau^{\text{new}}_{X_3}+\frac{\beta}{4},x^{\text{new}}_{X_3}\right)\right\rangle_{\mathbb{K}^2},
\end{split}
\ee}
where the four point function can be calculated by
{ \be\label{eq:Method-Image-Antipodal-Interval}
\begin{split}
    &\left\langle \sigma_n\left(\tau^{\text{new}}_{X_2}+\frac{\beta}{4},x^{\text{new}}_{X_2}\right)\bar{\sigma}_n\left(\tau^{\text{new}}_{X_1}+\frac{\beta}{4},x^{\text{new}}_{X_1}\right)\sigma_n\left(\tau^{\text{new}}_{X_4}+\frac{\beta}{4},x^{\text{new}}_{X_4}\right)\bar{\sigma}_n\left(\tau^{\text{new}}_{X_3}+\frac{\beta}{4},x^{\text{new}}_{X_3}\right)\right\rangle_{\mathbb{K}^2}=\\
    &\Bigg\{\Bigg\langle \sigma_n\left(\frac{\beta}{4}+\tau^{\text{new}}_{X_2},x^{\text{new}}_{X_2}\right)\bar{\sigma}_n\left(\frac{\beta}{4}+\tau^{\text{new}}_{X_1},x^{\text{new}}_{X_1}\right)\sigma_n\left(\frac{\beta}{4}+\tau^{\text{new}}_{X_4},x^{\text{new}}_{X_4}\right)\bar{\sigma}_n\left(\frac{\beta}{4}+\tau^{\text{new}}_{X_3},x^{\text{new}}_{X_3}\right) \times\\
    &\bar{\sigma}_n\left(\frac{3\beta}{4}-\tau^{\text{new}}_{X_2},x^{\text{new}}_{X_2}+\frac{L}{2}\right)\sigma_n\left(\frac{3\beta}{4}-\tau^{\text{new}}_{X_1},x^{\text{new}}_{X_1}+\frac{L}{2}\right)\bar{\sigma}_n\left(\frac{3\beta}{4}-\tau^{\text{new}}_{X_4},x^{\text{new}}_{X_4}+\frac{L}{2}\right)\times\\
    &\sigma_n\left(\frac{3\beta}{4}-\tau^{\text{new}}_{X_3},x^{\text{new}}_{X_3}+\frac{L}{2}\right)\Bigg\rangle_{\mathbb{T}^2} \Bigg\}^{\frac{1}{2}}.
\end{split}
\ee}
The entanglement entropy is obtained by taking the von Neumann limit $n\to 1$ and continuing to Lorentzian time $\tau\to it$
\be\label{eq:Von-Neumann-EE}
    S_{\mathcal{V}=A,A\cup B}(t)=\lim_{n\to 1}S_{\mathcal{V};\text{E}}^{(n)}|_{\tau\to it}.
\ee
In order to proceed, we note the following useful relations
\be\label{eq:Useful-Relations-1}
\begin{split}
    \lim_{n\to 1}&-\frac{h_n}{(1-n)}=\frac{c}{12},
    \quad 
    h_n=\frac{c(n^2-1)}{24n},
    \quad \bar{h}_n=h_n,\\
    \lim_{n\to 1}&\frac{1}{2(1-n)}\log\left\langle\sigma_n\left(\tau^{\text{new}}_1,x^{\text{new}}_1\right) \bar{\sigma}_n\left(\tau^{\text{new}}_2,x^{\text{new}}_2\right)\right\rangle_{\mathbb{T}^2}\\
    =&\frac{c}{6}\log\left(\frac{\beta}{\pi}\right)+\frac{c}{12}\log\left(\frac{1}{2}\left|\cosh\left(\frac{2\pi(x^{\text{new}}_1-x^{\text{new}}_2)}{\beta}\right)-\cos\left(\frac{2\pi(\tau^{\text{new}}_1-\tau^{\text{new}}_2)}{\beta}\right)\right|\right).
\end{split}
\ee
The above expressions are valid for the Euclidean signature; in order to guarantee that they also apply for the Lorentzian signature, we define the angular coordinates
\be\label{eq:Post-Quench-Angular-Coordinates}
    \left(w^{\text{new}}=\frac{iL\varphi}{2q\pi},\bar{w}^{\text{new}}=-\frac{iL\bar{\varphi}}{2q\pi}\right)\bigg|_{\tau=it}.
\ee
Recall that we have defined the post-quench space and time coordinates:
\be\label{eq:Post-Quench-Space-Time-Coordinates}
    x^{\text{new}}(t,x)=x^{\text{new}}_x=\frac{L}{4q\pi}\left(\varphi(t,x)+\bar{\varphi}(t,x)\right),~t^{\text{new}}(t,x)=t^{\text{new}}_x=\frac{L}{4q\pi}\left(\varphi(t,x)-\bar{\varphi}(t,x)\right).
\ee
Then, using $(t^{\text{new}},x^{\text{new}})$ together with~\eqref{eq:Cylinder-Two-Pt-Fn} and~\eqref{eq:Method-Image-Single-Interval}, the single-interval entanglement entropy is given by
\be\label{eq:Single-Interval-EE-Expressions}
\begin{split}
    S_{A}(t)&=S_{A}^{\text{U.}}+\text{Min}\left[S_{A;(a)}^{\text{con.}},S_{A;(b)}^{\text{con.}},S_{A}^{\text{dis.}}\right],\\
    S_{A}^{\text{U.}}&=\frac{c}{3}\log\left(\frac{\beta}{\pi}\right)-\frac{c}{12}\log\left(\prod_{j=1}^2\left[\frac{dw^{\text{new}}_{j}}{dw_{j}}\frac{d\bar{w}^{\text{new}}_{j}}{d\bar{w}_{j}}\right]\right)\bigg|_{\tau=it},\\
    S_{A;(a)}^{\text{con.}}&=\frac{c}{6}\log\left(\sinh\left[\frac{L}{2q\beta}(\varphi_{(t,X_1)}-\varphi_{(t,X_2)})\right]\sinh\left[\frac{L}{2q\beta}(\bar{\varphi}_{(t,X_1)}-\bar{\varphi}_{(t,X_2)})\right]\right),\\
    &= \frac{c}{6}\log\left(\frac{1}{2}\left|\cosh\left(\frac{2\pi}{\beta}(x_{X_1}^{\text{new}}-x_{X_2}^{\text{new}})\right)-\cosh\left(\frac{2\pi}{\beta}(t_{X_1}^{\text{new}}-t_{X_2}^{\text{new}})\right)\right|\right),\\
    S_{A;(b)}^{\text{con.}}&=\frac{c}{6}\log\left(\sinh\left[\frac{L}{2q\beta}\left(2q\pi-(\varphi_{(t,X_1)}-\varphi_{(t,X_2)})\right)\right]\sinh\left[\frac{L}{2q\beta}\left(2q\pi-(\bar{\varphi}_{(t,X_1)}-\bar{\varphi}_{(t,X_2)})\right)\right]\right)\\
    &= \frac{c}{6}\log\left(\frac{1}{2}\left|\cosh\left(\frac{2\pi}{\beta}(L-[x_{X_1}^{\text{new}}-x_{X_2}^{\text{new}}])\right)-\cosh\left(\frac{2\pi}{\beta}(t_{X_1}^{\text{new}}-t_{X_2}^{\text{new}})\right)\right|\right),\\
    S_{A}^{\text{dis.}}&=\frac{c}{12}\sum_{j=1}^2\log\left(\frac{1}{2}\left[\cosh\left(\frac{\pi L}{\beta}\right)+\cosh\left(\frac{4\pi}{\beta}t^{\text{new}}_{X_j}\right)\right]\right)\geq\text{Min}\left[S_{A;(a)}^{\text{con.}},S_{A;(b)}^{\text{con.}}\right],
\end{split}
\ee
where the labels ``con.'' and ``dis.'' denote the connected (thermal) and disconnected (wormhole) phases on the gravity side, respectively, as detailed in Sec.~\ref{sec:Holo-Dual} and Appendix~\ref{app:RTHRT}.
Similarly, for antipodal double interval $A\cup B$ with $B=\overline{A}$, the entanglement entropy can be obtained by
\be\label{eq:Antipodal-EE}
\begin{split}
    &S_{A_\cup B}(t)=S_{A\cup B}^{\text{U.}}+\text{Min}\left[S_{A\cup B}^{\text{dis.}},S_{A\cup B}^{\text{con.}}\right],~S_{A\cup B}^{\text{con.}}=\text{Min}\left[S_{A\cup B}^{a},S_{A\cup B}^{b}\right],\\
    &S_{A\cup B}^{\text{U.}}=-\frac{c}{12}\sum_{j=1}^4\log\left(\frac{dw^{\text{new}}_{j}}{dw_{j}}\cdot\frac{d\bar{w}^{\text{new}}_{j}}{d\bar{w}_{j}}\right)+\frac{2c}{3}\log\left(\frac{\beta}{\pi}\right),\\
    &S_{A\cup B}^a=\frac{c}{6}\log\left(\frac{1}{2}\left|\cosh\left[\frac{2\pi}{\beta}\left(x^{\text{new}}_{X_1}-x^{\text{new}}_{X_2}\right)\right]-\cosh\left[\frac{2\pi}{\beta}\left(t^{\text{new}}_{X_1}-t^{\text{new}}_{X_2}\right)\right]\right|\right)\\
    &\qquad +\frac{c}{6}\log\left(\frac{1}{2}\left|\cosh\left[\frac{2\pi}{\beta}\left(x^{\text{new}}_{X_3}-x^{\text{new}}_{X_4}\right)\right]-\cosh\left[\frac{2\pi}{\beta}\left(t^{\text{new}}_{X_3}-t^{\text{new}}_{X_4}\right)\right]\right|\right),\\
    &S_{A\cup B}^b=\frac{c}{6}\log\left(\frac{1}{2}\left|\cosh\left[\frac{2\pi}{\beta}\left(L-[x^{\text{new}}_{X_3}-x^{\text{new}}_{X_2}]\right)\right]-\cosh\left[\frac{2\pi}{\beta}\left(t^{\text{new}}_{X_3}-t^{\text{new}}_{X_2}\right)\right]\right|\right)\\
    &\qquad 
    +\frac{c}{6}\log\left(\frac{1}{2}\left|\cosh\left[\frac{2\pi}{\beta}\left(x^{\text{new}}_{X_4}-x^{\text{new}}_{X_1}\right)\right]-\cosh\left[\frac{2\pi}{\beta}\left(t^{\text{new}}_{X_4}-t^{\text{new}}_{X_1}\right)\right]\right|\right),\\
    &S^{\text{dis.}}_{A\cup B}=\text{Min}\left[S^{c}_{A\cup B},S^{d}_{A\cup B}\right],\\
    &S^{c}_{A\cup B}=\frac{c}{6}\sum_{j=1}^2\log\left(\frac{1}{2}\left|\cosh\left[\frac{2\pi}{\beta}\left(x^{\text{new}}_{X_{j+2}}-x^{\text{new}}_{X_j}-\frac{L}{2}\right)\right]+\cosh\left[\frac{2\pi}{\beta}\left(t^{\text{new}}_{X_{j+2}}+t^{\text{new}}_{X_j}\right)\right]\right|\right),\\
    &S^{d}_{A\cup B}=\frac{c}{12}\sum_{j=1}^4\log\left(\frac{1}{4}\left|\cosh\left[\frac{\pi L}{\beta}\right]+\cosh\left[\frac{4\pi t^{\text{new}}_{X_j}}{\beta}\right]\right|\right).\\ 
\end{split}
\ee
Once the angular functions $\varphi(t,x)$ and $\bar{\varphi}(t,x)$ or post-quench space and time coordinates $(t^{\text{new}},x^{\text{new}})$ are obtained, we substitute them into~\eqref{eq:Single-Interval-EE-Expressions} and~\eqref{eq:Antipodal-EE} to compute $S_{A}(t)$ and $S_{A\cup B}(t)$ in holographic CFTs, explicitly. We now present the details of angular functions/post-quench space and time coordinates case by case.

\subsection{$q$-M\"obius Evolution}
The derivatives contributing to Jacobian factors are given by
\be
\begin{split}
    &\frac{dw^{\text{new}}_x}{dw_x}=\frac{A(t)D(t)-B(t)C(t)}{\left[A(t)e^{i\frac{q\pi x}{L}}+B(t)e^{-i\frac{q\pi x}{L}}\right]\left[C(t)e^{i\frac{q\pi x}{L}}+D(t)e^{-i\frac{q\pi x}{L}}\right]},\\
    &\frac{d\bar{w}^{\text{new}}_x}{d\bar{w}_x}=\frac{A(t)D(t)-B(t)C(t)}{\left[A(t)e^{-i\frac{q\pi x}{L}}+B(t)e^{i\frac{q\pi x}{L}}\right]\left[C(t)e^{-i\frac{q\pi x}{L}}+D(t)e^{i\frac{q\pi x}{L}}\right]},
\end{split}
\ee
where $A(t),B(t),C(t),D(t)$ are periodic functions of $t$ and are given in~\eqref{eq:Post-Quench-Coordinate-Coe-Mobius}. In this case, the angular functions are very complicated and are periodic functions of $t$. Thus, we present the procedures to define them numerically. First, we define
\be
\begin{split}
    &z^{\text{new}}_x=\frac{A(t)z_x^q+B(t)}{C(t)z_x^q+D(t)},\quad \bar{z}^{\text{new}}_x=\frac{A(t)\bar{z}_x^q+B(t)}{C(t)\bar{z}_x^q+D(t)},
\end{split}
\ee
where $\left(z_x=e^{i\frac{2\pi x}{L}},\bar{z}_x=e^{-i\frac{2\pi x}{L}}\right)$, such that $w^{\text{new}}=\frac{L}{2q\pi}\log z^{\text{new}}$, $\bar{w}^{\text{new}}=\frac{L}{2q\pi}\log \bar{z}^{\text{new}}$. Second, we define the angular functions with the Lorentzian time by
\be
\begin{split}
    &z^{\text{new}}_x=e^{i\varphi(t,x)}~\Rightarrow~\cos\left(\varphi(t,x)\right)=\text{Re}\left[z^{\text{new}}_x\right],\quad \sin\left(\varphi(t,x)\right)=\text{Im}\left[z^{\text{new}}_x\right],\\
    &\bar{z}^{\text{new}}_x=e^{-i\bar{\varphi}(t,x)}~\Rightarrow~\cos\left(\bar{\varphi}(t,x)\right)=\text{Re}\left[\bar{z}^{\text{new}}_x\right],\quad \sin\left(\bar{\varphi}(t,x)\right)=-\text{Im}\left[\bar{z}^{\text{new}}_x\right],
\end{split}
\ee
which satisfy~\eqref{eq:Post-Quench-Angular-Coordinates}. In the end, the post-quench space and time coordinates are given by
\be\label{app-eq:Angular-Function-Def}
\begin{split}
    &\cos\left(\frac{4q\pi x^{\text{new}}_x}{L}\right)=\text{Re}\left[z^{\text{new}}_x/\bar{z}^{\text{new}}_x\right]=\cos\left(\varphi(t,x)\right)\cos\left(\bar{\varphi}(t,x)\right)-\sin\left(\varphi(t,x)\right)\sin\left(\bar{\varphi}(t,x)\right),\\
    &\sin\left(\frac{4q\pi x^{\text{new}}_x}{L}\right)=\text{Im}\left[z^{\text{new}}_x/\bar{z}^{\text{new}}_x\right]=\sin\left(\varphi(t,x)\right)\cos\left(\bar{\varphi}(t,x)\right)+\cos\left(\varphi(t,x)\right)\sin\left(\bar{\varphi}(t,x)\right),\\
    &\cos\left(\frac{4q\pi t^{\text{new}}_x}{L}\right)=\text{Re}\left[z^{\text{new}}_x\bar{z}^{\text{new}}_x\right]=\cos\left(\varphi(t,x)\right)\cos\left(\bar{\varphi}(t,x)\right)+\sin\left(\varphi(t,x)\right)\sin\left(\bar{\varphi}(t,x)\right),\\
    &\sin\left(\frac{4q\pi t^{\text{new}}_x}{L}\right)=\text{Im}\left[z^{\text{new}}_x\bar{z}^{\text{new}}_x\right]=\sin\left(\varphi(t,x)\right)\cos\left(\bar{\varphi}(t,x)\right)-\cos\left(\varphi(t,x)\right)\sin\left(\bar{\varphi}(t,x)\right).\\
\end{split}
\ee

\subsection{$q$-SSD Evolution}
The derivatives contributing to Jacobian factors are given by
\be\label{eq:q-SSD-Uni-Piece-Conformal-Factor}
\begin{split}
    &\frac{dw^{\text{new}}_x}{dw_x}=\frac{L^2}{L^2+4q^2\pi^2t^2\sin^2\left(\frac{q\pi x}{L}\right)-4q\pi t L\sin\left(\frac{q\pi x}{L}\right)\cos\left(\frac{q\pi x}{L}\right)},\\
    &\frac{d\bar{w}^{\text{new}}_x}{d\bar{w}_x}=\frac{L^2}{L^2+4q^2\pi^2t^2\sin^2\left(\frac{q\pi x}{L}\right)+4q\pi t L\sin\left(\frac{q\pi x}{L}\right)\cos\left(\frac{q\pi x}{L}\right)},\\
\end{split}
\ee
which reduce to $1$ when $t=0$. At the same time, the angular functions in the Lorentzian signature are given by
\be\label{app-eq:sin-cos-q-ssd}
\begin{split}
    &\sin\left(\varphi(t,x)\right)=\frac{L\left(L\sin\left[\frac{2q\pi x}{L}\right]-4q\pi t \sin^2\left[\frac{q\pi x}{L}\right]\right)}{r_x(t)},~\cos\left(\varphi(t,x)\right)=1-\frac{2L^2\sin^2\left[\frac{q\pi x}{L}\right]}{r_x(t)},\\
    &\sin\left(\bar{\varphi}(t,x)\right)=\frac{L\left(L\sin\left[\frac{2q\pi x}{L}\right]+4q\pi t \sin^2\left[\frac{q\pi x}{L}\right]\right)}{\overline{r}_x(t)},~\cos\left(\bar{\varphi}(t,x)\right)=1-\frac{2L^2\sin^2\left[\frac{q\pi x}{L}\right]}{\overline{r}_x(t)},\\
    &r_x(t)=L^2+4q^2\pi^2t^2\sin^2\left[\frac{q\pi x}{L}\right]-2q\pi t L\sin\left[\frac{2q\pi x}{L}\right],\\
    &\overline{r}_x(t)=L^2+4q^2\pi^2t^2\sin^2\left[\frac{q\pi x}{L}\right]+2q\pi t L\sin\left[\frac{2q\pi x}{L}\right],\\
\end{split}
\ee
where $\sin\left(\varphi(t,X_m^f)\right)=\sin\left(\bar{\varphi}(t,X_m^f)\right)=0$ and $\cos\left(\varphi(t,X_m^f)\right)=\cos\left(\bar{\varphi}(t,X_m^f)\right)=1$ forever at fixed points. For late-time clarity, endpoints are never chosen at fixed points and are separated from their nearest endpoint by a distance strictly greater than $\beta$. Then, the post-quench space and time coordinates are obtained from~\eqref{app-eq:Angular-Function-Def}. Note that at late times $t\gg L$, the numerators of ``$\sin$'' functions in \eqref{app-eq:sin-cos-q-ssd} are much larger than the denominators, such that the post-quench space and time coordinate differences are approximately given by
\be
\begin{split}
    &x^{\text{new}}_x-x^{\text{new}}_y\approx\begin{cases}
        \frac{L^3\sin\left[\frac{q\pi (x-y)}{L}\right]}{4q^3\pi^3 t^2\sin\left[\frac{q\pi x}{L}\right]\sin\left[\frac{q\pi y}{L}\right]}~&\text{if}~X_m^f<y<x<X_{m+1}^f\\
        \frac{L}{q}\cdot\left(\left\lfloor \frac{q(x-y)}{L}\right\rfloor\mod q\right)~&\text{if }~\left\lfloor \frac{q(x-y)}{L}\right\rfloor~\text{fixed points between } $x,y$\\
    \end{cases},\\
    &t^{\text{new}}_x\approx \frac{L}{4q\pi}\cdot \left(2\pi-\frac{2L}{q\pi t}\right)\approx \frac{L}{2q},~x,y\neq X_m^f~\&~\underset{m}{\text{Min}}\left[|x-X_m^f|,|y-X_m^f|\right]\gg \beta.
\end{split}
\ee
In addition, the Jacobian factors reduce to
\be
\begin{split}
    &\frac{dw^{\text{new}}_x}{dw_x}\approx\frac{d\bar{w}^{\text{new}}_x}{d\bar{w}_x}\approx\frac{L^2}{4q^2\pi^2t^2\sin^2\left(\frac{q\pi x}{L}\right)}.
\end{split}
\ee
Substituting the post-quench coordinates and the Jacobian factors back into~\eqref{eq:Single-Interval-EE-Expressions}, one can recover~\eqref{eq:Conformal-Cooling-Vacuum} for $X_m^f<y<x<X_{m+1}^f$.

As a next step, we now derive Eq.~\eqref{eq:qSSD-Asym-MI}. To do so, note that the Jacobian factors are canceled for $I_{A,B}(t)$, such that $S_{A\cup B}^{\text{U}}=S_{A}^{\text{U}}+S_{B}^{\text{U}}$. Suppose that $q>4$ and $A$ contains $p<\frac{q}{2}$ or $p+1<\left\lfloor\frac{q}{2}\right\rfloor$ fixed points with respect to even or odd $q$. Since $l[A]=l[B]<L/2$, $S_A(t)$ and $S_B(t)$ are determined by $S_{A;(a)}^{\text{con.}},S_{B;(a)}^{\text{con.}}$ as
\be
\begin{split}
    &S_{A}(t)-S_{A}^{\text{U.}}\approx\frac{c\pi}{3\beta}\frac{L}{q}\cdot\begin{cases}
        p~&\text{for even }q\\
        p+1~&\text{for odd }q\\
    \end{cases},~S_{B}(t)-S_{B}^{\text{U.}}\approx\frac{c\pi}{3\beta}\frac{L}{q}\cdot p,
\end{split}
\ee
while $S_{A\cup B}(t)$ is given by the connected piece $S_{A\cup B}^{c}$, i.e.,
\be
\begin{split}
    &S_{A\cup B}(t)-S_{A\cup B}^{\text{U}}\approx\frac{c\pi}{3\beta}\frac{L}{q}\cdot 2.
\end{split}
\ee
Therefore, using~\eqref{eq:Mutual-Information}, we successfully recover~\eqref{eq:qSSD-Asym-MI} analytically.

\subsection{$q$-Displacement Evolution}
Under the $q$-Displacement dynamics, the conformal Jacobian factor is
\be
    \frac{dw^{\text{new}}_x}{dw_x}\frac{d\bar{w}^{\text{new}}_x}{d\bar{w}_x}=\frac{1}{\cosh^2\left(\frac{2q\pi t}{L}\right)-\sinh^2\left(\frac{2q\pi t}{L}\right)\cos^2\left(\frac{2q\pi x}{L}\right)}\overset{t\gg L}{\approx}\frac{e^{-\frac{4q\pi t}{L}}}{\sin^2\left(\frac{q\pi x}{L}\right)\cos^2\left(\frac{q\pi x}{L}\right)}.
\ee
Furthermore, the post-quench time and space coordinates read
\be
\begin{split}
    &\cos\left(\frac{4q\pi t^{\text{new}}_x}{L}\right)=\frac{-2\sin\left(\frac{2q\pi x}{L}\right)\sinh\left(\frac{2q\pi t}{L}\right)}{r_x(t)},\\
    &\sin\left(\frac{4q\pi t^{\text{new}}_x}{L}\right)=\frac{5-\cos\left(\frac{4q\pi x}{L}\right)-\cosh\left(\frac{4q\pi t}{L}\right)+\cos\left(\frac{4q\pi x}{L}\right)\cosh\left(\frac{4q\pi t}{L}\right)}{4r_x(t)},\\
    &\cos\left(\frac{4q\pi x^{\text{new}}_x}{L}\right)=\frac{\sin\left(\frac{4q\pi x}{L}\right)\cosh\left(\frac{2q\pi t}{L}\right)}{r_x(t)},\\
    &\sin\left(\frac{4q\pi x^{\text{new}}_x}{L}\right)=\frac{1+2\cos\left(\frac{4q\pi x}{L}\right)-\cosh\left(\frac{4q\pi t}{L}\right)+\cos\left(\frac{4q\pi x}{L}\right)\cosh\left(\frac{4q\pi t}{L}\right)}{4r_x(t)},\\
    &r_x(t)=\cosh^2\left(\frac{2q\pi t}{L}\right)-\sinh^2\left(\frac{2q\pi t}{L}\right)\cos^2\left(\frac{2q\pi x}{L}\right).
\end{split}
\ee
Hence, suppose endpoints of subsystems $A$, $B$ are away from the fixed points $X_{\frac{m}{2}}^f$ in the late time limit $t\gg L$, the space coordinate difference and the post-quench time coordinate are approximated as
\be
\begin{split}
    &x^{\text{new}}_x-x^{\text{new}}_y\approx\begin{cases}
        \frac{2Le^{-\frac{2q\pi t}{L}}\sin\left[\frac{q\pi(x-y)}{L}\right]}{q\pi\sqrt{\left|\sin\left(\frac{2q\pi x}{L}\right)\cos\left(\frac{2q\pi x}{L}\right)\right|}}~&\text{if}~X_{\frac{m}{2}}^f<y<x<X_{\frac{m+1}{2}}^f\\
        \frac{L}{2q}\cdot\left\lfloor \frac{2q(x-y)}{L}\right\rfloor~&\text{if}~\left\lfloor \frac{2q(x-y)}{L}\right\rfloor~\text{fixed points in }(y,x)
    \end{cases},\\
    &t^{\text{new}}_x\approx(-1)^{\left\lfloor \frac{2q(x-y)}{L}\right\rfloor}\left[\frac{L}{4q}-\frac{Le^{-\frac{2q\pi t}{L}}}{q\pi}\csc\left(\frac{q\pi x}{L}\right)\right]\approx (-1)^{\left\lfloor \frac{2q(x-y)}{L}\right\rfloor}\frac{L}{4q},
\end{split}
\ee
where the extra phase $(-1)^{\left\lfloor \frac{2q(x-y)}{L}\right\rfloor}$ arises because adjacent intervals between fixed points carry opposite energy-density signs.
Thus, if we consider a subsystem $A$
with $X_{\frac{m}{2}}^f<X_2<X_1<X_{\frac{m+1}{2}}^f$,~\eqref{eq:Single-Interval-EE-Expressions} reduces to~\eqref{eq:Conformal-Cooling-Vacuum}. We emphasize that $A$ must include at least two fixed points. Otherwise, the holographic CFT computation becomes ill-defined, since $t_{X_1}^{\text{new}}-t_{X_2}^{\text{new}}\approx x_{X_1}^{\text{new}}-x_{X_2}^{\text{new}}\approx \frac{L}{2q}$, and both $S_{A;(a)}^{\text{con.}}$ and $S_{A\cup B}^{a}$ diverge to negative infinity.

Analogous to the $q$-SSD case, we can further derive~\eqref{eq:qDis-Asym-MI} from~\eqref{eq:Single-Interval-EE-Expressions} and~\eqref{eq:Antipodal-EE}. Again, because $S_{A}^{\text{U}}+S_{B}^{\text{U}}=S_{A\cup B}^{\text{U}}$, the linear-$t$ divergent terms do not affect the mutual information. Consider $q>2$, and that subsystems $A$ and $B=\overline{A}$ both include $1<p<q$ fixed points; their independent contributions are
\be
    S_{A}(t)-S_{A}^{\text{U.}}\approx S_{B}(t)-S_{B}^{\text{U.}}\approx\frac{c\pi}{3\beta}\frac{L}{2q}\cdot p,
\ee
which are identical for even and odd $q$. The contribution from the antipodal subsystem $A\cup B$ is approximately given by
\be
     S_{A\cup B}(t)-S_{A\cup B}^{\text{U.}}\approx\frac{c\pi}{3\beta}\frac{L}{2q}\cdot\begin{cases}
         2~&\text{for even}~q\\
         0~&\text{for odd}~q
     \end{cases},
\ee
which vanishes for odd $q$, since there are $q-1$ fixed points within $(X_j,X_j+L/2)$ with $j=1,2$. Combining the above two equations, we obtain~\eqref{eq:qDis-Asym-MI}.

\section{RT/HRT Calculations for Entanglement Entropies\label{app:RTHRT}}
In this appendix, we provide technical details for the entanglement entropy calculations using the RT/HRT formula in the geon geometry. In the field theory side, we are working in the Heisenberg picture, such that the metric is unchanged on the gravity side and is fixed by the initial state $\ket{\Psi(0)}$. As discussed in previous works~\cite{Maldacena:2001kr,Wei:2024zez,Wei:2024kkp,Louko:1998hc,Maloney:2016gsg,Pathak:2024cpo,Maxfield:2014kra}, the holographic dual of this state is an $\text{AdS}_3$ geon in the high temperature regime, $L/\beta>1$. The $\text{AdS}_3$ geon is described by the following maximally extended geometry, together with a $\mathbb{Z}_2$ identification
\be\label{eq:Maximally-Extended-Coordinate-Geon}
\begin{split}
    &ds^2=\sec^2\left(2X_c\right)\cdot\left[-4dT_c^2+4dX_c^2+{r^{\text{new}}_{h}}^2\cos^2\left(2T_c\right)d{x^{\text{new}}}^2\right],\quad r^{\text{new}}_{h}=\frac{2\pi}{\beta},\\
    &T_c,X_c\in\left(-\frac{\pi}{2},\frac{\pi}{2}\right),
    \quad \left(T_c,X_c,{x^{\text{new}}}\right)\sim\left(T_c,-X_c,{x^{\text{new}}}+L/2\right),
\end{split}
\ee
where the event horizon is located at $T_c=\pm X_c$, and the identification is along the $X_c=0$ slice (we call this the location of the crosscap), ${x^{\text{new}}}={x^{\text{new}}}(t,x)$ is the post-quench spatial coordinate introduced in~\eqref{eq:Post-Quench-Space-Time-Coordinates} and $T_c=T_c(t,x),X_c=X_c(t,x)$. We also set the $\AdS_3$ radius to be one. The two-sided eternal $\text{AdS}_3$ black hole is the double cover of the $\text{AdS}_3$ geon. Therefore, we can compute physical quantities in the two-sided picture using the doubling trick~\cite{Maxfield:2014kra}. To understand the relationship between $(t,x)$ and $(T_c,X_c)$, we work in the exterior coordinates
\be
\begin{split}
    ds^2&=-{r^{\text{new}}_h}^2\sinh^2(\rho)d{t^{\text{new}}}^2+d\rho^2+{r^{\text{new}}_h}^2\cosh^2(\rho)d{x^{\text{new}}}^2\\
    &= -\left({r^{\text{new}}}^2-{r^{\text{new}}_h}^2\right)d{t^{\text{new}}}^2+\frac{d{r^{\text{new}}}^2}{{r^{\text{new}}}^2-{r^{\text{new}}_h}^2}+{r^{\text{new}}}^2d{x^{\text{new}}}^2,
    \quad r^{\text{new}}=r^{\text{new}}_h\cosh(\rho),
    \quad \rho\geq 0,\\
    X_c&=\f{1}{2}\cdot\begin{cases}
         \arctan\left[\sqrt{\f{r-{r^{\text{new}}_h}}{r^{\text{new}}+{r^{\text{new}}_h}}}e^{{r^{\text{new}}_h}t^{\text{new}}}\right]+\arctan\left[\sqrt{\f{r^{\text{new}}-{r^{\text{new}}_h}}{r^{\text{new}}+{r^{\text{new}}_h}}}e^{-{r^{\text{new}}_h}t^{\text{new}}}\right]~&\text{right exterior}\\
         \pm\left(\arctan\left[\sqrt{\f{{r^{\text{new}}_h}-r^{\text{new}}}{r^{\text{new}}+{r^{\text{new}}_h}}}e^{{r^{\text{new}}_h}t_{\text{I.}}}\right]-\arctan\left[\sqrt{\f{{r^{\text{new}}_h}-r^{\text{new}}}{r^{\text{new}}+{r^{\text{new}}_h}}}e^{-{r^{\text{new}}_h}t_{\text{I.}}}\right]\right)~&\text{interiors}\\
         -\arctan\left[\sqrt{\f{r^{\text{new}}-{r^{\text{new}}_h}}{r^{\text{new}}+{r^{\text{new}}_h}}}e^{{r^{\text{new}}_h}t^{\text{new}}}\right]-\arctan\left[\sqrt{\f{r^{\text{new}}-{r^{\text{new}}_h}}{r^{\text{new}}+{r^{\text{new}}_h}}}e^{-{r^{\text{new}}_h}t^{\text{new}}}\right]~&\text{left exterior}\\
     \end{cases}\\
     &=\f{1}{2}\cdot\begin{cases}
         \arctan\left[\tanh\left(\f{\rho}{2}\right)e^{{r^{\text{new}}_h}t^{\text{new}}}\right]+\arctan\left[\tanh\left(\f{\rho}{2}\right)e^{-{r^{\text{new}}_h}t^{\text{new}}}\right]~&\text{right exterior}\\
         \pm\left(\arctan\left[\tan\left(\f{\alpha}{2}\right)e^{{r^{\text{new}}_h}t_{\text{I.}}}\right]-\arctan\left[\tan\left(\f{\alpha}{2}\right)e^{-{r^{\text{new}}_h}t_{\text{I.}}}\right]\right)~&\text{interiors}\\
         -\arctan\left[\tanh\left(\f{\rho}{2}\right)e^{{r^{\text{new}}_h}t^{\text{new}}}\right]-\arctan\left[\tanh\left(\f{\rho}{2}\right)e^{-{r^{\text{new}}_h}t^{\text{new}}}\right]~&\text{left exterior}\\
     \end{cases},\\
     T_c&=\f{1}{2}\cdot\begin{cases}
         \arctan\left[\sqrt{\f{r^{\text{new}}-{r^{\text{new}}_h}}{r^{\text{new}}+{r^{\text{new}}_h}}}e^{{r^{\text{new}}_h}t^{\text{new}}}\right]-\arctan\left[\sqrt{\f{r^{\text{new}}-{r^{\text{new}}_h}}{r^{\text{new}}+{r^{\text{new}}_h}}}e^{-{r^{\text{new}}_h}t^{\text{new}}}\right]~&\text{exteriors}\\
         \pm\left(\arctan\left[\sqrt{\f{{r^{\text{new}}_h}-r^{\text{new}}}{r^{\text{new}}+{r^{\text{new}}_h}}}e^{{r^{\text{new}}_h}t_{\text{I.}}}\right]+\arctan\left[\sqrt{\f{{r^{\text{new}}_h}-r^{\text{new}}}{r^{\text{new}}+{r^{\text{new}}_h}}}e^{-{r^{\text{new}}_h}t_{\text{I.}}}\right]\right)~&\text{interiors}\\
     \end{cases}\\
     &=\f{1}{2}\cdot\begin{cases}
         \arctan\left[\tanh\left(\f{\rho}{2}\right)e^{{r^{\text{new}}_h}t^{\text{new}}}\right]-\arctan\left[\tanh\left(\f{\rho}{2}\right)e^{-{r^{\text{new}}_h}t^{\text{new}}}\right]~&\text{exteriors}\\
         \pm\left(\arctan\left[\tan\left(\f{\alpha}{2}\right)e^{{r^{\text{new}}_h}t_{\text{I.}}}\right]+\arctan\left[\tan\left(\f{\alpha}{2}\right)e^{-{r^{\text{new}}_h}t_{\text{I.}}}\right]\right)~&\text{interiors}\\
     \end{cases}\\
\end{split}
\ee
where $t^{\text{new}}=t^{\text{new}}(t,x)$ is the post-quench time coordinate and $t_{\text{I.}},\alpha=i\rho$ are interior coordinates. Note that the (1+1)d holographic CFT is defined on the asymptotic boundary of the right exterior; its mirror theory lives on the opposite boundary~\cite{Wei:2024kkp}. In the exterior coordinates, the $\mathbb{Z}_2$ identification reads
\be
    (t^{\text{new}},r^{\text{new}},x^{\text{new}})_{\text{right exterior}}\sim \left(-t^{\text{new}},r^{\text{new}},x^{\text{new}}+\frac{L}{2}\right)_{\text{left exterior}}.
\ee
To compute the entanglement entropy holographically, we also introduce the embedding coordinates as
\be
    ds^2=-dY_{-1}^2-dY_0^2+dY_1^2+dY_2^2,
    \quad 
    -Y_{-1}^2-Y_0^2+Y_1^2+Y_2^2=-1,
\ee
such that the geodesic length, $D(P_1,P_2)=D(P_2,P_1)$, between two spacetime points $P_1=(Y^{A=-1,0,1,2}_{(1)})$ and $P_2=(Y^{A=-1,0,1,2}_{(2)})$ is given by
\be\label{eq:AdS3-Geodesic-Length}
\begin{split}
    \cosh\left[D(P_1,P_2)\right]&=-\eta^{AB}Y_A^{(1)}Y_B^{(2)} =Y_{-1}^{(1)}Y_{-1}^{(2)}+Y_{0}^{(1)}Y_{0}^{(2)}-Y_{1}^{(1)}Y_{1}^{(2)}-Y_{2}^{(1)}Y_{2}^{(2)},
    \nonumber \\
    \eta^{AB}&=\eta_{AB}=\text{diag}(-1,-1,1,1).
\end{split}
\ee
The embedding coordinate is related to the maximally extended coordinate via
\be
\begin{split}
    &Y_{-1}=\left[\frac{\cos(2T_c)}{\cos(2X_c)}\right]\cosh\left(r^{\text{new}}_hx^{\text{new}}\right),
    \quad Y_0=\left[\frac{\sin(2T_c)}{\cos(2X_c)}\right],\\
    &Y_{1}=\tan(2X_c),
    \quad Y_2=\left[\frac{\cos(2T_c)}{\cos(2X_c)}\right]\sinh\left(r^{\text{new}}_hx^{\text{new}}\right),
\end{split}
\ee
and is related to the exterior metric by
\be
\begin{split}
    &\text{Right Exterior: }\begin{cases}
        Y_{-1}=\cosh(\rho)\cosh\left(r^{\text{new}}_hx^{\text{new}}\right),~Y_0=\sinh(\rho)\sinh\left(r^{\text{new}}_ht^{\text{new}}\right),\\
        Y_1=\sinh(\rho)\cosh\left(r^{\text{new}}_ht^{\text{new}}\right),~Y_2=\cosh(\rho)\sinh\left(r^{\text{new}}_hx^{\text{new}}\right),
    \end{cases}\\
    &\text{Left Exterior: }\begin{cases}
        Y_{-1}=\cosh(\rho)\cosh\left(r^{\text{new}}_hx^{\text{new}}\right),~Y_0=\sinh(\rho)\sinh\left(r^{\text{new}}_ht^{\text{new}}\right),\\
        Y_1=-\sinh(\rho)\cosh\left(r^{\text{new}}_ht^{\text{new}}\right),~Y_2=\cosh(\rho)\sinh\left(r^{\text{new}}_hx^{\text{new}}\right).
    \end{cases}
\end{split}
\ee
Accordingly, in the embedding coordinate, the $\mathbb{Z}_2$ identification reads
\be
    (Y_{-1},Y_{0},Y_1,Y_2)\sim\left(Y_{-1}(x^{\text{new}}\to x^{\text{new}}+L/2),Y_0,-Y_1,Y_2(x^{\text{new}}\to x^{\text{new}}+L/2)\right).
\ee

To calculate entanglement entropies and mutual information in this Lorentzian spacetime\footnote{Specifically, the calculations are applied in the two-sided picture, where the spacetime is orientable. Once the lengths are determined, we incorporate the $\mathbb{Z}_2$ identification.}, two types of geodesic lengths are necessary, i.e., the geodesic length between two endpoints on the same asymptotic boundary and the length for a geodesic connecting a point on the asymptotic boundary and a point on the image asymptotic boundary. Let us consider them case by case. The two endpoints on the same asymptotic boundary are
\be
\begin{split}
    &P_1=\frac{r_{R;\text{cut,1}}}{r^{\text{new}}_h}\left(\cosh\left(r^{\text{new}}_hx^{\text{new}}_1\right),\sinh\left(r^{\text{new}}_ht^{\text{new}}_1\right),\cosh\left(r^{\text{new}}_ht^{\text{new}}_1\right),\sinh\left(r^{\text{new}}_hx^{\text{new}}_1\right)\right),\\
    &P_2=\frac{r_{R;\text{cut,2}}}{r^{\text{new}}_h}\left(\cosh\left(r^{\text{new}}_hx^{\text{new}}_2\right),\sinh\left(r^{\text{new}}_ht^{\text{new}}_2\right),\cosh\left(r^{\text{new}}_ht^{\text{new}}_2\right),\sinh\left(r^{\text{new}}_hx^{\text{new}}_2\right)\right),\\
\end{split}
\ee
in the embedding coordinates, and $r_{R;\text{cut,i=1,2}}=\lim_{r^{\text{new}}\to\infty}r^{\text{new}}$ are the local bulk IR-cut offs near the asymptotic boundary (on the right exterior for the two sided picture). Using~\eqref{eq:AdS3-Geodesic-Length}, the geodesic length connecting these two points is given by
\begin{align}
\label{eq:Same-Side-Geodesic-Length}
    \cosh[D(P_1,P_2)]&=\frac{r_{R;\text{cut,1}}r_{R;\text{cut,2}}}{{r^{\text{new}}_h}^2}\left(\cosh\left[r^{\text{new}}_h(x^{\text{new}}_1-x^{\text{new}}_2)\right]-\cosh\left[r^{\text{new}}_h(t^{\text{new}}_1-t^{\text{new}}_2)\right]\right)
    \nonumber \\
    &
    \approx \frac{e^{D(P_1,P_2)}}{2},
    \end{align}
where the approximation is always valid as $D(P_1,P_2)\gg 1$.
Then, we consider the second case. One endpoint is still $P_1$, and we use $P_3$ to represent the endpoint on the image asymptotic boundary (left boundary) as
\be
    P_3=\frac{r_{L;\text{cut,3}}}{r^{\text{new}}_h}\left(\cosh\left(r^{\text{new}}_hx^{\text{new}}_3\right),\sinh\left(r^{\text{new}}_ht^{\text{new}}_3\right),-\cosh\left(r^{\text{new}}_ht^{\text{new}}_3\right),\sinh\left(r^{\text{new}}_hx^{\text{new}}_3\right)\right).
\ee
Consequently, the geodesic length between $P_1$ and $P_3$ is given by
\begin{align}
\label{eq:Different-Side-Geodesic-Length}
    \cosh[D(P_1,P_3)]&=\frac{r_{R;\text{cut,1}}r_{L;\text{cut,3}}}{{r^{\text{new}}_h}^2}\left(\cosh\left[r^{\text{new}}_h(x^{\text{new}}_1-x^{\text{new}}_3)\right]+\cosh\left[r^{\text{new}}_h(t^{\text{new}}_1+t^{\text{new}}_3)\right]\right)
    \nonumber \\
    &\approx \frac{e^{D(P_1,P_3)}}{2} .
\end{align}

\subsection{Holographic Entanglement Entropies}
The holographic entanglement entropy for an arbitrary subsystem $\mathcal{V}$ (with the reduced density matrix $\rho_{\mathcal{V}}$) is given by~\eqref{eq:HRT-Formula}. In three dimensional spacetime, the HRT surfaces are space-like geodesics anchored at the boundary of subsystem, whose lengths are computed by~\eqref{eq:Same-Side-Geodesic-Length} and~\eqref{eq:Different-Side-Geodesic-Length}. 

\subsubsection{Single-Interval Cases}
Here, we derive the holographic entanglement entropy for single interval cases. The endpoints of $\mathcal{V}=A$ are at $(t,X_2),(t,X_1)$, respectively. Therefore, in the Embedding coordinate near asymptotic boundaries, the endpoints are
\be
\begin{split}
    &P_1=\frac{r_{R;\text{cut,1}}}{r^{\text{new}}_h}\left(\cosh\left(r^{\text{new}}_hx^{\text{new}}_{X_1}\right),\sinh\left(r^{\text{new}}_ht^{\text{new}}_{X_1}\right),\cosh\left(r^{\text{new}}_ht^{\text{new}}_{X_1}\right),\sinh\left(r^{\text{new}}_hx^{\text{new}}_{X_1}\right)\right),\\
    &P_2=\frac{r_{R;\text{cut,2}}}{r^{\text{new}}_h}\left(\cosh\left(r^{\text{new}}_hx^{\text{new}}_{X_2}\right),\sinh\left(r^{\text{new}}_ht^{\text{new}}_{X_2}\right),\cosh\left(r^{\text{new}}_ht^{\text{new}}_{X_2}\right),\sinh\left(r^{\text{new}}_hx^{\text{new}}_{X_2}\right)\right),\\
\end{split}
\ee
where $x^{\text{new}}_{X_j}=x^{\text{new}}(t,X_j)$ and $t^{\text{new}}_{X_j}=t^{\text{new}}(t,X_j)$, and their corresponding image points on the other side are given by
{\small
\be 
\begin{split}
&P_1^{I}=\frac{r_{L;\text{cut,1}}}{r^{\text{new}}_h}\left(\cosh\left(r^{\text{new}}_h[x^{\text{new}}_{X_1}+\frac{L}{2}]\right),\sinh\left(r^{\text{new}}_ht^{\text{new}}_{X_1}\right),-\cosh\left(r^{\text{new}}_ht^{\text{new}}_{X_1}\right),\sinh\left(r^{\text{new}}_h[x^{\text{new}}_{X_1}+\frac{L}{2}]\right)\right),\\
    &P_2^{I}=\frac{r_{L;\text{cut,1}}}{r^{\text{new}}_h}\left(\cosh\left(r^{\text{new}}_h[x^{\text{new}}_{X_2}+\frac{L}{2}]\right),\sinh\left(r^{\text{new}}_ht^{\text{new}}_{X_2}\right),-\cosh\left(r^{\text{new}}_ht^{\text{new}}_{X_2}\right),\sinh\left(r^{\text{new}}_h[x^{\text{new}}_{X_2}+\frac{L}{2}]\right)\right).\\
\end{split}
\ee}Taking the homology condition into account, the candidate extremal surfaces are
\be\label{eq:Single-Interval-HEE-Candidates}
\begin{split}
    &D_{\text{con.}}^{(a)}=D(P_1,P_2)\\
    &=\log\left(\frac{4r_{R;\text{cut,1}}r_{R;\text{cut,2}}}{{r^{\text{new}}_h}^2}\right)+\log\left(\frac{1}{2}\left|\cosh\left[r^{\text{new}}_h(x^{\text{new}}_{X_1}-x^{\text{new}}_{X_2})\right]-\cosh\left[r^{\text{new}}_h(t^{\text{new}}_{X_1}-t^{\text{new}}_{X_2})\right]\right|\right),\\
    &D_{\text{con.}}^{(b)}=D(P_1,P_2)\\
    &=\log\left(\frac{4r_{R;\text{cut,1}}r_{R;\text{cut,2}}}{{r^{\text{new}}_h}^2}\right)+\log\left(\frac{1}{2}\left|\cosh\left[r^{\text{new}}_h(L-[x^{\text{new}}_{X_1}-x^{\text{new}}_{X_2}])\right]-\cosh\left[r^{\text{new}}_h(t^{\text{new}}_{X_1}-t^{\text{new}}_{X_2})\right]\right|\right),\\
    &D_{\text{dis.}}=\frac{\sum_{k=1,2}D(P_j,P_j^I)}{2}\\
    &=\frac{1}{2}\sum_{k=1,2}\left[\log\left(\frac{4r_{R;\text{cut},j}r_{L;\text{cut},j}}{{r^{\text{new}}_h}^2}\right)+\log\left(\frac{1}{2}\left|\cosh\left[\frac{r^{\text{new}}_hL}{2}\right]+\cosh\left[2r^{\text{new}}_ht^{\text{new}}_{X_j}\right]\right|\right)\right],\\
\end{split}
\ee
where $r_h^{\text{new}}=\frac{2\pi}{\beta}$, and the subscripts ``con.'', ``dis.'' denote the so-called ``connected geodesics'' and ``disconnected geodesics''\footnote{In the two-sided picture, disconnected geodesics connect endpoints of $A$ and their image points in another exterior region, while connected geodesics just connect the endpoints of $A$. Therefore, disconnected geodesics have been proposed as a probe of non-traversable wormholes (or the growth of black hole interior)~\cite{Hartman:2013qma,Maldacena:2013xja,Numasawa:2018grg}.}. The bulk IR cut-off (at point $x=X$) can be re-written as the uniform expressions in the original coordinate $(t,x)$, i.e.,
\be\label{eq:Uniform-Bulk-IR-Cut-Off}
    r_{i=L,R;\text{cut},X}=\frac{r_{\text{uni;cut}}}{\sqrt{\left(\frac{dw^{\text{new}}_X}{dw_{X}}\right)\left(\frac{d\bar{w}^{\text{new}}_X}{d\bar{w}_{X}}\right)}\bigg|_{\tau=it}}.
\ee
Using~\eqref{eq:HRT-Formula} and the Brown-Henneaux relation $c=\frac{3}{2G_N}$~\cite{Brown:1986nw}, the single-interval entanglement entropy is given by
\be
    S_A(t)=\frac{c}{6}\cdot\text{Min}\left[D_{\text{con.}},D_{\text{dis.}}\right],
    \quad D_{\text{con.}}=\text{Min}\left[D_{\text{con.}}^{(a)},D_{\text{con.}}^{(b)}\right].
\ee
Applying~\eqref{eq:Single-Interval-HEE-Candidates} to the formula above and setting $r_{\text{uni;cut}}=1$, it is easy to check that the holographic single-entanglement entropy reproduces the expressions we obtained in~\eqref{eq:Single-Interval-EE-Expressions}.

However, in order to understand the growth of the black hole interior (or the growth of the wormhole), we need to divide the contributions from the exterior and interior regions, and go through further subtleties for the interior contribution. We will come back to this topic in the next section.

\subsubsection{Antipodal Double-Interval Cases}
Next, we study the holographic entanglement entropy and the mutual information for a antipodally distributed subsystem, $A\cup B$ with $B=\overline{A}$. If the subsystem has endpoints
\be
\begin{split}
    &P_1=\frac{r_{R;\text{cut,1}}}{r^{\text{new}}_h}\left(\cosh\left(r^{\text{new}}_hx^{\text{new}}_{X_1}\right),\sinh\left(r^{\text{new}}_ht^{\text{new}}_{X_1}\right),\cosh\left(r^{\text{new}}_ht^{\text{new}}_{X_1}\right),\sinh\left(r^{\text{new}}_hx^{\text{new}}_{X_1}\right)\right),\\
    &P_2=\frac{r_{R;\text{cut,2}}}{r^{\text{new}}_h}\left(\cosh\left(r^{\text{new}}_hx^{\text{new}}_{X_2}\right),\sinh\left(r^{\text{new}}_ht^{\text{new}}_{X_2}\right),\cosh\left(r^{\text{new}}_ht^{\text{new}}_{X_2}\right),\sinh\left(r^{\text{new}}_hx^{\text{new}}_{X_2}\right)\right),\\
    &P_3=\frac{r_{R;\text{cut,3}}}{r^{\text{new}}_h}\left(\cosh\left(r^{\text{new}}_hx^{\text{new}}_{X_3}\right),\sinh\left(r^{\text{new}}_ht^{\text{new}}_{X_3}\right),\cosh\left(r^{\text{new}}_ht^{\text{new}}_{X_3}\right),\sinh\left(r^{\text{new}}_hx^{\text{new}}_{X_3}\right)\right),\\
    &P_4=\frac{r_{R;\text{cut,4}}}{r^{\text{new}}_h}\left(\cosh\left(r^{\text{new}}_hx^{\text{new}}_{X_4}\right),\sinh\left(r^{\text{new}}_ht^{\text{new}}_{X_4}\right),\cosh\left(r^{\text{new}}_ht^{\text{new}}_{X_4}\right),\sinh\left(r^{\text{new}}_hx^{\text{new}}_{X_4}\right)\right).\\
\end{split}
\ee
Then, their image points are given by
\be
\begin{split}
    &P_1^I=\frac{r_{L;\text{cut,1}}}{r^{\text{new}}_h}\left(\cosh\left(r^{\text{new}}_hx^{\text{new};I}_{X_1}\right),\sinh\left(r^{\text{new}}_ht^{\text{new}}_{X_1}\right),-\cosh\left(r^{\text{new}}_ht^{\text{new}}_{X_1}\right),\sinh\left(r^{\text{new}}_hx^{\text{new};I}_{X_1}\right)\right),\\
    &P_2^I=\frac{r_{L;\text{cut,2}}}{r^{\text{new}}_h}\left(\cosh\left(r^{\text{new}}_hx^{\text{new};I}_{X_2}\right),\sinh\left(r^{\text{new}}_ht^{\text{new}}_{X_2}\right),-\cosh\left(r^{\text{new}}_ht^{\text{new}}_{X_2}\right),\sinh\left(r^{\text{new}}_hx^{\text{new};I}_{X_2}\right)\right),\\
    &P_3^I=\frac{r_{L;\text{cut,3}}}{r^{\text{new}}_h}\left(\cosh\left(r^{\text{new}}_hx^{\text{new};I}_{X_3}\right),\sinh\left(r^{\text{new}}_ht^{\text{new}}_{X_3}\right),-\cosh\left(r^{\text{new}}_ht^{\text{new}}_{X_3}\right),\sinh\left(r^{\text{new}}_hx^{\text{new};I}_{X_3}\right)\right),\\
    &P_4^I=\frac{r_{L;\text{cut,4}}}{r^{\text{new}}_h}\left(\cosh\left(r^{\text{new}}_hx^{\text{new};I}_{X_4}\right),\sinh\left(r^{\text{new}}_ht^{\text{new}}_{X_4}\right),-\cosh\left(r^{\text{new}}_ht^{\text{new}}_{X_4}\right),\sinh\left(r^{\text{new}}_hx^{\text{new};I}_{X_4}\right)\right),\\
\end{split}
\ee
where we define the imaged space coordinate as $x^{\text{new};I}=x^{\text{new}}+L/2\mod L$. Accordingly, considering the homologous condition, the possible extremized surfaces have lengths as follows:
{\footnotesize
\be\label{eq:Antipodal-Interval-HEE-Candidates}
\begin{split}
    &D_{\text{con.}}^{(a)}=D(P_1,P_2)+D(P_3,P_4)\\
    &=\log\left(\frac{4r_{R;\text{cut,1}}r_{R;\text{cut,2}}}{{r^{\text{new}}_h}^2}\right)+\log\left(\frac{1}{2}\left|\cosh\left[r^{\text{new}}_h(x^{\text{new}}_{X_1}-x^{\text{new}}_{X_2})\right]-\cosh\left[r^{\text{new}}_h(t^{\text{new}}_{X_1}-t^{\text{new}}_{X_2})\right]\right|\right)\\
    &+\log\left(\frac{4r_{R;\text{cut,3}}r_{R;\text{cut,4}}}{{r^{\text{new}}_h}^2}\right)+\log\left(\frac{1}{2}\left|\cosh\left[r^{\text{new}}_h(x^{\text{new}}_{X_3}-x^{\text{new}}_{X_4})\right]-\cosh\left[r^{\text{new}}_h(t^{\text{new}}_{X_3}-t^{\text{new}}_{X_4})\right]\right|\right),\\
    &D_{\text{con.}}^{(b)}=D(P_1,P_4)+D(P_2,P_3)\\
    &=\log\left(\frac{4r_{R;\text{cut,1}}r_{R;\text{cut,4}}}{{r^{\text{new}}_h}^2}\right)+\log\left(\frac{1}{2}\left|\cosh\left[r^{\text{new}}_h(x^{\text{new}}_{X_1}-x^{\text{new}}_{X_4})\right]-\cosh\left[r^{\text{new}}_h(t^{\text{new}}_{X_1}-t^{\text{new}}_{X_4})\right]\right|\right)\\
    &+\log\left(\frac{4r_{R;\text{cut,2}}r_{R;\text{cut,3}}}{{r^{\text{new}}_h}^2}\right)+\log\left(\frac{1}{2}\left|\cosh\left[r^{\text{new}}_h\left(\frac{L}{2}-[x^{\text{new}}_{X_3}-x^{\text{new}}_{X_2}]\right)\right]-\cosh\left[r^{\text{new}}_h(t^{\text{new}}_{X_3}-t^{\text{new}}_{X_2})\right]\right|\right),\\
    &D_{\text{dis.}}^{(a)}=\frac{D(P_1,P_3^I)+D(P_1^I,P_3)+D(P_2,P_4^I)+D(P_2^I,P_4)}{2}\\
    &=\frac{1}{2}\sum_{j=1}^{4}\log\left(\frac{4r_{R;\text{cut},j}r_{L;\text{cut},j}}{{r^{\text{new}}_h}^2}\right)+\frac{1}{2}\sum_{j=1}^{2}\bigg[\log\left(\frac{1}{2}\left|\cosh\left[r^{\text{new}}_h(x^{\text{new};I}_{X_k+L/2}-x^{\text{new}}_{X_j})\right]+\cosh\left[r^{\text{new}}_h(t^{\text{new}}_{X_k+\frac{L}{2}}+t^{\text{new}}_{X_j})\right]\right|\right)\\
    &+\log\left(\frac{1}{2}\left|\cosh\left[r^{\text{new}}_h(x^{\text{new};I}_{X_k}-x^{\text{new}}_{X_k+\frac{L}{2}})\right]+\cosh\left[r^{\text{new}}_h(t^{\text{new}}_{X_k+\frac{L}{2}}+t^{\text{new}}_{X_j})\right]\right|\right) \bigg],\\
    &D_{\text{dis.}}^{(b)}=\frac{\sum_{j=1}^{4}D(P_j,P_j^I)}{2}\\
    &=\frac{1}{2}\sum_{j=1}^{4}\log\left(\frac{4r_{R;\text{cut},j}r_{L;\text{cut},j}}{{r^{\text{new}}_h}^2}\right)+\frac{1}{2}\sum_{j=1}^{2}\bigg[\log\left(\frac{1}{2}\left|\cosh\left[\frac{r^{\text{new}}_hL}{2}\right]+\cosh\left[2r^{\text{new}}_ht^{\text{new}}_{X_j}\right]\right|\right)\\
    &+\log\left(\frac{1}{2}\left|\cosh\left[\frac{r^{\text{new}}_hL}{2}\right]+\cosh\left[2r^{\text{new}}_ht^{\text{new}}_{X_k+\frac{L}{2}}\right]\right|\right)\bigg].
\end{split}
\ee
}
Again, we apply~\eqref{eq:Uniform-Bulk-IR-Cut-Off}, $r_{\text{uni;cut}}=1$, the Brown-Henneaux formula and substitute~\eqref{eq:Antipodal-Interval-HEE-Candidates} to the following minimization over extremized surfaces:
\be
\begin{split}
    &S_{A\cup B}(t)=\frac{c}{6}\cdot\text{Min}\left[D_{\text{con.}},D_{\text{dis.}}\right],
    \quad D_{\text{con.}}=\text{Min}\left[D_{\text{con.}}^{(a)},D_{\text{con.}}^{(b)}\right],
    \quad D_{\text{dis.}}=\text{Min}\left[D_{\text{dis.}}^{(a)},D_{\text{dis.}}^{(b)}\right],
\end{split}
\ee
which gives rise to the same expression as~\eqref{eq:Antipodal-EE}.

\section{Inhomogeneous Growth of the Geon Interior: Geodesic Mismatch\label{app:GM}}

In this appendix, we analyze the interior portions of geodesics in the geon spacetime linking a boundary point of the subsystem to its mirror image point, and of geodesics forming the disconnected extremized surface. In typical examples (boundary-state quenches, uniform crosscap quench), these interior pieces have identical length. However, for inhomogeneous crosscap quenches that break antipodal symmetry, they differ in a striking way. We claim that this mismatch is a consequence of the interplay between the crosscap initial state and the inhomogeneous dynamics, similar to the quasiparticle graph-like patterns in Sec.~\ref{sec:QP-EP}.

\begin{figure}
    \centering
    \includegraphics[width=0.5\linewidth]{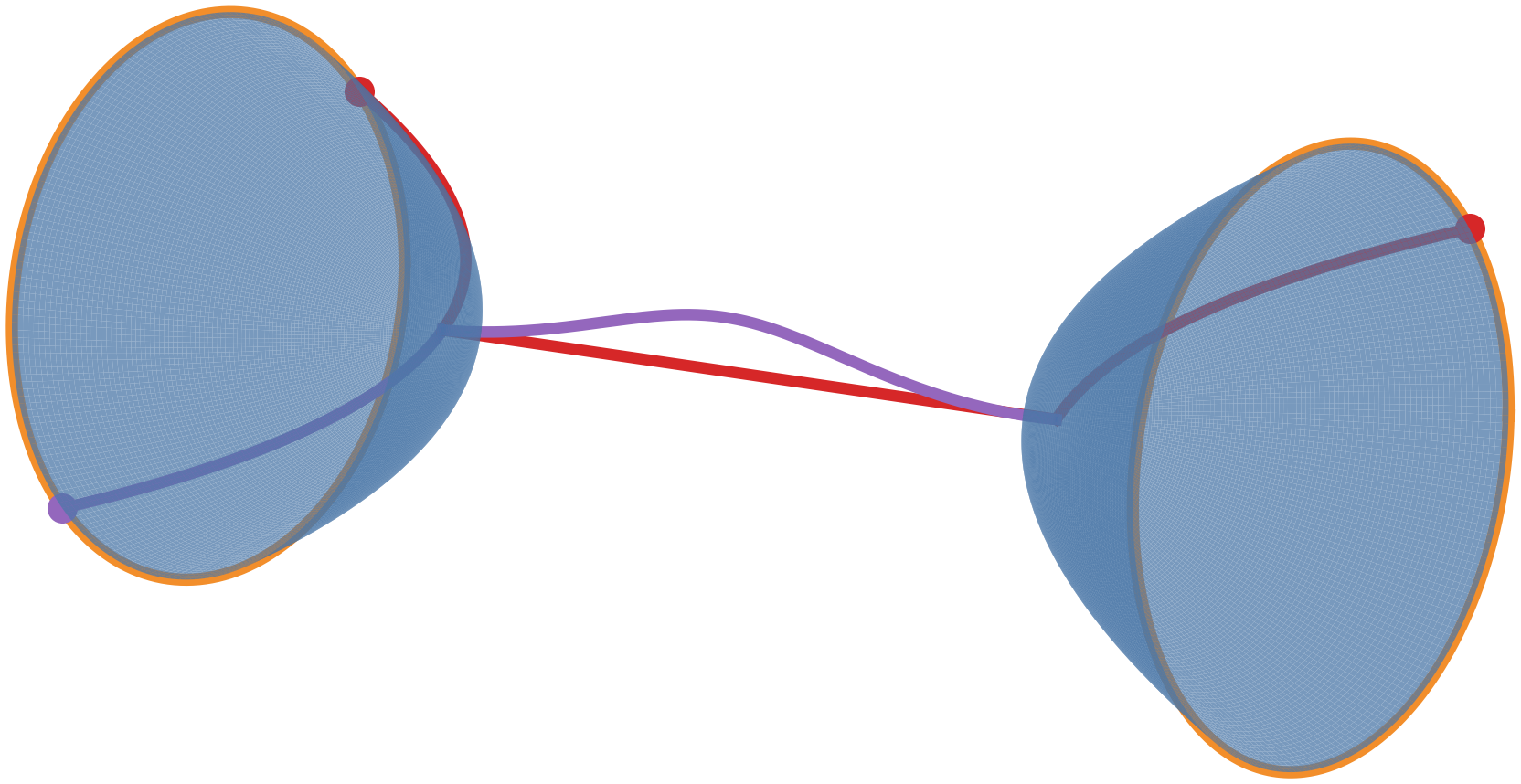}
    \caption{Schematic of interior–geodesic mismatch on a $t$-slice. Orange circles depict the spatial circles of the holographic CFT (right) and its image (left) at fixed time $t$; blue shading marks the exterior regions. Red lines are the disconnected extremal geodesics, while purple lines connect a subsystem endpoint to its mirror image. The segments joining the two “bowls” lie inside the black hole interior. Typically (boundary state quenches, homogeneous crosscap quenches, or inhomogeneous crosscap quenches with antipodal symmetry), the red and purple lines coincide; if the evolution breaks antipodal symmetry, they differ.}
    \label{fig:geodesic mismatch}
\end{figure}

Working in the two-sided spacetime, if a disconnected geodesic connects $P_1=(t^{\text{new}}_1,x^{\text{new}}_1)$ and $P_2=(-t^{\text{new}}_2,x^{\text{new}}_2)$ on the right and left exteriors (in the coordinate $(t^{\text{new}},x^{\text{new}})$), its interior part has length
\be\label{eq:Interior-Geodesic-Length-Two-Sided}
    D_{\text{int.}}^{\text{two-side}}(P_1,P_2)=r^{\text{new}}_h\cdot(t^{\text{new}}_1+t^{\text{new}}_2),
\ee
whose derivation will be provided in the forthcoming work~\cite{BaiNozakiMiyataMao_inprep}. If the time evolution is generated by $H_0$, $ D_{\text{int.}}^{\text{two-side}}$ describes the linear increase of the black-hole interior, which in turn explains the linear growth of the entanglement entropy~\cite{Hartman:2013qma}. Since geon is obtained from the two-sided black hole from a $\mathbb{Z}_2$ quotient, after applying the $\mathbb{Z}_2$ identification, the interior geodesic length is
\be\label{eq:Interior-Geodesic-Length}
    D_{\text{int.}}(P_1,P_2)=r^{\text{new}}_h\cdot\frac{t^{\text{new}}_1+t^{\text{new}}_2}{2},
\ee
which is valid both for the single-sided black hole with a tensionless End-of-World (EoW) brane~\cite{Hartman:2013qma} and the geon spacetime. 

In contrast to the two-sided case, where endpoints can be placed on both asymptotic boundaries, the quotient constrains us in choosing geodesic endpoints $P_j,j=1,2,\cdots,$ on a single side (right exterior); the corresponding mirror points $P_j^I$ are thereby determined. The interior part of such a geodesic connecting $P_j$ and its mirror image $P_j^I$ has length
\be\label{eq:Interior-Mirror-Geodesic-Length-After-Quotient}
        D_{\text{int.}}^{\text{mirror}}(P_j)=\frac{D_{\text{int.}}^{\text{two-sided}}(P_j,P_j^I)}{2}=r^{\text{new}}_h\cdot\frac{t^{\text{new}}(t,X_j)+t^{\text{new}}(t,X_{j}+L/2)}{2},
\ee
where $X_{j}+L/2=X_{j}+L/2\mod L$ due to PBC. 
In general, this is equal to the interior length of the disconnected extremized surface. By contrast, under an inhomogeneous quench that violates antipodal symmetry, 
\be
    t^{\text{new}}(t,x)\neq t^{\text{new}}(t,x+L/2\mod L),
\ee
the interior length given by~\eqref{eq:Interior-Mirror-Geodesic-Length-After-Quotient} may become larger than that of the disconnected phase anchored on $P_j$. Recall that $S_{A\cup B}(t)$ selects the disconnected phase when the sum of the two geodesic lengths is minimal:
\be
    D_{\text{dis.}}=\text{Min}\left[D_{\text{dis.}}^{(a)},D_{\text{dis.}}^{(b)}\right]=D_{\text{dis.}}^{(a)},
\ee
where $D_{\text{dis.}}^{(a)},D_{\text{dis.}}^{(b)}$ are given in~\eqref{eq:Antipodal-Interval-HEE-Candidates}. Four disconnected geodesics, anchored at $X_1$, $X_2$, $X_1+L/2$ and $X_2+L/2$, respectively, contribute to the disconnected extremized surface. Their lengths are:
\be\label{eq:Interior-Extremal-Geodesic-Length-After-Quotient}
    \begin{split}
        &D_{\text{int.}}^{\text{extr.}}(P_j)=\frac{D_{\text{int.}}^{\text{two-sided}}(P_j,P_{j+2}^I)}{2}=r^{\text{new}}_h\f{t^{\text{new}}(t,X_j)+t^{\text{new}}(t,X_{j+2}+L/2)}{2}=r^{\text{new}}_ht^{\text{new}}(t,X_j),
    \end{split}
\ee
where $j=1,2,3,4$ with $j+2=j+2\mod 4$, as $X_3=X_1+L/2,X_4=X_2+L/2$ and PBC.

\begin{figure}[htbp]
  \centering
  \begin{subfigure}[b]{0.48\textwidth}
    \includegraphics[width=\textwidth]{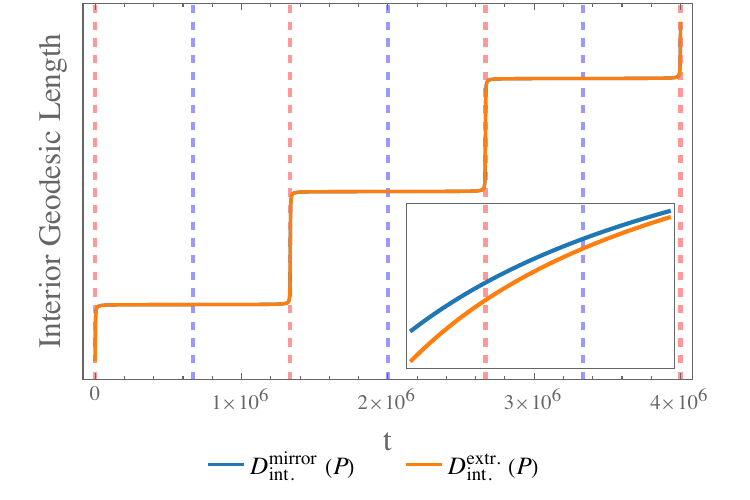}
    \caption{$q=3,~x=\frac{L}{3q},~\cosh(2\theta)=200$}
    \label{fig:qMobius Interior Growth q=3}
  \end{subfigure}
  \hfill
  \begin{subfigure}[b]{0.48\textwidth}
    \includegraphics[width=\textwidth]{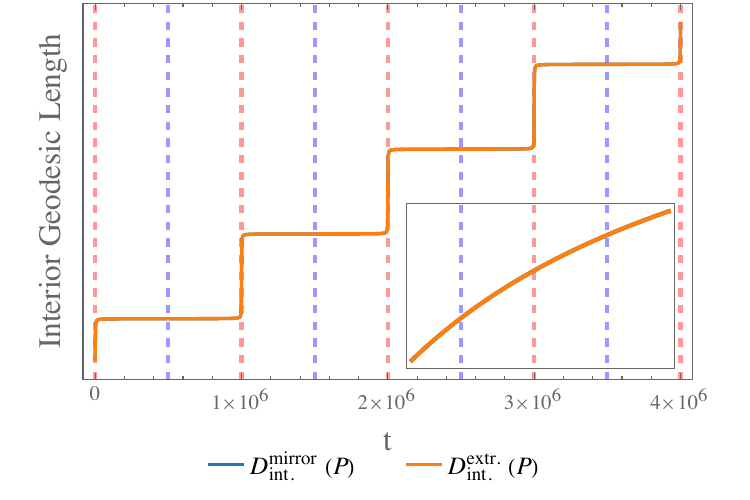}
    \caption{$q=4,~x=\frac{L}{3q},~\cosh(2\theta)=200$}
    \label{fig:qMobius Interior Growth q=4}
  \end{subfigure}
  
  \caption{Length of the interior segments of geodesics anchored at a boundary endpoint $P$ versus $q$-M\"obius time. Blue and orange dashed lines mark $t=\frac{mL_{\text{eff}}}{q}+\frac{L_{\text{eff}}}{2q},t=\frac{mL_{\text{eff}}}{q}$ time slices, respectively. The insets display a zoomed view of a short time window around the midpoint. Left panel: odd $q$ with $t^{\text{new}}(t,x)\neq t^{\text{new}}(t,x+L/2)$; right panel: even $q$ with $t^{\text{new}}(t,x)= t^{\text{new}}(t,x+L/2)$.}
  \label{fig:qMobius Interior Growth}
\end{figure}

Eqs.~\eqref{eq:Interior-Mirror-Geodesic-Length-After-Quotient} and~\eqref{eq:Interior-Extremal-Geodesic-Length-After-Quotient} agree iff the time evolution is antipodally symmetric; otherwise, they differ. As an illustration example, we depict their time dependence under a $q$-M\"obius evolution in Fig.~\ref{fig:qMobius Interior Growth}. 
Even $q$ yields $t_{\text{new}}(t,x)=t_{\text{new}}(t,x+L/2)$ and preserves the antipodal symmetry during dynamics, hence the two geodesics share the same interior length. 
Odd $q$ implies $t_{\text{new}}(t,x)\neq t_{\text{new}}(t,x+L/2)$ and their interior lengths are no longer identical; see Fig.~\ref{fig:geodesic mismatch} for a schematic.

We end this appendix by comparing with the inhomogeneous boundary state quenches~\cite{Nozaki:2023fkx,Bai:2024azk,BaiNozakiMiyataMao_inprep}. Given that~\eqref{eq:Regularized-Bdy-State} is short-range entangled, early-time behavior is usually controlled by the disconnected extremized surface for both $A$ and $A\cup B$. Holding an endpoint at the subsystem boundary, the interior pieces of the two options match exactly, each with interior length~\eqref{eq:Interior-Mirror-Geodesic-Length-After-Quotient}. Therefore, the interior lengths of mirror-image geodesics and those of the disconnected extremized surface agree. In summary, the interior-length mismatch between~\eqref{eq:Interior-Mirror-Geodesic-Length-After-Quotient} and~\eqref{eq:Interior-Extremal-Geodesic-Length-After-Quotient} appears to be a joint consequence of antipodal pairing in the crosscap state and inhomogeneous time evolution, analogous to the graph-like pattern of Sec.~\ref{sec:QP-EP}. Unfortunately, the precise relation between the two phenomena remains unknown.

\csname phantomsection\endcsname%
\addcontentsline{toc}{section}{References}

\bibliographystyle{./phd_thesis_abbrv_v2.1_unsrt}
\bibliography{Draftnew.bib}

\begin{thebibliography}{100}

\bibitem{Calabrese:2005in}
P.~Calabrese and J.~L. Cardy.
\newblock {Evolution of entanglement entropy in one-dimensional systems}.
\newblock \href{http://dx.doi.org/10.1088/1742-5468/2005/04/P04010}{J. Stat. Mech.  {\bf0504}, P04010 (2005)}, \href{http://arxiv.org/abs/cond-mat/0503393}{{\ttfamily arXiv:cond-mat/0503393}}.

\bibitem{Cheneau_2012}
M.~Cheneau, P.~Barmettler, D.~Poletti, M.~Endres, P.~Schauß, T.~Fukuhara, C.~Gross, I.~Bloch, C.~Kollath, and S.~Kuhr.
\newblock Light-cone-like spreading of correlations in a quantum many-body system.
\newblock \href{http://dx.doi.org/10.1038/nature10748}{Nature  {\bf481}(7382), 484–487 (2012)}.

\bibitem{Langen_2015}
T.~Langen, S.~Erne, R.~Geiger, B.~Rauer, T.~Schweigler, M.~Kuhnert, W.~Rohringer, I.~E. Mazets, T.~Gasenzer, and J.~Schmiedmayer.
\newblock Experimental observation of a generalized gibbs ensemble.
\newblock \href{http://dx.doi.org/10.1126/science.1257026}{Science  {\bf348}(6231), 207–211 (2015)}.

\bibitem{Kaufman_2016}
A.~M. Kaufman, M.~E. Tai, A.~Lukin, M.~Rispoli, R.~Schittko, P.~M. Preiss, and M.~Greiner.
\newblock Quantum thermalization through entanglement in an isolated many-body system.
\newblock \href{http://dx.doi.org/10.1126/science.aaf6725}{Science  {\bf353}(6301), 794–800 (2016)}.

\bibitem{Brydges_2019}
T.~Brydges, A.~Elben, P.~Jurcevic, B.~Vermersch, C.~Maier, B.~P. Lanyon, P.~Zoller, R.~Blatt, and C.~F. Roos.
\newblock Probing rényi entanglement entropy via randomized measurements.
\newblock \href{http://dx.doi.org/10.1126/science.aau4963}{Science  {\bf364}(6437), 260–263 (2019)}.

\bibitem{PhysRevLett.124.063601}
V.~Borish, O.~Markovi\ifmmode~\acute{c}\else \'{c}\fi{}, J.~A. Hines, S.~V. Rajagopal, and M.~Schleier-Smith.
\newblock Transverse-field ising dynamics in a rydberg-dressed atomic gas.
\newblock \href{http://dx.doi.org/10.1103/PhysRevLett.124.063601}{Phys. Rev. Lett.  {\bf124}, 063601 (2020)}.

\bibitem{Tajik_2023}
M.~Tajik, M.~Gluza, N.~Sebe, P.~Schüttelkopf, F.~Cataldini, J.~Sabino, F.~Møller, S.-C. Ji, S.~Erne, G.~Guarnieri, S.~Sotiriadis, J.~Eisert, and J.~Schmiedmayer.
\newblock Experimental observation of curved light-cones in a quantum field simulator.
\newblock \href{http://dx.doi.org/10.1073/pnas.2301287120}{Proceedings of the National Academy of Sciences  {\bf120}(21) (2023)}.

\bibitem{Dubail_2017_inhCFT}
J.~Dubail, J.-M. Stéphan, J.~Viti, and P.~Calabrese.
\newblock Conformal field theory for inhomogeneous one-dimensional quantum systems: the example of non-interacting fermi gases.
\newblock \href{http://dx.doi.org/10.21468/scipostphys.2.1.002}{SciPost Physics  {\bf2}(1) (2017)}.

\bibitem{Allegra_2016}
N.~Allegra, J.~Dubail, J.-M. Stéphan, and J.~Viti.
\newblock Inhomogeneous field theory inside the arctic circle.
\newblock \href{http://dx.doi.org/10.1088/1742-5468/2016/05/053108}{Journal of Statistical Mechanics: Theory and Experiment  {\bf2016}(5), 053108 (2016)}.

\bibitem{Dubail_2017}
J.~Dubail, J.-M. Stéphan, and P.~Calabrese.
\newblock Emergence of curved light-cones in a class of inhomogeneous luttinger liquids.
\newblock \href{http://dx.doi.org/10.21468/scipostphys.3.3.019}{SciPost Physics  {\bf3}(3) (2017)}.

\bibitem{Gaw_dzki_2018}
K.~Gawędzki, E.~Langmann, and P.~Moosavi.
\newblock Finite-time universality in nonequilibrium cft.
\newblock \href{http://dx.doi.org/10.1007/s10955-018-2025-x}{Journal of Statistical Physics  {\bf172}(2), 353–378 (2018)}.

\bibitem{Wen_2018_SSD}
X.~Wen and J.-Q. Wu.
\newblock Quantum dynamics in sine-square deformed conformal field theory: Quench from uniform to nonuniform conformal field theory.
\newblock \href{http://dx.doi.org/10.1103/physrevb.97.184309}{Physical Review B  {\bf97}(18) (2018)}.

\bibitem{PhysRevLett.122.020201}
E.~Langmann and P.~Moosavi.
\newblock Diffusive heat waves in random conformal field theory.
\newblock \href{http://dx.doi.org/10.1103/PhysRevLett.122.020201}{Phys. Rev. Lett.  {\bf122}, 020201 (2019)}.

\bibitem{Lapierre:2019rwj}
B.~Lapierre, K.~Choo, C.~Tauber, A.~Tiwari, T.~Neupert, and R.~Chitra.
\newblock {Emergent black hole dynamics in critical Floquet systems}.
\newblock \href{http://dx.doi.org/10.1103/PhysRevResearch.2.023085}{Phys. Rev. Res.  {\bf2}(2), 023085 (2020)}, \href{http://arxiv.org/abs/1909.08618}{{\ttfamily arXiv:1909.08618 [cond-mat.str-el]}}.

\bibitem{cn3z-vfgr}
B.~Lapierre, T.~Numasawa, T.~Neupert, and S.~Ryu.
\newblock Floquet engineered inhomogeneous quantum chaos in critical systems.
\newblock \href{http://dx.doi.org/10.1103/cn3z-vfgr}{Phys. Rev. B  {\bf112}, 104317 (2025)}.

\bibitem{Wen:2020wee}
X.~Wen, R.~Fan, A.~Vishwanath, and Y.~Gu.
\newblock {Periodically, quasiperiodically, and randomly driven conformal field theories}.
\newblock \href{http://dx.doi.org/10.1103/PhysRevResearch.3.023044}{Phys. Rev. Res.  {\bf3}(2), 023044 (2021)}, \href{http://arxiv.org/abs/2006.10072}{{\ttfamily arXiv:2006.10072 [cond-mat.stat-mech]}}.

\bibitem{Moosavi_2021}
P.~Moosavi.
\newblock Inhomogeneous conformal field theory out of equilibrium.
\newblock \href{http://dx.doi.org/10.1007/s00023-021-01118-0}{Annales Henri Poincaré  {\bf25}(1), 1083–1122 (2021)}.

\bibitem{Wen:2021mlv}
X.~Wen, Y.~Gu, A.~Vishwanath, and R.~Fan.
\newblock {Periodically, Quasi-periodically, and Randomly Driven Conformal Field Theories (II): Furstenberg's Theorem and Exceptions to Heating Phases}.
\newblock \href{http://dx.doi.org/10.21468/SciPostPhys.13.4.082}{SciPost Phys.  {\bf13}(4), 082 (2022)}, \href{http://arxiv.org/abs/2109.10923}{{\ttfamily arXiv:2109.10923 [cond-mat.stat-mech]}}.

\bibitem{Wen:2018agb}
X.~Wen and J.-Q. Wu.
\newblock \href{http://dx.doi.org/10.48550/arXiv.1805.00031}{{Floquet conformal field theory}}.
\newblock arXiv:1805.00031 [cond-mat.str-el] (2018).

\bibitem{Fang:2025rie}
J.~Fang, Q.~Zhou, and X.~Wen.
\newblock {Phase transitions in quasiperiodically driven quantum critical systems: Analytical results}.
\newblock \href{http://dx.doi.org/10.1103/PhysRevB.111.094304}{Phys. Rev. B  {\bf111}(9), 094304 (2025)}, \href{http://arxiv.org/abs/2501.04795}{{\ttfamily arXiv:2501.04795 [cond-mat.stat-mech]}}.

\bibitem{Fan:2020orx}
R.~Fan, Y.~Gu, A.~Vishwanath, and X.~Wen.
\newblock {Floquet conformal field theories with generally deformed Hamiltonians}.
\newblock \href{http://dx.doi.org/10.21468/SciPostPhys.10.2.049}{SciPost Phys.  {\bf10}(2), 049 (2021)}, \href{http://arxiv.org/abs/2011.09491}{{\ttfamily arXiv:2011.09491 [hep-th]}}.

\bibitem{Liu:2023tiq}
X.~Liu, A.~McDonald, T.~Numasawa, B.~Lian, and S.~Ryu.
\newblock {Inhomogeneous Quantum Quenches of Conformal Field Theory with Boundaries}.
\newblock \href{http://dx.doi.org/10.1103/PhysRevLett.134.220404}{Phys. Rev. Lett.  {\bf134}(22), 220404 (2025)}, \href{http://arxiv.org/abs/2309.04540}{{\ttfamily arXiv:2309.04540 [cond-mat.stat-mech]}}.

\bibitem{chen2025symmetryresolvedentanglemententropy}
H.-H. Chen, X.-L. Zhou, J.~Yin, and M.~Zhang.
\newblock {Symmetry resolved entanglement entropy after an inhomogeneous quench}.
\newblock \href{http://dx.doi.org/10.1007/JHEP06(2025)224}{JHEP  {\bf06}, 224 (2025)}, \href{http://arxiv.org/abs/2504.14661}{{\ttfamily arXiv:2504.14661 [hep-th]}}.

\bibitem{goto2023spatialdeformationmanybodyquantum}
K.~Goto, T.~Guo, T.~Nosaka, M.~Nozaki, S.~Ryu, and K.~Tamaoka.
\newblock {Spatial deformation of many-body quantum chaotic systems and quantum information scrambling}.
\newblock \href{http://dx.doi.org/10.1103/PhysRevB.109.054301}{Phys. Rev. B  {\bf109}(5), 054301 (2024)}, \href{http://arxiv.org/abs/2305.01019}{{\ttfamily arXiv:2305.01019 [quant-ph]}}.

\bibitem{Fan:2019upv}
R.~Fan, Y.~Gu, A.~Vishwanath, and X.~Wen.
\newblock {Emergent Spatial Structure and Entanglement Localization in Floquet Conformal Field Theory}.
\newblock \href{http://dx.doi.org/10.1103/PhysRevX.10.031036}{Phys. Rev. X  {\bf10}(3), 031036 (2020)}, \href{http://arxiv.org/abs/1908.05289}{{\ttfamily arXiv:1908.05289 [cond-mat.str-el]}}.

\bibitem{Wen:2022pyj}
X.~Wen, R.~Fan, and A.~Vishwanath.
\newblock \href{http://dx.doi.org/10.48550/arXiv.2211.00040}{Floquet's refrigerator: Conformal cooling in driven quantum critical systems}.
\newblock arXiv:2211.00040 [cond-mat.str-el] (2022).

\bibitem{Bai:2024azk}
C.~Bai, A.~Miyata, and M.~Nozaki.
\newblock {Entanglement dynamics in 2d HCFTs on the curved background: the case of q-M{\"o}bius Hamiltonian}.
\newblock \href{http://dx.doi.org/10.1007/JHEP12(2024)208}{JHEP  {\bf12}, 208 (2024)}, \href{http://arxiv.org/abs/2408.06594}{{\ttfamily arXiv:2408.06594 [hep-th]}}.

\bibitem{Miyata:2024gvr}
A.~Miyata, M.~Nozaki, K.~Tamaoka, and M.~Watanabe.
\newblock {Hawking-Page and entanglement phase transition in 2d CFT on curved backgrounds}.
\newblock \href{http://dx.doi.org/10.1007/JHEP08(2024)190}{JHEP  {\bf08}, 190 (2024)}, \href{http://arxiv.org/abs/2406.06121}{{\ttfamily arXiv:2406.06121 [hep-th]}}.

\bibitem{Das:2024lra}
J.~Das and A.~Kundu.
\newblock {Flowery horizons {\&} bulk observers: sl$^{(q)}$(2,{\,}{\ensuremath{\mathbb{R}}}), drive in 2d holographic CFT}.
\newblock \href{http://dx.doi.org/10.1007/JHEP05(2025)035}{JHEP  {\bf05}, 035 (2025)}, \href{http://arxiv.org/abs/2412.18536}{{\ttfamily arXiv:2412.18536 [hep-th]}}.

\bibitem{deBoer:2023lrd}
J.~de~Boer, V.~Godet, J.~Kastikainen, and E.~Keski-Vakkuri.
\newblock {Quantum information geometry of driven CFTs}.
\newblock \href{http://dx.doi.org/10.1007/JHEP09(2023)087}{JHEP  {\bf09}, 087 (2023)}, \href{http://arxiv.org/abs/2306.00099}{{\ttfamily arXiv:2306.00099 [hep-th]}}.

\bibitem{Nozaki:2023fkx}
M.~Nozaki, K.~Tamaoka, and M.~T. Tan.
\newblock {Inhomogeneous quenches as state preparation in two-dimensional conformal field theories}.
\newblock \href{http://dx.doi.org/10.1103/PhysRevD.109.126014}{Phys. Rev. D  {\bf109}(12), 126014 (2024)}, \href{http://arxiv.org/abs/2310.19376}{{\ttfamily arXiv:2310.19376 [hep-th]}}.

\bibitem{Goto:2021sqx}
K.~Goto, M.~Nozaki, S.~Ryu, K.~Tamaoka, and M.~T. Tan.
\newblock {Non-equilibrating a black hole with inhomogeneous quantum quench}.
\newblock \href{http://dx.doi.org/10.1007/JHEP08(2025)186}{JHEP  {\bf08}, 186 (2025)}, \href{http://arxiv.org/abs/2112.14388}{{\ttfamily arXiv:2112.14388 [hep-th]}}.

\bibitem{Goto:2023wai}
K.~Goto, M.~Nozaki, S.~Ryu, K.~Tamaoka, and M.~T. Tan.
\newblock {Scrambling and recovery of quantum information in inhomogeneous quenches in two-dimensional conformal field theories}.
\newblock \href{http://dx.doi.org/10.1103/PhysRevResearch.6.023001}{Phys. Rev. Res.  {\bf6}(2), 023001 (2024)}, \href{http://arxiv.org/abs/2302.08009}{{\ttfamily arXiv:2302.08009 [hep-th]}}.

\bibitem{Mao:2025hkp}
W.~Mao and M.~Nozaki.
\newblock \href{http://dx.doi.org/10.48550/arXiv.2508.07645}{{Time Ordering Effects and Destruction of Quasiparticles in Two-dimensional Holographic CFTs}}.
\newblock arXiv:2508.07645 [hep-th] (2025).

\bibitem{Mao:2024cnm}
W.~Mao, M.~Nozaki, K.~Tamaoka, and M.~T. Tan.
\newblock {Local operator quench induced by two-dimensional inhomogeneous and homogeneous CFT Hamiltonians}.
\newblock \href{http://dx.doi.org/10.1007/JHEP07(2024)200}{JHEP  {\bf07}, 200 (2024)}, \href{http://arxiv.org/abs/2403.15851}{{\ttfamily arXiv:2403.15851 [hep-th]}}.

\bibitem{Das:2023xaw}
D.~Das, S.~R. Das, A.~Kundu, and K.~Sengupta.
\newblock {Exactly solvable floquet dynamics for conformal field theories in dimensions greater than two}.
\newblock \href{http://dx.doi.org/10.1007/JHEP09(2024)095}{JHEP  {\bf09}, 095 (2024)}, \href{http://arxiv.org/abs/2311.13468}{{\ttfamily arXiv:2311.13468 [hep-th]}}.

\bibitem{bernamonti2024boundaryinducedtransitionsmobiusquenches}
A.~Bernamonti, F.~Galli, and D.~Ge.
\newblock {Boundary-induced transitions in M{\"o}bius quenches of holographic BCFT}.
\newblock \href{http://dx.doi.org/10.1007/JHEP06(2024)184}{JHEP  {\bf06}, 184 (2024)}, \href{http://arxiv.org/abs/2402.16555}{{\ttfamily arXiv:2402.16555 [hep-th]}}.

\bibitem{erdmenger2025driveninhomogeneouscfttheory}
J.~Erdmenger, J.~Kastikainen, and T.~Schuhmann.
\newblock \href{http://dx.doi.org/10.48550/arXiv.2508.18350}{{Driven inhomogeneous CFT as a theory in curved space-time}}.
\newblock arXiv:2508.18350 [hep-th] (2025).

\bibitem{de_Boer_2022}
J.~de~Boer, R.~Espíndola, B.~Najian, D.~Patramanis, J.~van~der Heijden, and C.~Zukowski.
\newblock Virasoro entanglement berry phases.
\newblock \href{http://dx.doi.org/10.1007/jhep03(2022)179}{Journal of High Energy Physics  {\bf2022}(3) (2022)}.

\bibitem{Blumenhagen:2009zz}
R.~Blumenhagen and E.~Plauschinn.
\newblock \href{http://dx.doi.org/10.1007/978-3-642-00450-6}{{\em {Introduction to conformal field theory}: {with applications to String theory}}}, volume 779 (2009.
\newblock ).

\bibitem{DiFrancesco:1997nk}
P.~Di~Francesco, P.~Mathieu, and D.~Senechal.
\newblock \href{http://dx.doi.org/10.1007/978-1-4612-2256-9}{{\em {Conformal Field Theory}}}.
\newblock Graduate Texts in Contemporary Physics. Springer-Verlag, New York (1997).

\bibitem{Calabrese_2007}
P.~Calabrese and J.~Cardy.
\newblock Entanglement and correlation functions following a local quench: a conformal field theory approach.
\newblock \href{http://dx.doi.org/10.1088/1742-5468/2007/10/p10004}{Journal of Statistical Mechanics: Theory and Experiment  {\bf2007}(10), P10004–P10004 (2007)}.

\bibitem{PhysRevLett.112.111602}
M.~Nozaki, T.~Numasawa, and T.~Takayanagi.
\newblock Quantum entanglement of local operators in conformal field theories.
\newblock \href{http://dx.doi.org/10.1103/PhysRevLett.112.111602}{Phys. Rev. Lett.  {\bf112}, 111602 (2014)}.

\bibitem{Yoneta_2024}
Y.~Yoneta.
\newblock Thermal pure states for systems with antiunitary symmetries and their tensor network representations.
\newblock \href{http://dx.doi.org/10.1103/physrevresearch.6.l042062}{Physical Review Research  {\bf6}(4) (2024)}.

\bibitem{Chiba:2024dch}
Y.~Chiba and Y.~Yoneta.
\newblock {Exact Thermal Eigenstates of Nonintegrable Spin Chains at Infinite Temperature}.
\newblock \href{http://dx.doi.org/10.1103/PhysRevLett.133.170404}{Phys. Rev. Lett.  {\bf133}(17), 170404 (2024)}, \href{http://arxiv.org/abs/2403.12330}{{\ttfamily arXiv:2403.12330 [cond-mat.stat-mech]}}.

\bibitem{mestyán2025crosscapstatestunableentanglement}
M.~Mestyán and B.~Pozsgay.
\newblock \href{http://dx.doi.org/10.48550/arXiv.2503.15640}{{Crosscap states with tunable entanglement as exact eigenstates of local spin chain Hamiltonians}}.
\newblock arXiv:2503.15640 [cond-mat.stat-mech] (2025).

\bibitem{Caetano:2021dbh}
J.~Caetano and S.~Komatsu.
\newblock {Crosscap States in Integrable Field Theories and Spin Chains}.
\newblock \href{http://dx.doi.org/10.1007/s10955-022-02914-6}{J. Statist. Phys.  {\bf187}(3), 30 (2022)}, \href{http://arxiv.org/abs/2111.09901}{{\ttfamily arXiv:2111.09901 [hep-th]}}.

\bibitem{PhysRevLett.134.210403}
S.~Mohapatra, S.~Moudgalya, and A.~C. Balram.
\newblock Exact volume-law entangled zero-energy eigenstates in a large class of spin models.
\newblock \href{http://dx.doi.org/10.1103/PhysRevLett.134.210403}{Phys. Rev. Lett.  {\bf134}, 210403 (2025)}.

\bibitem{Mori_2018}
T.~Mori, T.~N. Ikeda, E.~Kaminishi, and M.~Ueda.
\newblock Thermalization and prethermalization in isolated quantum systems: a theoretical overview.
\newblock \href{http://dx.doi.org/10.1088/1361-6455/aabcdf}{Journal of Physics B: Atomic, Molecular and Optical Physics  {\bf51}(11), 112001 (2018)}.

\bibitem{Takahashi:1996zn}
Y.~Takahashi and H.~Umezawa.
\newblock {Thermo field dynamics}.
\newblock \href{http://dx.doi.org/10.1142/S0217979296000817}{Int. J. Mod. Phys. B  {\bf10}, 1755--1805 (1996)}.

\bibitem{Chalas:2024yts}
K.~Chalas, P.~Calabrese, and C.~Rylands.
\newblock \href{http://dx.doi.org/10.48550/arXiv.2412.04187}{{Quench dynamics of entanglement from crosscap states}}.
\newblock arXiv:2412.04187 [cond-mat.stat-mech] (2024).

\bibitem{Wei:2024kkp}
Z.~Wei and Y.~Yoneta.
\newblock \href{http://dx.doi.org/10.48550/arXiv.2412.18610}{{Crosscap Quenches and Entanglement Evolution}}.
\newblock arXiv:2412.18610 [hep-th] (2024).

\bibitem{Brown:1986nw}
J.~D. Brown and M.~Henneaux.
\newblock {Central Charges in the Canonical Realization of Asymptotic Symmetries: An Example from Three-Dimensional Gravity}.
\newblock \href{http://dx.doi.org/10.1007/BF01211590}{Commun. Math. Phys.  {\bf104}, 207--226 (1986)}.

\bibitem{Hartman_2014}
T.~Hartman, C.~A. Keller, and B.~Stoica.
\newblock Universal spectrum of 2d conformal field theory in the large c limit.
\newblock \href{http://dx.doi.org/10.1007/jhep09(2014)118}{Journal of High Energy Physics  {\bf2014}(9) (2014)}.

\bibitem{Hartman:2013qma}
T.~Hartman and J.~Maldacena.
\newblock {Time Evolution of Entanglement Entropy from Black Hole Interiors}.
\newblock \href{http://dx.doi.org/10.1007/JHEP05(2013)014}{JHEP  {\bf05}, 014 (2013)}, \href{http://arxiv.org/abs/1303.1080}{{\ttfamily arXiv:1303.1080 [hep-th]}}.

\bibitem{PhysRevLett.112.011601}
H.~Liu and S.~J. Suh.
\newblock Entanglement tsunami: Universal scaling in holographic thermalization.
\newblock \href{http://dx.doi.org/10.1103/PhysRevLett.112.011601}{Phys. Rev. Lett.  {\bf112}, 011601 (2014)}.

\bibitem{PhysRevX.7.031016}
A.~Nahum, J.~Ruhman, S.~Vijay, and J.~Haah.
\newblock Quantum entanglement growth under random unitary dynamics.
\newblock \href{http://dx.doi.org/10.1103/PhysRevX.7.031016}{Phys. Rev. X  {\bf7}, 031016 (2017)}.

\bibitem{jonay2018coarsegraineddynamicsoperatorstate}
C.~Jonay, D.~A. Huse, and A.~Nahum.
\newblock \href{http://dx.doi.org/10.48550/arXiv.1803.00089}{{Coarse-grained dynamics of operator and state entanglement}}.
\newblock arXiv:1803.00089 [cond-mat.stat-mech] (2018).

\bibitem{Mezei_2018}
M.~Mezei.
\newblock Membrane theory of entanglement dynamics from holography.
\newblock \href{http://dx.doi.org/10.1103/physrevd.98.106025}{Physical Review D  {\bf98}(10) (2018)}.

\bibitem{Zhou_2019}
T.~Zhou and A.~Nahum.
\newblock Emergent statistical mechanics of entanglement in random unitary circuits.
\newblock \href{http://dx.doi.org/10.1103/physrevb.99.174205}{Physical Review B  {\bf99}(17) (2019)}.

\bibitem{Kudler_Flam_2020}
J.~Kudler-Flam, M.~Nozaki, S.~Ryu, and M.~T. Tan.
\newblock Quantum vs. classical information: operator negativity as a probe of scrambling.
\newblock \href{http://dx.doi.org/10.1007/jhep01(2020)031}{Journal of High Energy Physics  {\bf2020}(1) (2020)}.

\bibitem{Cardy:1984bb}
J.~L. Cardy.
\newblock {Conformal Invariance and Surface Critical Behavior}.
\newblock \href{http://dx.doi.org/10.1016/0550-3213(84)90241-4}{Nucl. Phys. B  {\bf240}, 514--532 (1984)}.

\bibitem{Ishibashi:1988kg}
N.~Ishibashi.
\newblock {The Boundary and Crosscap States in Conformal Field Theories}.
\newblock \href{http://dx.doi.org/10.1142/S0217732389000320}{Mod. Phys. Lett. A  {\bf4}, 251 (1989)}.

\bibitem{Zhang:2024rnh}
Y.~Zhang, Y.-H. Wu, L.~Wang, and H.-H. Tu.
\newblock \href{http://dx.doi.org/10.48550/arXiv.2409.11046}{{Crosscap states and duality of Ising field theory in two dimensions}}.
\newblock arXiv:2409.11046 [cond-mat.str-el] (2024).

\bibitem{Calabrese:2016xau}
P.~Calabrese and J.~Cardy.
\newblock {Quantum quenches in 1 + 1 dimensional conformal field theories}.
\newblock \href{http://dx.doi.org/10.1088/1742-5468/2016/06/064003}{J. Stat. Mech.  {\bf1606}(6), 064003 (2016)}, \href{http://arxiv.org/abs/1603.02889}{{\ttfamily arXiv:1603.02889 [cond-mat.stat-mech]}}.

\bibitem{Calabrese:2009qy}
P.~Calabrese and J.~Cardy.
\newblock {Entanglement entropy and conformal field theory}.
\newblock \href{http://dx.doi.org/10.1088/1751-8113/42/50/504005}{J. Phys. A  {\bf42}, 504005 (2009)}, \href{http://arxiv.org/abs/0905.4013}{{\ttfamily arXiv:0905.4013 [cond-mat.stat-mech]}}.

\bibitem{Asplund:2015eha}
C.~T. Asplund, A.~Bernamonti, F.~Galli, and T.~Hartman.
\newblock {Entanglement Scrambling in 2d Conformal Field Theory}.
\newblock \href{http://dx.doi.org/10.1007/JHEP09(2015)110}{JHEP  {\bf09}, 110 (2015)}, \href{http://arxiv.org/abs/1506.03772}{{\ttfamily arXiv:1506.03772 [hep-th]}}.

\bibitem{Wei:2024zez}
Z.~Wei.
\newblock \href{http://dx.doi.org/10.48550/arXiv.2405.03755}{{Holographic Dual of Crosscap Conformal Field Theory}}.
\newblock arXiv:2405.03755 [hep-th] (2024).

\bibitem{Han:2020kwp}
B.~Han and X.~Wen.
\newblock {Classification of $SL_2$ deformed Floquet conformal field theories}.
\newblock \href{http://dx.doi.org/10.1103/PhysRevB.102.205125}{Phys. Rev. B  {\bf102}(20), 205125 (2020)}, \href{http://arxiv.org/abs/2008.01123}{{\ttfamily arXiv:2008.01123 [cond-mat.stat-mech]}}.

\bibitem{Ishibashi:2015jba}
N.~Ishibashi and T.~Tada.
\newblock {Infinite circumference limit of conformal field theory}.
\newblock \href{http://dx.doi.org/10.1088/1751-8113/48/31/315402}{J. Phys. A  {\bf48}(31), 315402 (2015)}, \href{http://arxiv.org/abs/1504.00138}{{\ttfamily arXiv:1504.00138 [hep-th]}}.

\bibitem{Ishibashi:2016bey}
N.~Ishibashi and T.~Tada.
\newblock {Dipolar quantization and the infinite circumference limit of two-dimensional conformal field theories}.
\newblock \href{http://dx.doi.org/10.1142/S0217751X16501700}{Int. J. Mod. Phys. A  {\bf31}(32), 1650170 (2016)}, \href{http://arxiv.org/abs/1602.01190}{{\ttfamily arXiv:1602.01190 [hep-th]}}.

\bibitem{Lapierre:2020ftq}
B.~Lapierre and P.~Moosavi.
\newblock {Geometric approach to inhomogeneous Floquet systems}.
\newblock \href{http://dx.doi.org/10.1103/PhysRevB.103.224303}{Phys. Rev. B  {\bf103}, 224303 (2021)}, \href{http://arxiv.org/abs/2010.11268}{{\ttfamily arXiv:2010.11268 [cond-mat.stat-mech]}}.

\bibitem{Goto:2021gve}
K.~Goto, A.~Mollabashi, M.~Nozaki, K.~Tamaoka, and M.~T. Tan.
\newblock {Information scrambling versus quantum revival through the lens of operator entanglement}.
\newblock \href{http://dx.doi.org/10.1007/JHEP06(2022)100}{JHEP  {\bf06}, 100 (2022)}, \href{http://arxiv.org/abs/2112.00802}{{\ttfamily arXiv:2112.00802 [hep-th]}}.

\bibitem{Jiang:2024hgt}
H.~Jiang and M.~Mezei.
\newblock {New horizons for inhomogeneous quenches and Floquet CFT}.
\newblock \href{http://dx.doi.org/10.1007/JHEP04(2025)025}{JHEP  {\bf04}, 025 (2025)}, \href{http://arxiv.org/abs/2404.07884}{{\ttfamily arXiv:2404.07884 [hep-th]}}.

\bibitem{Calabrese:2004eu}
P.~Calabrese and J.~L. Cardy.
\newblock {Entanglement entropy and quantum field theory}.
\newblock \href{http://dx.doi.org/10.1088/1742-5468/2004/06/P06002}{J. Stat. Mech.  {\bf0406}, P06002 (2004)}, \href{http://arxiv.org/abs/hep-th/0405152}{{\ttfamily arXiv:hep-th/0405152}}.

\bibitem{Cardy:2007mb}
J.~L. Cardy, O.~A. Castro-Alvaredo, and B.~Doyon.
\newblock {Form factors of branch-point twist fields in quantum integrable models and entanglement entropy}.
\newblock \href{http://dx.doi.org/10.1007/s10955-007-9422-x}{J. Statist. Phys.  {\bf130}, 129--168 (2008)}, \href{http://arxiv.org/abs/0706.3384}{{\ttfamily arXiv:0706.3384 [hep-th]}}.

\bibitem{Ryu:2006bv}
S.~Ryu and T.~Takayanagi.
\newblock {Holographic derivation of entanglement entropy from AdS/CFT}.
\newblock \href{http://dx.doi.org/10.1103/PhysRevLett.96.181602}{Phys. Rev. Lett.  {\bf96}, 181602 (2006)}, \href{http://arxiv.org/abs/hep-th/0603001}{{\ttfamily arXiv:hep-th/0603001}}.

\bibitem{Hubeny:2007xt}
V.~E. Hubeny, M.~Rangamani, and T.~Takayanagi.
\newblock {A Covariant holographic entanglement entropy proposal}.
\newblock \href{http://dx.doi.org/10.1088/1126-6708/2007/07/062}{JHEP  {\bf07}, 062 (2007)}, \href{http://arxiv.org/abs/0705.0016}{{\ttfamily arXiv:0705.0016 [hep-th]}}.

\bibitem{Tsiares:2020ewp}
I.~Tsiares.
\newblock {Universal dynamics in non-orientable CFT$_{2}$}.
\newblock \href{http://dx.doi.org/10.1007/JHEP09(2025)095}{JHEP  {\bf09}, 095 (2025)}, \href{http://arxiv.org/abs/2011.09250}{{\ttfamily arXiv:2011.09250 [hep-th]}}.

\bibitem{Miyaji:2014mca}
M.~Miyaji, S.~Ryu, T.~Takayanagi, and X.~Wen.
\newblock {Boundary States as Holographic Duals of Trivial Spacetimes}.
\newblock \href{http://dx.doi.org/10.1007/JHEP05(2015)152}{JHEP  {\bf05}, 152 (2015)}, \href{http://arxiv.org/abs/1412.6226}{{\ttfamily arXiv:1412.6226 [hep-th]}}.

\bibitem{Okunishi:2016zat}
K.~Okunishi.
\newblock {Sine-square deformation and M{\"o}bius quantization of 2D conformal field theory}.
\newblock \href{http://dx.doi.org/10.1093/ptep/ptw060}{PTEP  {\bf2016}(6), 063A02 (2016)}, \href{http://arxiv.org/abs/1603.09543}{{\ttfamily arXiv:1603.09543 [hep-th]}}.

\bibitem{Caputa:2021sib}
P.~Caputa, J.~M. Magan, and D.~Patramanis.
\newblock {Geometry of Krylov complexity}.
\newblock \href{http://dx.doi.org/10.1103/PhysRevResearch.4.013041}{Phys. Rev. Res.  {\bf4}(1), 013041 (2022)}, \href{http://arxiv.org/abs/2109.03824}{{\ttfamily arXiv:2109.03824 [hep-th]}}.

\bibitem{Lapierre:2025zsg}
B.~Lapierre, P.~Pelliconi, S.~Ryu, and J.~Sonner.
\newblock {Driven nonunitary dynamics of quantum critical systems}.
\newblock \href{http://dx.doi.org/10.1103/lwrz-jxrr}{Phys. Rev. B  {\bf112}(10), 104322 (2025)}, \href{http://arxiv.org/abs/2505.01508}{{\ttfamily arXiv:2505.01508 [cond-mat.str-el]}}.

\bibitem{GodsilRoyle2001AGT}
C.~Godsil and G.~F. Royle.
\newblock \href{http://dx.doi.org/10.1007/978-1-4613-0163-9}{{\em Algebraic Graph Theory}}, volume 207 of {\em Graduate Texts in Mathematics}.
\newblock Springer, New York, NY (2001).
\newblock 1st edition.

\bibitem{Diestel2025GraphTheory}
R.~Diestel.
\newblock \href{http://dx.doi.org/10.1007/978-3-662-70107-2}{{\em Graph Theory}}, volume 173 of {\em Graduate Texts in Mathematics}.
\newblock Springer, Berlin, Heidelberg, 6 edition (2025).
\newblock eBook published 20 Jan 2025; Hardcover ISBN 978-3-662-70106-5.

\bibitem{Louko:1998hc}
J.~Louko and D.~Marolf.
\newblock {Single exterior black holes and the AdS / CFT conjecture}.
\newblock \href{http://dx.doi.org/10.1103/PhysRevD.59.066002}{Phys. Rev. D  {\bf59}, 066002 (1999)}, \href{http://arxiv.org/abs/hep-th/9808081}{{\ttfamily arXiv:hep-th/9808081}}.

\bibitem{Maldacena:2001kr}
J.~M. Maldacena.
\newblock {Eternal black holes in anti-de Sitter}.
\newblock \href{http://dx.doi.org/10.1088/1126-6708/2003/04/021}{JHEP  {\bf04}, 021 (2003)}, \href{http://arxiv.org/abs/hep-th/0106112}{{\ttfamily arXiv:hep-th/0106112}}.

\bibitem{Maloney:2016gsg}
A.~Maloney and S.~F. Ross.
\newblock {Holography on Non-Orientable Surfaces}.
\newblock \href{http://dx.doi.org/10.1088/0264-9381/33/18/185006}{Class. Quant. Grav.  {\bf33}(18), 185006 (2016)}, \href{http://arxiv.org/abs/1603.04426}{{\ttfamily arXiv:1603.04426 [hep-th]}}.

\bibitem{Maldacena:1997re}
J.~M. Maldacena.
\newblock {The Large $N$ limit of superconformal field theories and supergravity}.
\newblock \href{http://dx.doi.org/10.4310/ATMP.1998.v2.n2.a1}{Adv. Theor. Math. Phys.  {\bf2}, 231--252 (1998)}, \href{http://arxiv.org/abs/hep-th/9711200}{{\ttfamily arXiv:hep-th/9711200}}.

\bibitem{Harada:2025uhh}
W.~Harada, J.~Kaidi, Y.~Kusuki, and Y.~Liu.
\newblock \href{http://dx.doi.org/10.48550/arXiv.2508.18357}{{New Crosscap States}}.
\newblock arXiv:2508.18357 [hep-th] (2025).

\bibitem{PhysRevB.105.L140301}
A.~Pocklington, Y.-X. Wang, Y.~Yanay, and A.~A. Clerk.
\newblock Stabilizing volume-law entangled states of fermions and qubits using local dissipation.
\newblock \href{http://dx.doi.org/10.1103/PhysRevB.105.L140301}{Phys. Rev. B  {\bf105}, L140301 (2022)}.

\bibitem{Witten:1987ty}
E.~Witten.
\newblock {Coadjoint Orbits of the Virasoro Group}.
\newblock \href{http://dx.doi.org/10.1007/BF01218287}{Commun. Math. Phys.  {\bf114}, 1 (1988)}.

\bibitem{Li:2025rzl}
Z.~Li, Z.~Li, and J.~Tian.
\newblock {The holography of the 2D inhomogeneously deformed CFT}.
\newblock \href{http://dx.doi.org/10.1007/JHEP05(2025)161}{JHEP  {\bf05}, 161 (2025)}, \href{http://arxiv.org/abs/2502.11135}{{\ttfamily arXiv:2502.11135 [hep-th]}}.

\bibitem{Chen:2025wzd}
H.-H. Chen.
\newblock \href{http://dx.doi.org/10.48550/arXiv.2510.05346}{{Exact Quench Dynamics from Thermal Pure Quantum States}}.
\newblock arXiv:2510.05346 [cond-mat.stat-mech] (2025).

\bibitem{Takayanagi:2010wp}
T.~Takayanagi and T.~Ugajin.
\newblock {Measuring Black Hole Formations by Entanglement Entropy via Coarse-Graining}.
\newblock \href{http://dx.doi.org/10.1007/JHEP11(2010)054}{JHEP  {\bf11}, 054 (2010)}, \href{http://arxiv.org/abs/1008.3439}{{\ttfamily arXiv:1008.3439 [hep-th]}}.

\bibitem{Senechal:1999us}
D.~Senechal.
\newblock {An Introduction to bosonization}.
\newblock In {\em {CRM Workshop on Theoretical Methods for Strongly Correlated Fermions}} (1999).

\bibitem{Polchinski:1998rq}
J.~Polchinski.
\newblock \href{http://dx.doi.org/10.1017/CBO9780511816079}{{\em {String theory. Vol. 1: An introduction to the bosonic string}}}.
\newblock Cambridge Monographs on Mathematical Physics. Cambridge University Press (2007).

\bibitem{Tan:2024dcd}
B.-Y. Tan, Y.~Zhang, H.-C. Zhang, W.~Tang, L.~Wang, H.-H. Tu, and Y.-H. Wu.
\newblock {Extracting the Luttinger Parameter from a Single Wave Function}.
\newblock \href{http://dx.doi.org/10.1103/PhysRevLett.134.076501}{Phys. Rev. Lett.  {\bf134}(7), 076501 (2025)}, \href{http://arxiv.org/abs/2402.18364}{{\ttfamily arXiv:2402.18364 [cond-mat.str-el]}}.

\bibitem{Pathak:2024cpo}
S.~Pathak and L.~K. Kovalsky.
\newblock {Non-orientable AdS$_{3}$ and its holography}.
\newblock \href{http://dx.doi.org/10.1007/JHEP06(2025)141}{JHEP  {\bf06}, 141 (2025)}, \href{http://arxiv.org/abs/2404.12546}{{\ttfamily arXiv:2404.12546 [hep-th]}}.

\bibitem{Maxfield:2014kra}
H.~Maxfield.
\newblock {Entanglement entropy in three dimensional gravity}.
\newblock \href{http://dx.doi.org/10.1007/JHEP04(2015)031}{JHEP  {\bf04}, 031 (2015)}, \href{http://arxiv.org/abs/1412.0687}{{\ttfamily arXiv:1412.0687 [hep-th]}}.

\bibitem{Maldacena:2013xja}
J.~Maldacena and L.~Susskind.
\newblock {Cool horizons for entangled black holes}.
\newblock \href{http://dx.doi.org/10.1002/prop.201300020}{Fortsch. Phys.  {\bf61}, 781--811 (2013)}, \href{http://arxiv.org/abs/1306.0533}{{\ttfamily arXiv:1306.0533 [hep-th]}}.

\bibitem{Numasawa:2018grg}
T.~Numasawa.
\newblock {Holographic Complexity for disentangled states}.
\newblock \href{http://dx.doi.org/10.1093/ptep/ptz156}{PTEP  {\bf2020}(3), 033B02 (2020)}, \href{http://arxiv.org/abs/1811.03597}{{\ttfamily arXiv:1811.03597 [hep-th]}}.

\bibitem{BaiNozakiMiyataMao_inprep}
C.~Bai, M.~Nozaki, A.~Miyata, and W.~Mao.
\newblock Upcoming work.
\newblock in preparation.

\end{thebibliography}

\end{document}